\documentclass[11pt,a4paper]{article}
\usepackage{graphicx,amsmath,amssymb,slashed}
\usepackage{jheppub}

\newcommand{\nn}{\nonumber}
\newcommand{\bs}[1]{\boldsymbol{{#1}}}
\newcommand\sss{\scriptscriptstyle}
\newcommand{\kt}{k_T}

\preprint{\\[2mm]
 CERN-PH-TH/2013-148\\
 CP3-13-30\\
 NIKHEF-2013-020\\ 
 RECAPP-HRI-2013-015\\
 ZU-TH 13/13}

\title{A framework for Higgs characterisation}

\author[a]{P.~Artoisenet,}
\author[b]{P.~de Aquino,} 
\author[c]{F.~Demartin,} 
\author[d]{R.~Frederix,} 
\author[d,e]{S.~Frixione,} 
\author[c]{F.~Maltoni,}
\author[f]{M.~K.~Mandal,} 
\author[g]{P.~Mathews,}
\author[b]{K.~Mawatari,}
\author[h]{V.~Ravindran,} 
\author[g]{S.~Seth,} 
\author[i]{P.~Torrielli,} 
\author[c]{M.~Zaro}

\affiliation[a]{Nikhef Theory Group, Science Park 105, 1098 XG Amsterdam, 
 The Netherlands}
\affiliation[b]{Theoretische Natuurkunde and IIHE/ELEM, 
 Vrije Universiteit Brussel,\\
 and International Solvay Institutes, Pleinlaan 2, 
 B-1050 Brussels, Belgium}
\affiliation[c]{Centre for Cosmology, 
 Particle Physics and Phenomenology (CP3),\\
 Universit\'e Catholique de Louvain, B-1348 Louvain-la-Neuve, Belgium}
\affiliation[d]{PH Department, TH Unit, CERN, CH-1211 Geneva 23, Switzerland}
\affiliation[e]{ITPP, EPFL, CH-1015 Lausanne, Switzerland}
\affiliation[f]{ Regional Centre for Accelerator-based Particle Physics,\\
 Harish-Chandra Research Institute Chhatnag Road, Jhunsi, Allahabad 211019, 
 India}
\affiliation[g]{Saha Institute of Nuclear Physics, 1/AF Bidhan Nagar, 
 Kolkata 700 064, India}
\affiliation[h]{Institute of Mathematical Sciences, CIT Campus, Taramani, 
 Chennai-113, India}
\affiliation[i]{Institut f\"ur Theoretische Physik, Universit\"at Z\"urich, 
 Winterthurerstrasse 190, CH-8057 Z\"urich, Switzerland}

\abstract{We introduce a  framework, based on an effective field
  theory approach, that allows one to perform characterisation studies of the
  boson recently discovered at the LHC, for all the relevant channels and in a
  consistent, systematic and accurate way. The production and decay of such 
  a boson with various spin and parity assignments can be simulated by means
  of multi-parton, tree-level matrix elements and of next-to-leading order 
  QCD calculations, both matched with parton showers. Several sample 
  applications are presented which show, in particular, that
  beyond-leading-order effects in QCD have non-trivial phenomenological
  implications.
  }


\begin{document}

\maketitle

\section{Introduction}\label{sec:intro}

Any major discovery is the beginning of a new journey. With the luminosity
accumulated by the LHC the existence of a new boson with a mass of about
125 GeV has by now reached an overwhelming
evidence~\cite{Aad:2012tfa,Chatrchyan:2012ufa}. In addition, several
independent, yet preliminary, studies~\cite{atlas2013,cms2013} give rather
strong indications that the new particle is indeed a parity-even scalar, with
the properties predicted by the Standard Model
(SM)~\cite{Weinberg:1975gm}. Were this the case, we would have the first
evidence for the actual relevance of the Brout-Englert-Higgs
mechanism~\cite{Englert:1964et,Higgs:1964pj}, together with the discovery of
the first elementary scalar particle. Furthermore, and maybe even more far
reaching, this would mean that a genuinely different short-range interaction of
Yukawa type, i.e. not under the spell of a gauge principle, is at work in the
Universe.

Assessing beyond any reasonable doubt that the new boson {\em is} the 
scalar particle predicted by the SM is therefore an endeavour of 
utmost importance.

The questions to be addressed can be organised into two levels, assuming that
only one resonance has been observed. At the lowest level (let us call it
Level 1) there are the questions on the very nature of the new particle, 
such as what are its spin and parity. At the next level (Level 2) one must
investigate the interactions of such a resonance with SM particles. Several
approaches have been proposed to address questions at both levels, with
different degrees of observable/model dependence and generality.

The first possibility is that of focussing on specific processes and
observables, and of analysing their sensitivity to a given hypothesis, such as
the spin or the parity of the resonance. This approach has the advantage that
it can give useful indications to experimental analyses on the most sensitive
observables and, in some cases, can be made really model independent (see for
instance ref.~\cite{Choi:2012yg}).  On the other hand, it is normally limited
to predictions at the lowest order, and by construction it does not provide a
framework where all information regarding the resonance can be collected and
globally analysed.

A more general approach is that of writing all vertices which enter a given set
of processes (as for example the three-particle vertices in $pp \to X(J^P) \to
VV, f\bar f$ with $V=\gamma,W,Z$) in the most general form compatible with 
the desired symmetries (which implies non-SM, i.e. anomalous,
couplings~\cite{Gao:2010qx,Bolognesi:2012mm,Gainer:2013rxa}). A positive
aspect of this approach is that it gives the possibility to promote
couplings to form factors, since no assumptions are made on where new physics
might lie. 
Such form factors also allow one to reinstate unitarity in case
of need, at the price of introducing an explicit model dependence.  As a possible
shortcoming of this approach, higher orders in QCD and electroweak (EW) couplings are, 
in general, more difficult to include.  Finally, a plethora of new parameters are needed 
without the possibility of establishing any hierarchy among them.
 
A third very common and powerful approach is that of relying on an effective
field theory valid up to a scale $\Lambda$, that features only one new state
$X(J^P)$ at the EW scale $v$; furthermore, one assumes that any
other new (i.e., non-SM) state resides at scales larger than or equal to
$\Lambda$.  One can show that a theory of this kind is renormalizable order by
order in the $\sqrt{\hat s}/\Lambda$ expansion, that it can be systematically
improved by adding operators of higher dimensions and QCD/EW corrections, and
that in general it depends only on a few free parameters, since the gauge
symmetries and the hierarchy in terms of the number of canonical dimensions 
of the relevant operators drastically reduce the number of allowed terms. The
drawback of any effective field theory is that by construction it assumes no
new physics below $\Lambda$, and that it violates unitarity and loses
predictivity for $\sqrt{\hat s}>\Lambda$; still, it remains an exceedingly
viable approach to the problem of new-particle characterisation, in particular
to Level-2 questions, and as such it has been widely advocated in the context
of the Higgs discovery (see for example refs.~\cite{Hagiwara:1993qt,
  Giudice:2007fh,Gripaios:2009pe,Low:2009di,Morrissey:2009tf,
  Contino:2010mh,Espinosa:2010vn,Azatov:2012bz,Espinosa:2012ir,Ellis:2012rx,
  Low:2012rj,Montull:2012ik,Espinosa:2012im,Carmi:2012in,Corbett:2012dm,
  Ellis:2012hz,Passarino:2012cb,Corbett:2012ja,Cheung:2013kla,Falkowski:2013dza,Contino:2013kra}, and more
in general refs.~\cite{Buchmuller:1985jz,Grzadkowski:2010es}).

The goal of this work is that of presenting the implementation of
a simple effective-field lagrangian below the electroweak symmetry breaking
(EWSB) scale, devised with Level-1 issues in mind, yet perfectly suitable to
also address questions on the strength of the Higgs couplings.  
The framework we propose is minimal, yet it has the advantage of being
systematically improvable through the inclusion of higher-order corrections,
notably those coming from QCD; in the following, we shall amply exploit this
feature, and the opportunities for accurate simulations it provides, in the
context both of multi-parton tree-level ({\sc MadGraph}\,5) and of
next-to-leading order ({\sc aMC@NLO}) computations.

In a nutshell, our assumptions are simply that the resonance structure
observed in data corresponds to one bosonic state ($X(J^P)$ with $J=0$, $1$,
or $2$, and a mass of about $125$ GeV), and that no other new state below the
cutoff $\Lambda$ coupled to such a resonance exists. We also follow the
principle that any new physics is dominantly described by the lowest
dimensional operators.  This means, for example, that for the spin-0 $CP$-even
case (which corresponds to the SM scalar) we include all effects coming from
the set of dimension-six operators relevant to the Higgs three-point
couplings.\footnote{The extension of our effective Lagrangian to include operators generating 
new four-point interactions  is straightforward.} 
Given that our goal is that of providing a simulation framework
in terms of mass eigenstates, and consistently with the general guidelines
outlined above, we construct an effective lagrangian below the EWSB
scale, where $SU(2)_L \times U(1)_Y$ reduces to $U(1)_{EM}$; moreover, we do
not require $CP$ conservation, and we leave open the possibility that the new
boson might be a scalar with no definite $CP$ properties.

The paper is organised as follows. In sect.~\ref{sec:eft} we introduce the
effective lagrangians for spin 0, 1, and 2 which form the basis of this work,
and discuss in detail their characteristics. In sect.~\ref{sec:hosim} we deal
with the implementation of these lagrangians in {\sc FeynRules} and 
{\sc MadGraph\,5} and its subsequent validation, and with the
capabilities of the resulting framework for simulations that emphasise
accuracy, namely {\sc aMC@NLO} and tree-level matrix element/parton shower
merging (ME+PS). In sect.~\ref{sec:apps} we turn to give a few sample
applications: the high-energy behaviour of a spin-2 hypothesis with
non-universal couplings to SM particles, the effects of initial-state QCD
radiation on spin-correlation observables, and an application of the matrix
element method to the determination of the amount of $CP$-mixing of a spin-0
resonance. In sect.~\ref{sec:summary} we present our conclusions and give a
brief outlook on future prospects. Some technical details are collected in the
appendices.

\section{Effective lagrangian}\label{sec:eft}

Our effective field theory consists of the SM (except for the Higgs itself),
expressed through the physical degrees of freedom present below the EWSB
scale, plus a new bosonic state $X(J^P)$ with spin/parity assignments
$J^P=0^+$, $0^-$, $1^+$, $1^-$, and $2^+$. The new state can couple to SM
particles via interactions of the lowest possible dimensions. In addition,
the state $0^+$ is allowed to mix with the $0^-$ one, and can interact 
with SM particles with higher-dimensional operators beyond those of the
SM. Technically, the implementation of the lagrangian is performed following
the path outlined in ref.~\cite{Christensen:2009jx}, i.e., starting from {\sc
  FeynRules}~\cite{Christensen:2008py} by extending and completing an earlier
version of the model used in ref.~\cite{Englert:2012xt}. The model particle
content and its Feynman rules can be exported to any matrix element generator
in the UFO~\cite{Degrande:2011ua}. We dub it \emph{Higgs Characterisation
  model}; it is publicly available at the {\sc FeynRules} on-line
database~\cite{HCweb}.

The lagrangian of our model relevant to the physics of the 
boson $X(J^P)$ is written as follows:
\begin{align}
{\cal  L}_{HC,J} ={\cal  L}_{SM-H} + {\cal  L}_{J}\,,
\end{align}
where the first term on the r.h.s.~describes the SM degrees of freedom except
for the Higgs, and ${\cal L}_J$ contains the kinetic and interaction 
terms (with SM particles) of the new bosonic state.

\subsection{Spin 0}

The construction of the effective lagrangian for the spin-0 state is obtained
by requiring that the parametrisation: i) allows one to recover the SM case
easily; ii) has the possibility to include all possible interactions that are generated by
gauge-invariant dimension-six operators above the EW scale; iii) includes
$0^-$ state couplings typical of SUSY or of generic two-Higgs-doublet models
(2HDM); and iv) allows $CP$-mixing between $0^+$ and $0^-$ states (which we
parametrise in terms of an angle $\alpha$).   Let us comment on the second
requirement, which is an important one. Our aim is that of using a formulation
which is general enough to include all effects coming from dimension-six
operators invariant under $SU(2)_L \times U(1)_Y$, i.e. above the EW
scale. This results in a limited subset of all possible dimension-six
operators~\cite{Buchmuller:1985jz,Grzadkowski:2010es} that govern Higgs
interactions. In addition, as a first step, we limit ourselves to include the operators
that modify the three-point Higgs interactions. For the fermions, there is only one operator that
modifies the Yukawa interaction, e.g.~for the top
quark, ${\cal L}^{\rm dim=6}_Y = (\phi^\dagger \phi) Q_L \tilde \phi t_R$, 
where $Q_L$ is the $SU(2)_L$ doublet $(t_L, b_L)$. 
As far as the interactions to vector bosons are concerned, a larger number
of dimension-six operators can be written down; the framework we adopt
is general enough to account for them all, even though for practical reasons at this stage
the implementation includes only those affecting all possible three-point interactions
with exactly one Higgs field. We point out that, for a $CP$-even 
state, this parametrisation is in one-to-one correspondence with those of
refs.~\cite{Giudice:2007fh,Contino:2013kra} (see e.g. eq.~(3.46) of
ref.~\cite{Contino:2013kra}) not including the terms in ${\cal L}_{F_1}$ and ${\cal L}_{F_2}$ which modify 
four-point interactions, and equivalent to eq.~(3) of ref.~\cite{Corbett:2012dm}. 
For a $CP$-odd state this is equivalent  to eq.~(A.98) of ref.~\cite{Contino:2013kra}.

Let us start with the interaction lagrangian relevant to fermions which,
while being extremely simple, illustrates our philosophy well. 
Such a lagrangian is:
\begin{align}
 {\cal L}_0^f 
   = -\sum_{f=t,b,\tau}\bar\psi_f\big( c_{\alpha}\kappa_{\sss Hff}g_{\sss Hff}\,
        +i s_{\alpha}\kappa_{\sss Aff}g_{\sss Aff}\, \gamma_5 \big)
\psi_f X_0 \,,
\label{eq:0ff}
\end{align}
where we use the notation:
\begin{align}
c_\alpha\equiv \cos \alpha\,, \quad s_\alpha\equiv \sin \alpha\,,
\end{align}
and denote by $g_{Hff}=m_f/v$ $(g_{Aff}=m_f/v)$ the strength of the scalar 
(pseudoscalar) coupling in the SM (in a 2HDM with $\tan\beta=1$). We point 
out that the constants $\kappa_i$ can be taken real without any loss of
generality, except $\kappa_{H\partial W}$ in eq.~(\ref{eq:0vv}). For
simplicity, we have assumed that only the third-generation 
of fermions couple to the scalar state; extensions to the other families and
flavour-changing structures are trivial to implement, which can be directly 
done by users of {\sc FeynRules}.  As mentioned above, the interaction of
eq.~(\ref{eq:0ff}) can also parametrise the effects of a ${\cal L}^{\rm
  dim=6}_Y = (\phi^\dagger \phi) Q_L \tilde \phi t_R$ operator. Note 
also that all requirements listed above are satisfied at the
price of a small redundancy in the number of parameters. The SM is obtained
when $c_{\alpha}=1$ and $\kappa_{Hff}=1$. The pseudoscalar state of a type-II
$CP$-conserving 2HDM or SUSY is obtained by setting $s_{\alpha}=1$ and
$\kappa_{Aff}=\cot \beta$ or $\kappa_{Aff}=\tan \beta$ for up or down
components of the $SU(2)$ fermion doublet, respectively. The parametrisation
of $CP$ mixing is entirely realised in terms of the angle $\alpha$, i.e.
independently of the parameters $\kappa_i$, so that many interesting cases,
such as again $CP$-violation in generic 2HDM, can be covered.

The effective lagrangian for the interaction of scalar and pseudoscalar states
with vector bosons can be written as follows:
\begin{align}
 {\cal L}_0^V =\bigg\{&
  c_{\alpha}\kappa_{\rm SM}\big[\frac{1}{2}g_{\sss HZZ}\, Z_\mu Z^\mu 
                                +g_{\sss HWW}\, W^+_\mu W^{-\mu}\big] \nn\\
  &-\frac{1}{4}\big[c_{\alpha}\kappa_{\sss H\gamma\gamma}
 g_{\sss H\gamma\gamma} \, A_{\mu\nu}A^{\mu\nu}
        +s_{\alpha}\kappa_{\sss A\gamma\gamma}g_{ \sss A\gamma\gamma}\,
 A_{\mu\nu}\widetilde A^{\mu\nu}
 \big] \nn\\
  &-\frac{1}{2}\big[c_{\alpha}\kappa_{\sss HZ\gamma}g_{\sss HZ\gamma} \, 
 Z_{\mu\nu}A^{\mu\nu}
        +s_{\alpha}\kappa_{\sss AZ\gamma}g_{\sss AZ\gamma}\,Z_{\mu\nu}\widetilde A^{\mu\nu} \big] \nn\\
  &-\frac{1}{4}\big[c_{\alpha}\kappa_{\sss Hgg}g_{\sss Hgg} \, G_{\mu\nu}^aG^{a,\mu\nu}
        +s_{\alpha}\kappa_{\sss Agg}g_{\sss Agg}\,G_{\mu\nu}^a\widetilde G^{a,\mu\nu} \big] \nn\\
  &-\frac{1}{4}\frac{1}{\Lambda}\big[c_{\alpha}\kappa_{\sss HZZ} \, Z_{\mu\nu}Z^{\mu\nu}
        +s_{\alpha}\kappa_{\sss AZZ}\,Z_{\mu\nu}\widetilde Z^{\mu\nu} \big] \nn\\
  &-\frac{1}{2}\frac{1}{\Lambda}\big[c_{\alpha}\kappa_{\sss HWW} \, W^+_{\mu\nu}W^{-\mu\nu}
        +s_{\alpha}\kappa_{\sss AWW}\,W^+_{\mu\nu}\widetilde W^{-\mu\nu}\big] \nn\\ 
  &-\frac{1}{\Lambda}c_{\alpha} 
    \big[ \kappa_{\sss H\partial\gamma} \, Z_{\nu}\partial_{\mu}A^{\mu\nu}
         +\kappa_{\sss H\partial Z} \, Z_{\nu}\partial_{\mu}Z^{\mu\nu}
         +\big(\kappa_{\sss H\partial W} \, W_{\nu}^+\partial_{\mu}W^{-\mu\nu}+h.c.\big)
 \big]
 \bigg\} X_0\,,
 \label{eq:0vv}
\end{align}
where the (reduced) field strength tensors are defined as follows:
\begin{align}
 V_{\mu\nu} &=\partial_{\mu}V_{\nu}-\partial_{\nu}V_{\mu}\quad (V=A,Z,W^{\pm})\,, \\
 G_{\mu\nu}^a &=\partial_{\mu}^{}G_{\nu}^a-\partial_{\nu}^{}G_{\mu}^a
  +g_sf^{abc}G_{\mu}^bG_{\nu}^c\,,
\end{align}
and the dual tensor is:
\begin{align}
 \widetilde V_{\mu\nu} =\frac{1}{2}\epsilon_{\mu\nu\rho\sigma}V^{\rho\sigma}\,.
\end{align}
The parametrisation of the couplings to vectors follows the same principles as
that of the couplings to fermions. In particular, the mixing angle $\alpha$
allows for a completely general description of $CP$-mixed states. We stress
here that while in general in a given model $CP$ violation depends on the
whole set of possible interactions among the physical states and cannot be
established by looking only at a sub sector~\cite{CPV}, in our parametrisation
$\alpha\neq 0$ or $\alpha\neq\pi/2$ (and non-vanishing $\kappa_{Hff},
\kappa_{Aff},\kappa_{HVV},\kappa_{AVV}$) implies $CP$ violation.  This can be
easily understood by first noting that in eq.~(\ref{eq:0ff}) $\alpha \neq 0$
or $\alpha\neq\pi/2$ always leads to $CP$ violation and that the corresponding 
terms in eq.~(\ref{eq:0vv}) are generated via a fermion loop by the 
$X_0 ff$ interaction. The $CP$-odd analogues of the operators in the last 
line of eq.~(\ref{eq:0vv}) do vanish. 
 
In our implementation, the parameters listed in table~\ref{tab:param} can be
directly set by the user.  
The dimensionful couplings $g_{Xyy'}$ shown in table~\ref{tab:gXaa} are
set so as to reproduce a SM Higgs and a pseudoscalar one in a 2HDM with
$\tan\beta=1$. Note that
in this case we have chosen $v$ as a reference scale instead of $\Lambda$. The
main reason is simply that such operators appear at one-loop in the SM and
therefore their values are non-zero even in absence of new physics.
More precisely, the forms of the $g_{XVV'}$ couplings given in
table~\ref{tab:gXaa} are the same as those which are loop-induced in the SM,
when computed by retaining only the top-quark and the $W$ boson contributions
to the loops, and in the limit where their masses tend to infinity. These
settings are adopted essentially because of their extremely simple analytic
expressions (which, in fact, turn out to be excellent approximations for all
the true loop-induced form factors, except for $g_{HZ\gamma}$ one, which
underestimates the correct value of the full loop computation by a factor
slightly larger than two).   It is obvious that any generic value of these
couplings, and in particular those induced by mass or higher-order corrections
and by new-physics deviations from the SM predictions, can be accounted 
for by setting $\kappa_i \ne 1$.\footnote{Note, however, that for the sake of simplicity and to
normalize our results to the SM, we use $g^{\rm NLO}_{Hgg}=-\frac{\alpha_s}{3\pi} \left(1+ \frac{11}{4} \frac{\alpha_S}{\pi}\right)$
in our simulations at NLO in QCD (while no finite renormalisation is  needed for the pseudoscalar, $g^{\rm NLO}_{Agg}=g^{\rm LO}_{Hgg}$).}

\begin{table}
\center
\begin{tabular}{lll}
\hline
 parameter\hspace*{5mm} & reference value\hspace*{5mm} & description \\
\hline
 $\Lambda$ [GeV] & $10^3$ & cutoff scale \\
 $c_{\alpha}(\equiv \cos\alpha$) & 1 & mixing between $0^+$ and
	 $0^-$ \\
 $\kappa_i$ & 0 , 1 & dimensionless coupling parameter \\
\hline
\end{tabular}
\caption{Model parameters. }
\label{tab:param}
\end{table}

\begin{table}
\center
\begin{tabular}{rrccccc}
\hline
 $g_{Xyy'} \times v $ &\qquad & $ff$ & $ZZ/WW$ &$\gamma\gamma$ & $Z\gamma$ & $gg$ \\
\hline
 $H$ & & $m_f$ & $2m_{Z/W}^2$ & $47\alpha_{\rm EM}/18\pi$ & $ C (94 \cos^2\theta_W-13)/9\pi$ & $-\alpha_s/3\pi$ \\
 $A$ & & $m_f$ & 0        & $4\alpha_{\rm EM}/3\pi$ & $2 C (8\cos^2\theta_W-5)/3\pi$ & $\alpha_s/2\pi$\\ 
\hline
\end{tabular}
\caption{Values in units of $v$ taken by the couplings $g_{Xyy'}$. $C=\sqrt{\frac{\alpha_{\rm EM}G_F m_Z^2}{8\sqrt{2}\pi}}$.}
\label{tab:gXaa}
\end{table}

\subsection{Spin 1}

We now discuss how to build the most general interactions of a spin-1
resonance with SM particles.  One way to proceed would be that of assigning
$SU(2)_L \times U(1)_Y$ quantum numbers to the new vector, of writing all
possible operators up to dimension six with SM fields, and then of
re-expressing them in terms of the physical states below the EW scale,
following exactly the same procedure as was used for the scalars above. To be
fully general, however, one should consider different $SU(2)_L \times U(1)_Y$ 
gauge representations and mixings with the SM gauge bosons. 
A simpler approach is that of just writing the most general interactions at
the weak scale, and of considering only those with the lowest canonical
dimension. For simplicity we follow the latter approach.

The interaction lagrangian for the spin-1 boson with fermions is 
written as follows:
\begin{align}
 {\cal L}_1^f = \sum_{f=q,\ell} 
      \bar\psi_f \gamma_{\mu}(\kappa_{f_a}a_f - \kappa_{f_b}
           b_f\gamma_5)\psi_f X_{1}^{\mu}\,.
\end{align}
The $a_f$ and $b_f$ are the SM vector and
axial-vector couplings, i.e. for the quarks:
\begin{align} 
 a_u &= \frac{g}{2\cos\theta_W}\Big(\frac{1}{2}- \frac{4}{3} \sin^2\theta_W\Big)\,,\quad
 b_u = \frac{g}{2\cos\theta_W}\frac{1}{2}\,, \\
 a_d &= \frac{g}{2\cos\theta_W}\Big(-\frac{1}{2}+ \frac{2}{3} \sin^2\theta_W\Big)\,,\quad
 b_d = -\frac{g}{2\cos\theta_W}\frac{1}{2}\,,
\end{align}
and similarly for the leptons.
The most general $X_1WW$ interaction at the lowest dimension can be 
written as follows (see ref.~\cite{Hagiwara:1986vm}):
\begin{align}
 {\cal L}_1^W =
    &\,i\kappa_{\sss W_1}g_{\sss WWZ} (W^+_{\mu\nu}W^{- \mu} - W^-_{\mu\nu}W^{+\mu})
 X_{1}^{\nu} 
   + i\kappa_{\sss W_2}g_{\sss WWZ} W^+_\mu W^-_\nu X_{1}^{\mu\nu}  \nn\\
  &-\kappa_{\sss W_3} W^+_\mu W^-_\nu(\partial^\mu X_{1}^{\nu} +
 \partial^\nu X_{1}^{\mu})\nn\\
  &+ i\kappa_{\sss W_4} W^+_\mu W^-_\nu \widetilde X_{1}^{\mu\nu}
  - \kappa_{\sss W_5} \epsilon_{\mu\nu\rho\sigma}
  [{W^+}^\mu ({\partial}^\rho {W^-}^\nu)-({\partial}^\rho {W^+}^\mu){W^-}^\nu] X_{1}^{\sigma}\,,
\label{eq:1ww}
\end{align}
where $g_{WWZ}=-e\cot\theta_W$. Note, once again, that our effective field
theory description lives at energy scales where EW symmetry $SU(2)_L \times
U(1)_Y$ is broken to $U(1)_{EM}$. This approach does not require to specify
the transformation properties of $X_1$ with respect to the EW symmetry.  The
parametrisation above could also be used for describing $X_1Z\gamma$
interactions which, however, have not been implemented. In the case of $ZZ$,
Bose symmetry implies a reduction of the possible terms and the interaction
lagrangian reduces to~\cite{Hagiwara:1986vm,Keung:2008ve}:
\begin{align}
 {\cal L}_1^Z =
  -\kappa_{\sss Z_1} Z_{\mu\nu}Z^{\mu} X_{1}^{\nu}
  -\kappa_{\sss Z_3} X_{1}^{\mu}({\partial}^{\nu} Z_{\mu})Z_{\nu} 
  -\kappa_{\sss Z_5} \epsilon_{\mu\nu\rho\sigma}  X_{1}^{\mu}
  Z^{\nu} ({\partial}^\rho Z^{\sigma})\,. 
\label{eq:1zz-onshell}
\end{align}
The first term can be rewritten in terms of the second one plus a term that
vanishes if $\partial_\mu X_1^\mu =0$, which we do not assume (for example in
the SM $\partial_\mu Z^\mu \ne 0$ for non-vanishing fermion masses). No
effective lagrangian ${\cal L}_1^\gamma$ is introduced. Due to the Landau-Yang
theorem~\cite{Landau:1948kw,Yang:1950rg} no transition can occur between an
on-shell vector and two massless identical vectors.  However, for
completeness, we discuss the possibility of an off-shell spin-1 state
contributing to the $gg \to \gamma \gamma$ amplitude in appendix~\ref{app:spin1}.  Parity
conservation implies that for $X_1=1^-$
\begin{align}
 \kappa_{\sss f_b}= \kappa_{\sss V_4}=\kappa_{\sss V_5}= 0\,,
\end{align}
while for $X_1=1^+$ 
\begin{align}
 \kappa_{\sss f_a}=\kappa_{\sss V_1}=\kappa_{\sss V_2}=\kappa_{\sss V_3} = 0\,. 
\end{align}
Note that the conditions on $\kappa_{\sss V_2}$ and $\kappa_{\sss V_4}$
are trivial when $V=Z$ (see eq.~(\ref{eq:1zz-onshell})).

\subsection{Spin 2}

The interaction lagrangian for the spin-2 boson proceeds via the
energy-momentum (E-M) tensor of the SM fields and starts at
dimension five~\cite{Giudice:1998ck,Han:1998sg}. For a colour, weak and
electromagnetic singlet spin-2 resonance such an interaction is unique. For
the fermions we have
\begin{align}
 {\cal L}_2^f = -\frac{1}{\Lambda}\sum_{f=q,\ell} 
  \kappa_{f}\,T^f_{\mu\nu}X_2^{\mu\nu}\,,
\label{eq:grav}
\end{align}
and analogously for the vector bosons
\begin{align}
 {\cal L}_2^V = -\frac{1}{\Lambda}\sum_{V=Z,W,\gamma,g} 
  \kappa_{V}\,T^V_{\mu\nu}X_2^{\mu\nu}\,.
\end{align}
The coupling parameters $\kappa_f$ and $\kappa_V$ are 
introduced~\cite{Ellis:2012jv,Englert:2012xt} in full analogy with what
has been done in the spin-0 and -1 cases.
All of the E-M tensors $T^{f,V}_{\mu\nu}$ are given e.g. in
refs.~\cite{Han:1998sg,Hagiwara:2008jb}.  For the sake of later discussion, 
we explicitly present the E-M tensor for QED:
\begin{align}
  T^f_{\mu\nu}
  =&-g_{\mu\nu}
     \Big[ \bar{\psi}_f (i\gamma^\rho D_\rho-m_f)\psi_f
          -\frac{1}{2}\partial^\rho(\bar{\psi}_fi\gamma_\rho\psi_f)\Big]
 \nn\\
  &+\Big[ \frac{1}{2}\bar{\psi}_fi\gamma_\mu D_\nu\psi_f
               -\frac{1}{4}\partial_\mu(\bar{\psi}_fi\gamma_\nu\psi_f)
               +(\mu\leftrightarrow\nu)
          \Big]\,, \label{EMf}\\
 T^{\gamma}_{\mu\nu}
  =&-g_{\mu\nu}
      \Big[-\frac{1}{4}A^{\rho\sigma}A_{\rho\sigma}
           +\partial^\rho\partial^\sigma A_\sigma A_\rho
           +\frac{1}{2}(\partial^\rho A_\rho)^2\Big] \nn\\
  &-A^{\ \rho}_\mu A_{\nu\rho}
         +\partial_\mu\partial^\rho A_\rho A_\nu
         +\partial_\nu \partial^\rho A_\rho A_\mu\,, 
\label{EMa}
\end{align}
where $D_{\mu}=\partial_{\mu}-ieQ_fA_{\mu}$ and
$A_{\mu\nu}=\partial_{\mu}A_{\nu}-\partial_{\nu}A_{\mu}$.  For $X_2=2^+$ in
the minimal RS-like graviton scenario~\cite{Randall:1999ee}, i.e. the
universal coupling strength to the matter and gauge fields, the parameters
should be chosen as follows:
\begin{align}
\kappa_{f}=\kappa_{V} \qquad \forall \quad f,V\,.
\label{eq:RSspin2}
\end{align}

\section{Validation and comparisons}
\label{sec:hosim}

The implementation of the lagrangian ${\cal L}_{HC,J}$ in 
{\sc FeynRules}~\cite{Christensen:2008py} allows
the automated generation of the corresponding Feynman rules which can in turn
be exported to the {\sc MadGraph\,5} \cite{Alwall:2011uj} framework via the
{\sc UFO} model file~\cite{Degrande:2011ua,deAquino:2011ub}. This opens the
possibility of automatically creating event-generator codes for {\it any}
production and decay channel (including interferences between such two
mechanisms) at the tree level, which can then be used standalone (i.e.,
at the parton level) or interfaced with parton-shower MCs (ME+PS). 
The same automated generation can be achieved to NLO accuracy (where
the matching with showers is done according to the MC@NLO 
formalism~\cite{Frixione:2002ik}), with the present exception:
for a user-defined lagrangian, as is the case here, 
one-loop corrections in some cases have to be provided externally 
-- we shall give more details on this point in sect.~\ref{sec:HO}.

\subsection{Leading-order parton-level results}

We start by considering the most elementary type of predictions our approach
is capable of giving, namely those at the Born level without parton showers
(i.e., processes that do not feature any final-state particle either 
different from $X(J^P)$, or not resulting from the $X(J^P)$ decay).
Thus, this only involves the {\sc FeynRules} -- {\sc UFO} -- {\sc MadGraph\,5} 
chain, which by now has been applied to hundreds of processes and is therefore
extremely well tested. Still, it is appropriate to check the results
of the Higgs Characterisation model, in particular in view of other 
implementations available in the literature that aim at describing 
the same leading-order physics, and specifically that of 
JHU~\cite{Bolognesi:2012mm}. 

In table~\ref{tab:JHU} we give the choices of parameters to be made in 
order to obtain the benchmarks defined in ref.~\cite{Bolognesi:2012mm}.  For
all scenarios listed in that table we have found complete agreement in
the mass and angular distributions of the $X(J^P)$ decay products: 
$ZZ$, $WW$, and $\gamma\gamma$. For further studies
to be made here, we employ the process $pp \to X (\to ZZ^*) \to 4\ell$
to be definite, and we do not impose any final-state kinematical cuts.

\begin{table}
\center
\begin{tabular}{cccc}
\hline
JHU scenario &\hspace*{5mm}& \multicolumn{2}{c}{HC parameter choice} \\
&& $X$ production & $X$ decay \\
\hline
$0^+_m$ && $\kappa_{Hgg}\ne0$ & $\kappa_{\rm SM}\ne0\ (c_{\alpha}=1)$\\
$0^+_h$ && $\kappa_{Hgg}\ne0$ & $\kappa_{H\gamma\gamma,HZZ,HWW}\ne0\ (c_{\alpha}=1)$\\
$0^-$ && $\kappa_{Agg}\ne0$ &  $\kappa_{A\gamma\gamma,AZZ,AWW}\ne0\ (c_{\alpha}=0)$\\
$1^+$ && $\kappa_{f_a,f_b}\ne0$ & $\kappa_{Z_5,W_5}\ne0\ $ \\
$1^-$ && $\kappa_{f_a,f_b}\ne0$ & $\kappa_{Z_3,W_3}\ne0\ $\\
$2^+_m$ && $\kappa_{g}\ne0\ $ & $\kappa_{\gamma,Z,W}\ne0\ $\\
\hline 
\end{tabular}
\caption{Parameter correspondence to the benchmark scenarios defined in Table
  I of ref.~\cite{Bolognesi:2012mm}. In each scenario, the $\kappa_i$ 
couplings  that are not explicitly mentioned are understood to be 
equal to zero.}
\label{tab:JHU}
\end{table}

\begin{figure}
\center
 \includegraphics[width=0.40\textwidth,clip=true]{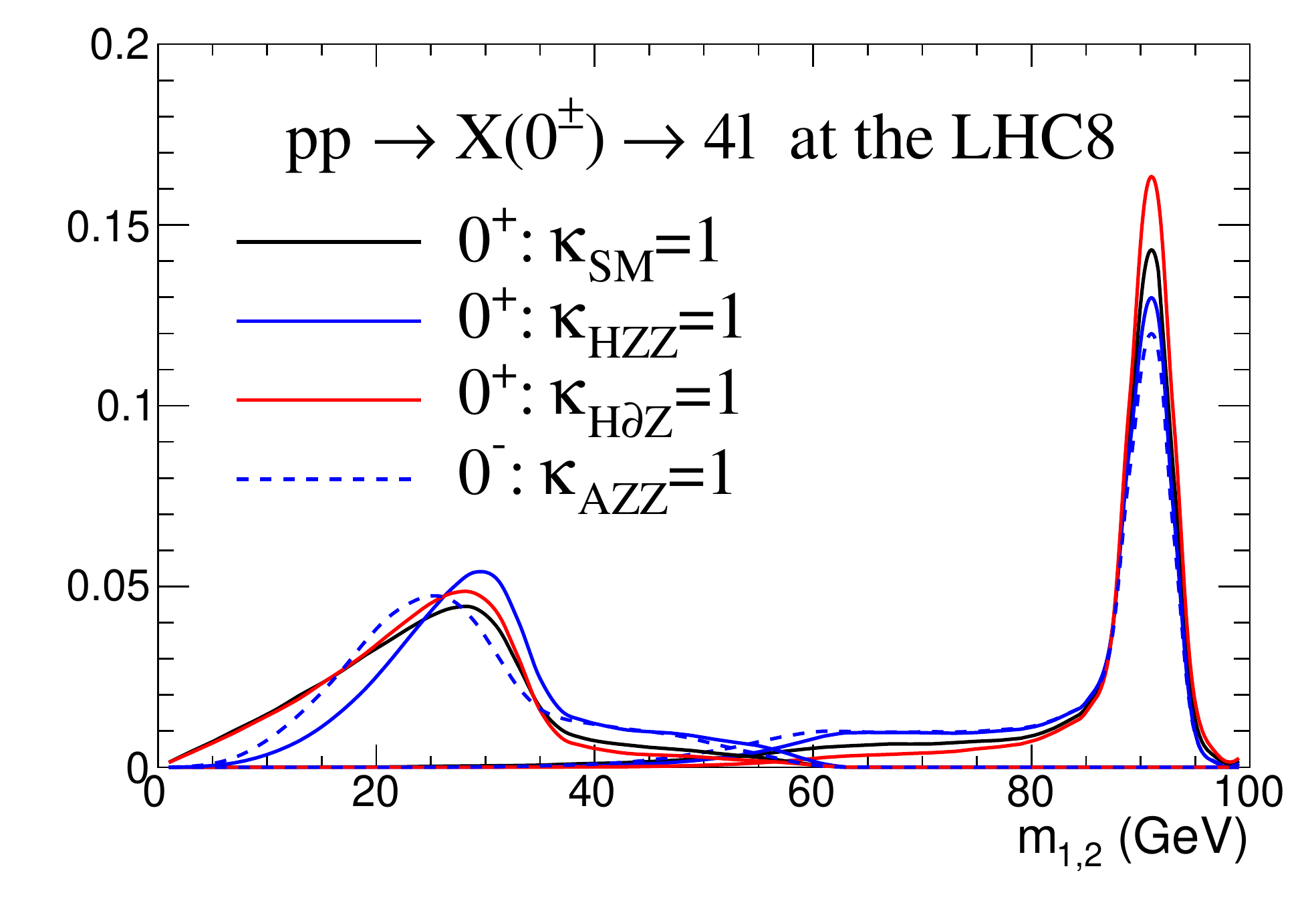}\quad
 \includegraphics[width=0.40\textwidth,clip=true]{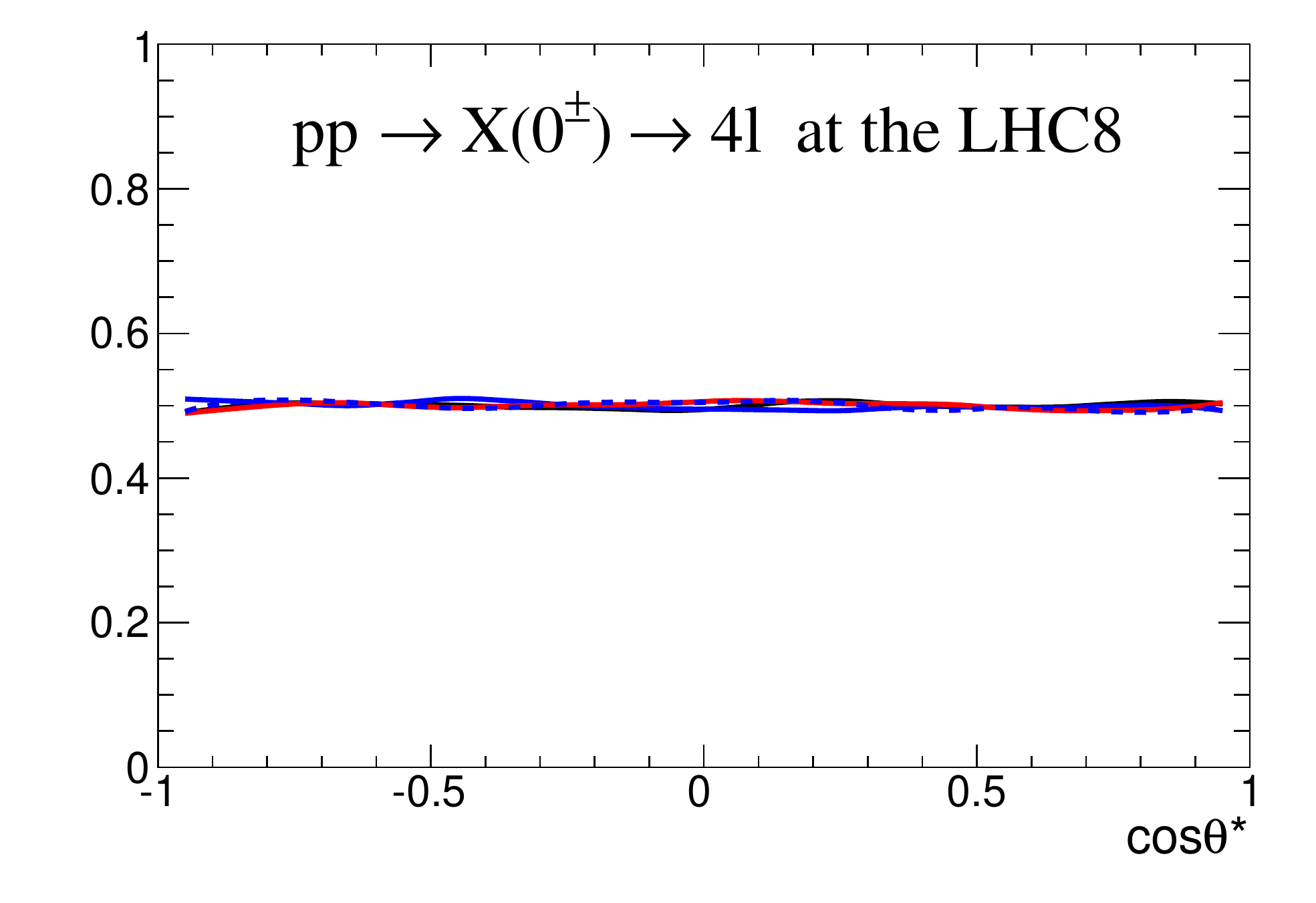}
 \includegraphics[width=0.40\textwidth,clip=true]{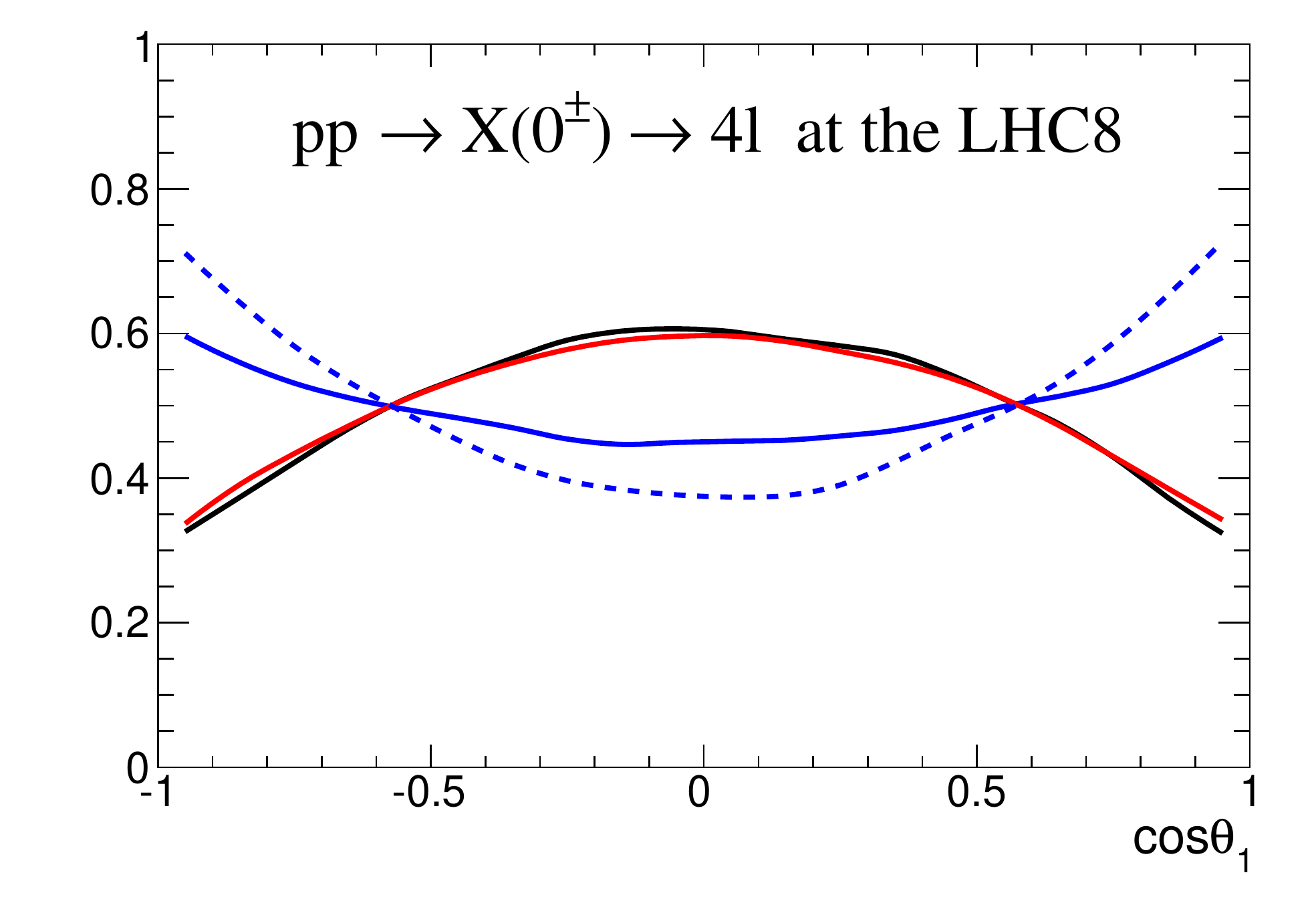}\quad
 \includegraphics[width=0.40\textwidth,clip=true]{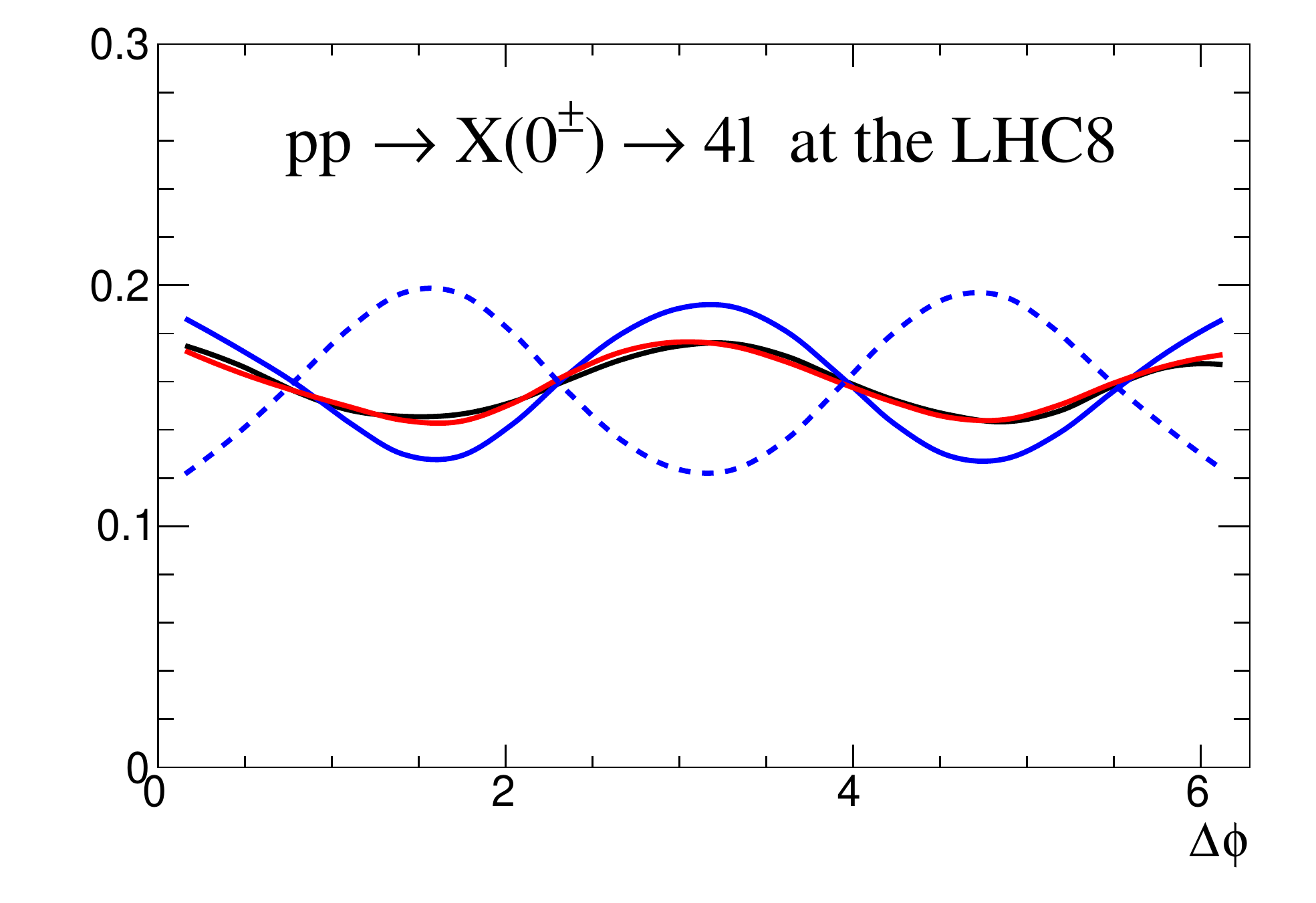}
 \caption{Normalised distributions in $pp\to X_0\to\mu^+\mu^-e^+e^-$ for
   different choices of $X_0ZZ$ couplings: the invariant masses of the two
   lepton pairs $m_1$, $m_2$ (with $m_1>m_2$), $\cos\theta^*$, $\cos\theta_1$,
   and $\Delta\phi$, as defined in ref.~\cite{Bolognesi:2012mm}. Event
   simulation performed at the leading order, parton level only (no
   shower/hadronisation).}
 \label{fig:x0decay_dis}
\end{figure}

We note that our $CP$-even spin-0 parametrisation also includes the so-called
``derivative operators" (last line of eq.~(\ref{eq:0vv})), that are absent in
the parametrisation of ref.~\cite{Bolognesi:2012mm}, and that give non-trivial
contributions to $X_0 \to VV$ decays. In fact, by using the equations of
motion, it can be easily seen that these operators can be related to contact
$X_0 V ff$ operators of the kind recently discussed in
refs.~\cite{Isidori:2013cla, Grinstein:2013vsa}. A representative set of
distributions for key spin-correlation observables is shown in
fig.~\ref{fig:x0decay_dis}.  One notices that the higher-dimensional operators
corresponding to $\kappa_{HZZ}$ ($CP$-even) and $\kappa_{AZZ}$ ($CP$-odd) have
dramatic effects on angular distributions, such as those of $\cos\theta_1$,
$\Delta\phi$, while the derivative operators corresponding to
$\kappa_{H\partial Z}$ only (mildly) affect the lepton invariant mass
distributions $m_1$ and $m_2$.

\begin{figure}
\center
 \includegraphics[width=0.40\textwidth,clip=true]{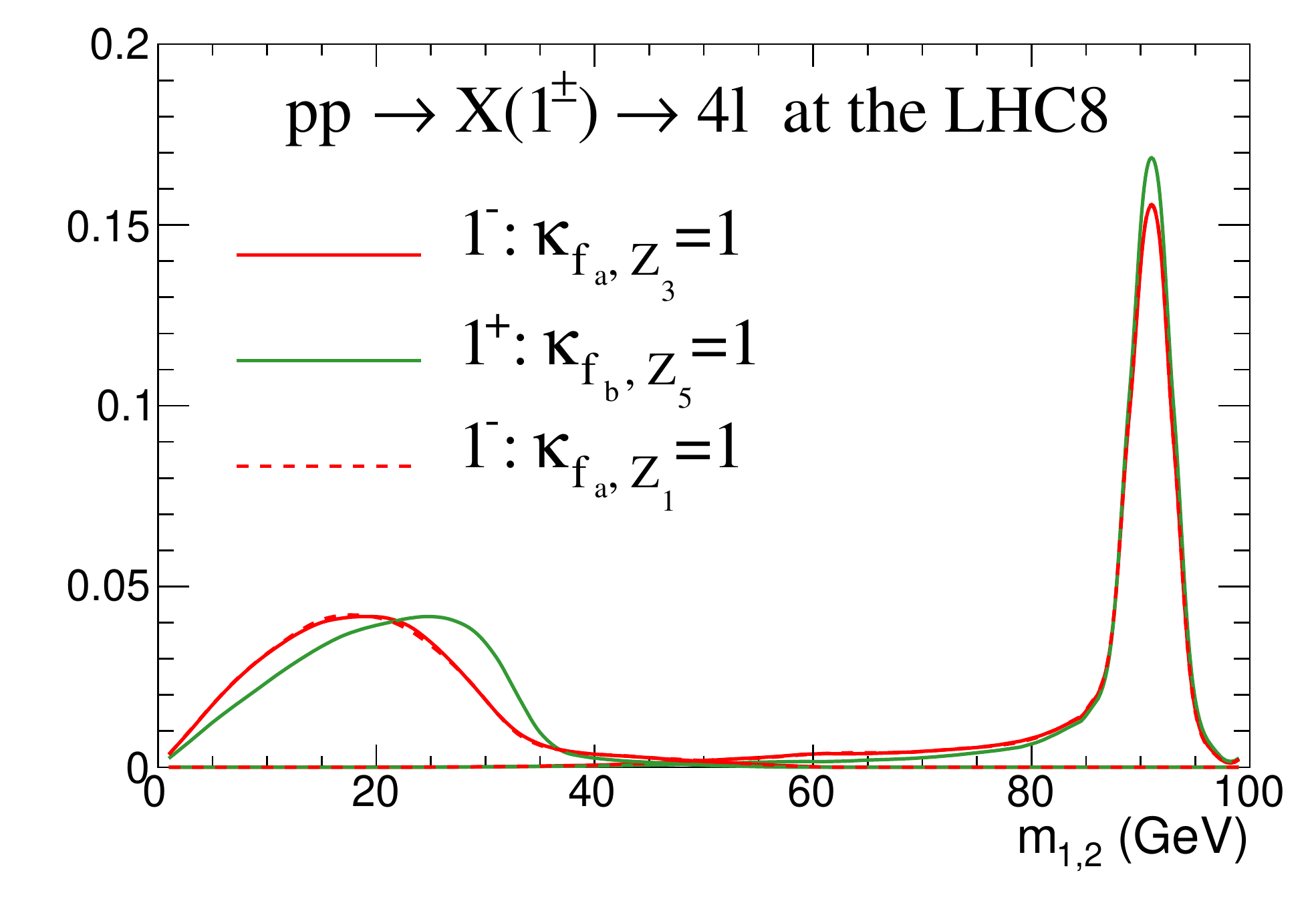}\quad
 \includegraphics[width=0.40\textwidth,clip=true]{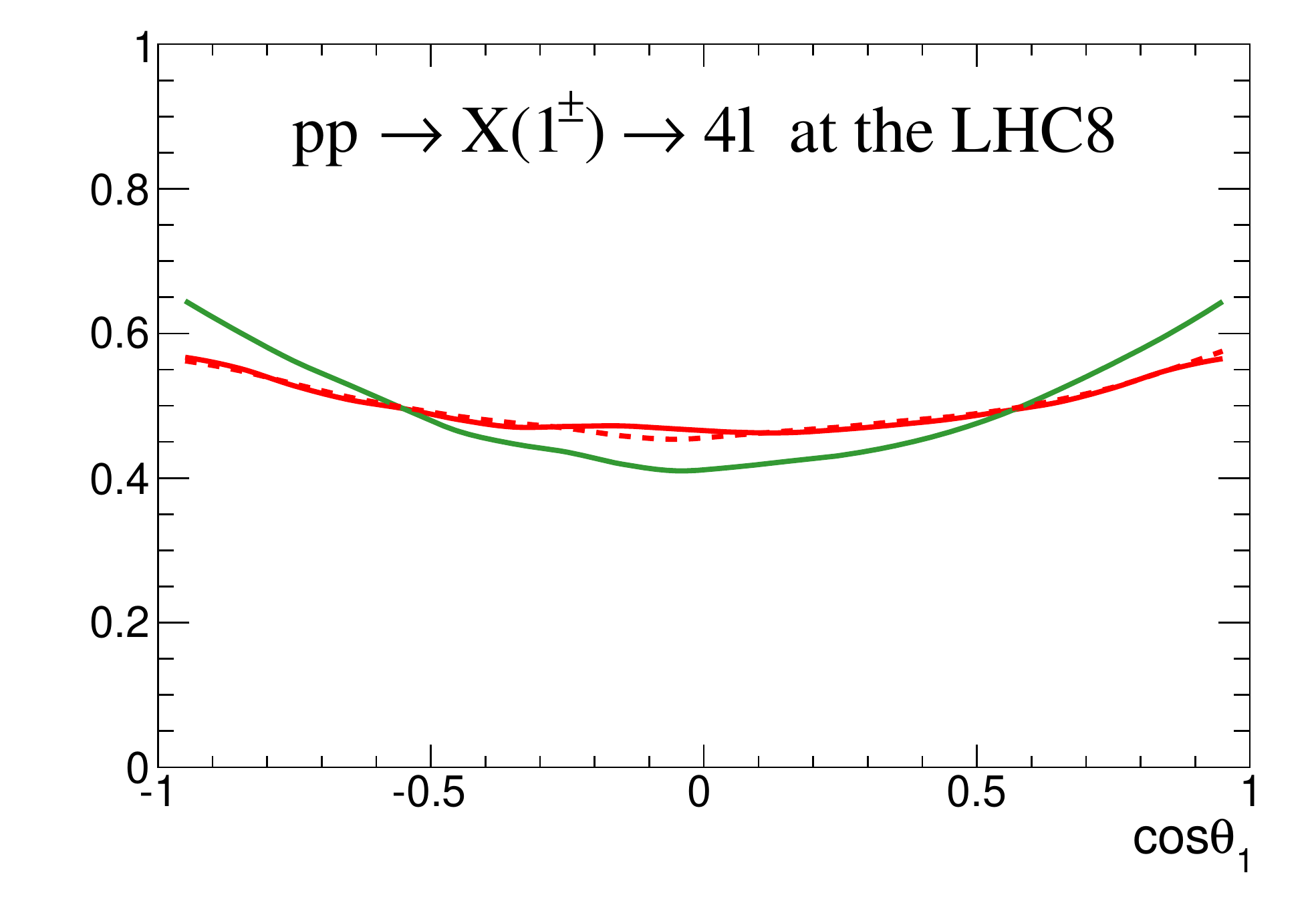}
 \includegraphics[width=0.40\textwidth,clip=true]{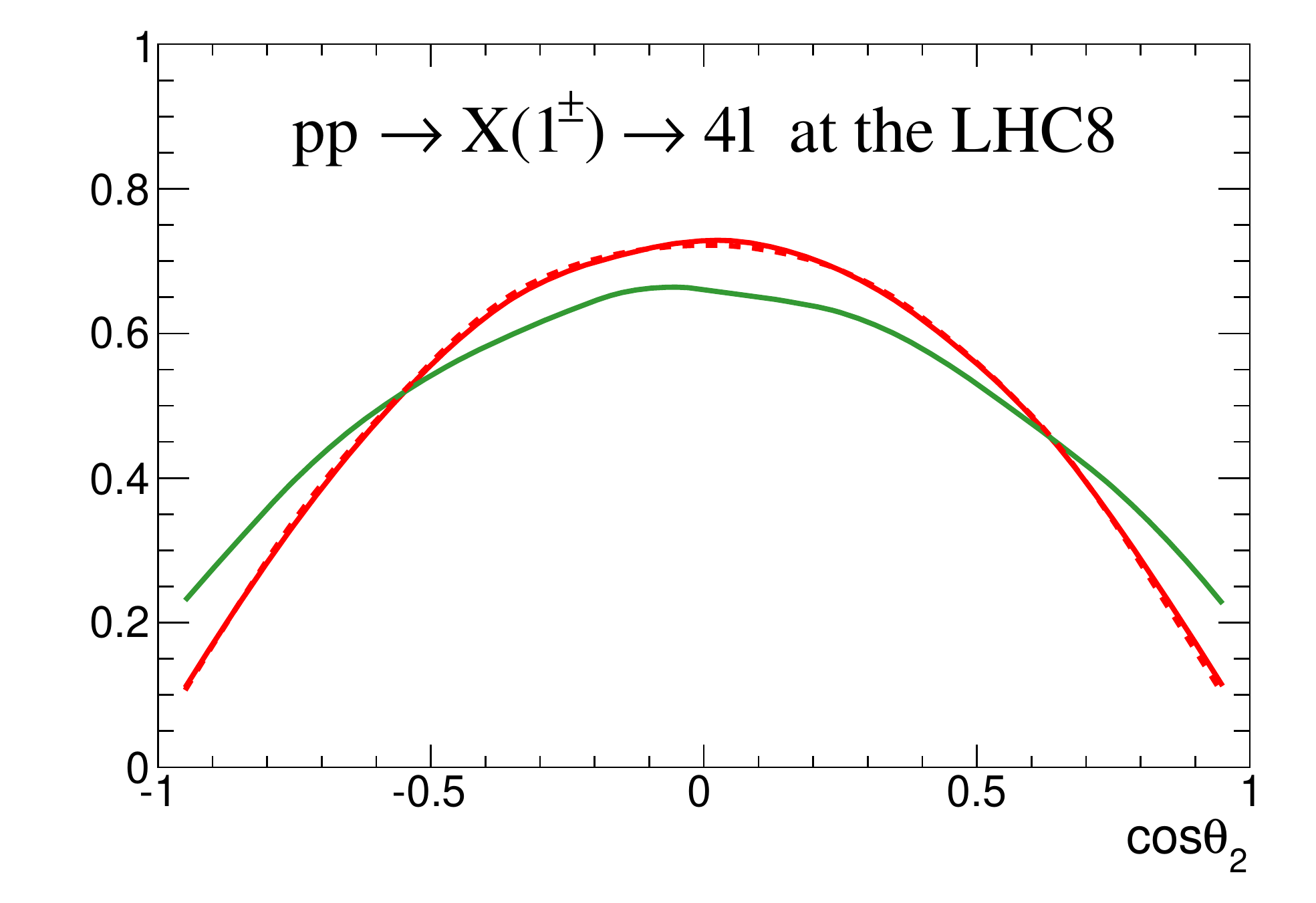}\quad
 \includegraphics[width=0.40\textwidth,clip=true]{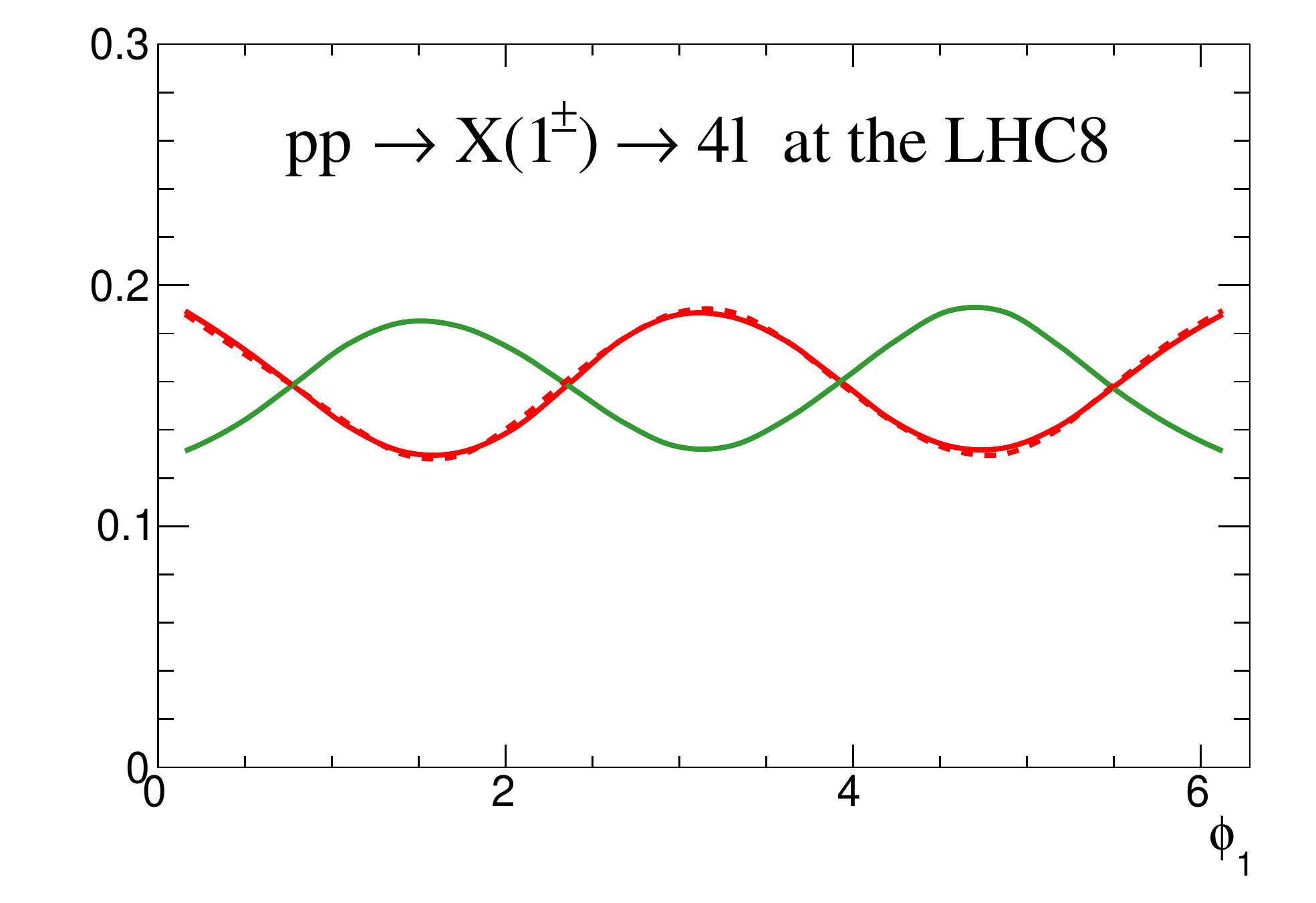}
 \caption{Normalised distributions in $pp\to X_1\to\mu^+\mu^-e^+e^-$
 for different choices
 of $X_1ZZ$ couplings:
 the invariant masses of the two lepton pairs $m_1$, $m_2$ (with $m_1>m_2$), 
 $\cos\theta_1$, $\cos\theta_2$, and $\phi_1$, as defined in 
 ref.~\cite{Bolognesi:2012mm}. Event simulation performed at the leading 
 order, parton level only (no shower/hadronisation). }
 \label{fig:x1decay_dis}
\end{figure}

For spin 1, we remark that the $X_1VV$ interactions defined in
ref.~\cite{Bolognesi:2012mm}  have one-to-one correspondence with the
$\kappa_{V_3}$ and $\kappa_{V_5}$ terms for both the $X_1WW$ and
$X_1ZZ$ cases. However, eqs.~(\ref{eq:1ww}) and (\ref{eq:1zz-onshell}) show
that in general the $X_1VV$ vertices can have a richer structure.
Sample distributions are shown in fig.~\ref{fig:x1decay_dis}
for $X_1\to ZZ$, where the difference between $1^+$ and $1^-$ are manifest.

\begin{figure}
\center
 \includegraphics[height=2.75cm,clip=true]{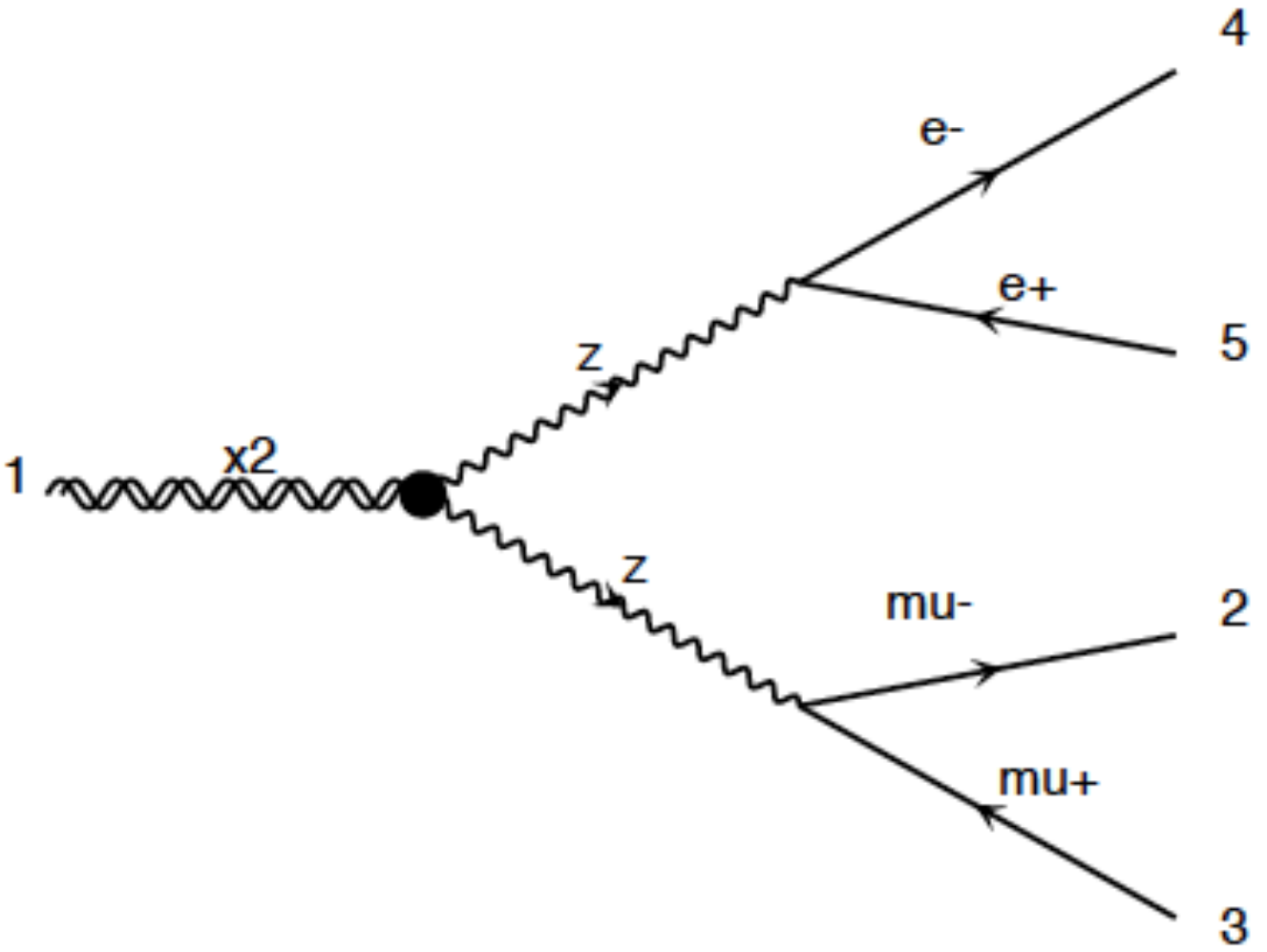}\quad
 \includegraphics[height=2.75cm,clip=true]{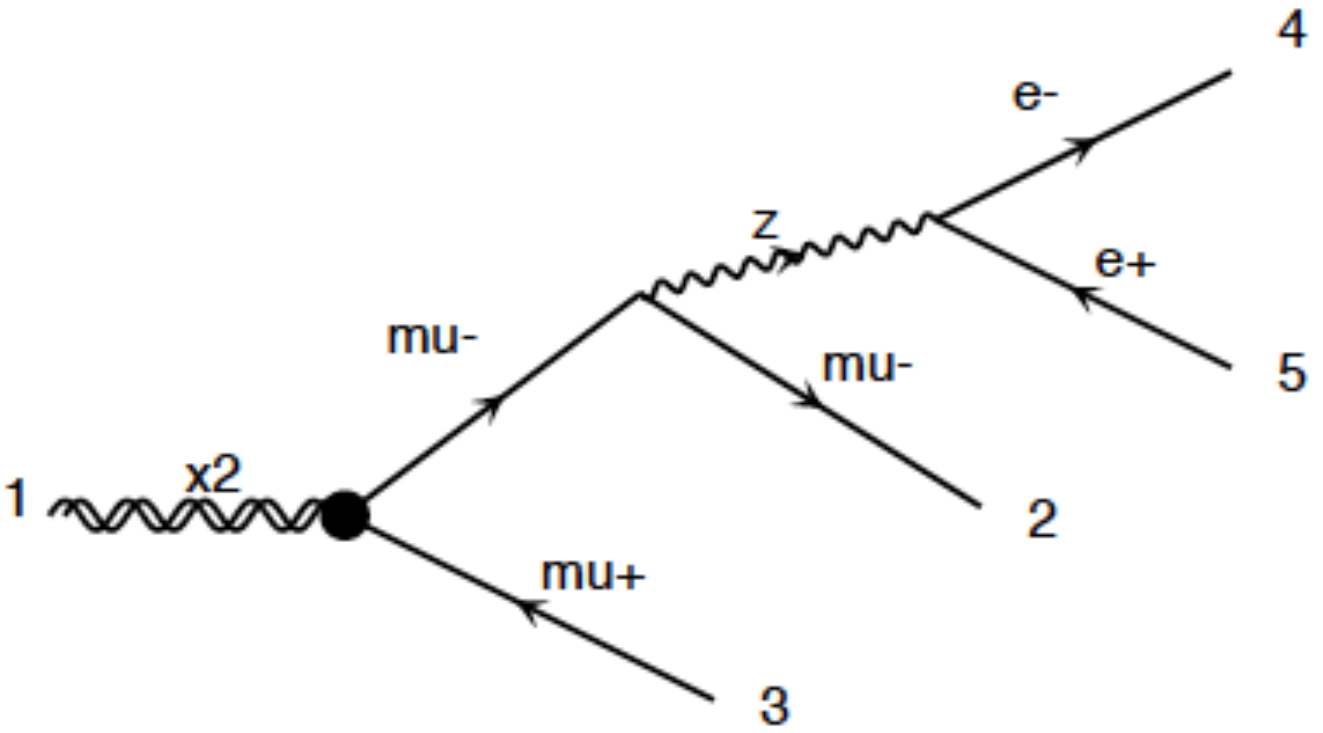}\quad 
 \includegraphics[height=2.75cm,clip=true]{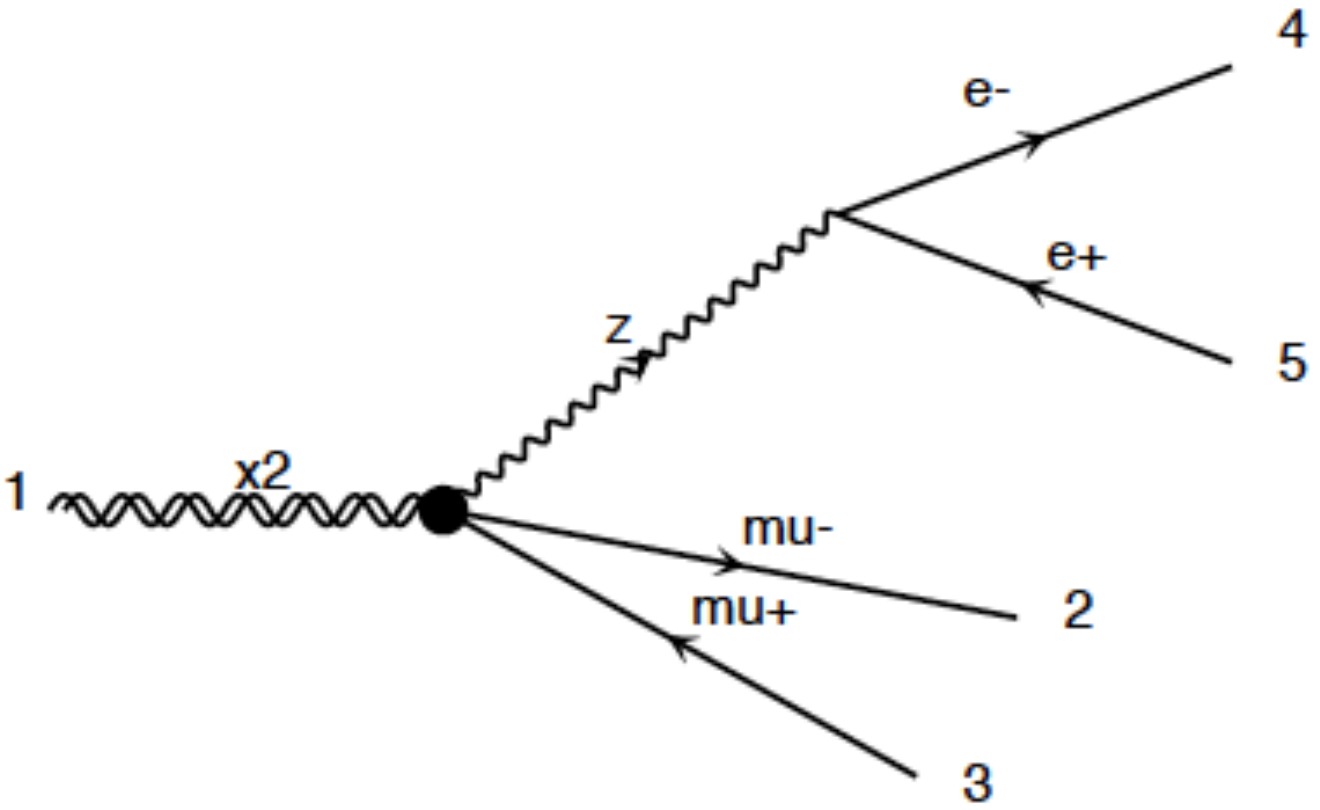} 
 \caption{Representative diagrams for the decay of $X_2\to 4\ell$.} 
 \label{fig:x2decay_dia}
\end{figure}

\begin{figure}
\center
 \includegraphics[width=0.40\textwidth,clip=true]{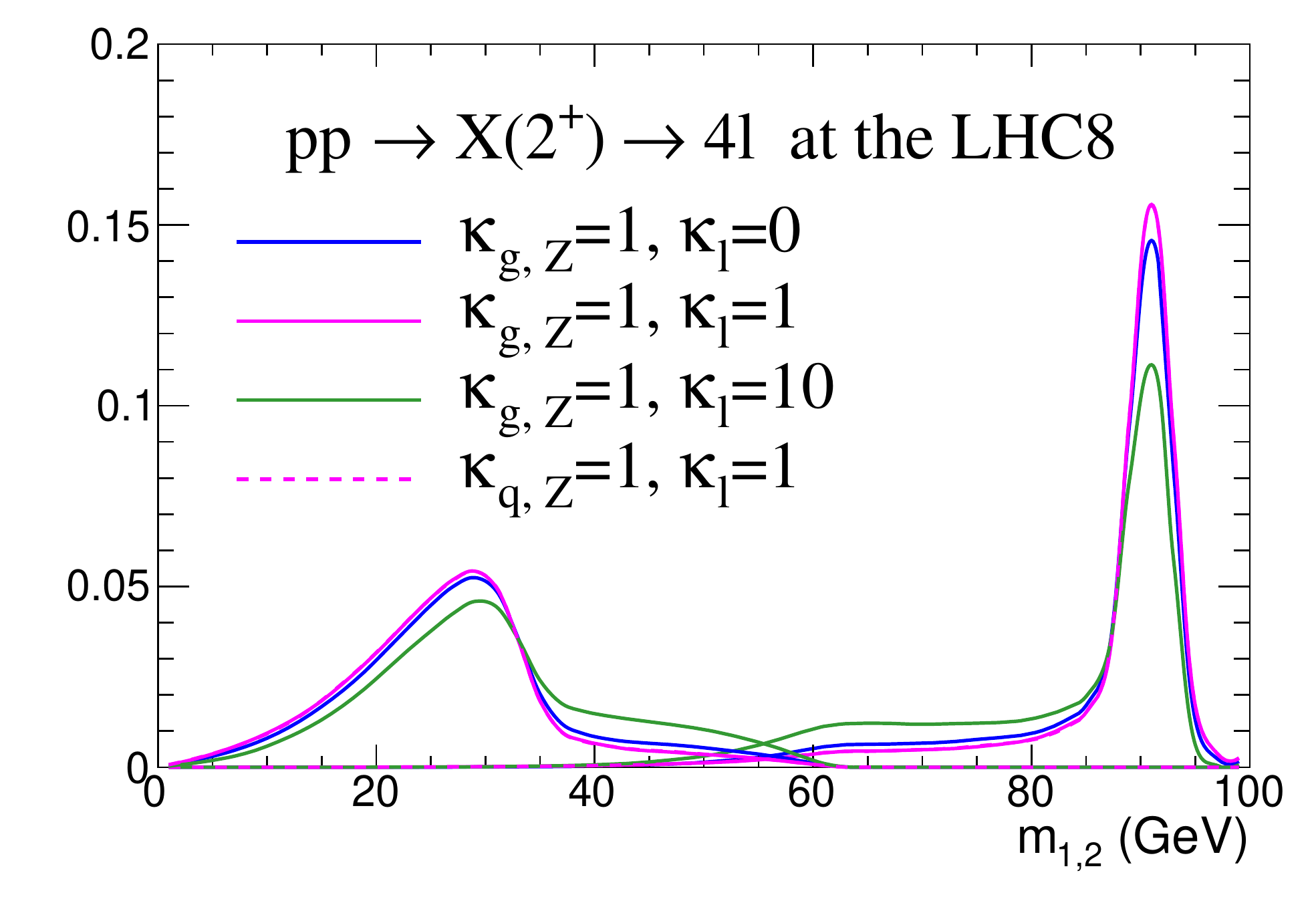}\quad
 \includegraphics[width=0.40\textwidth,clip=true]{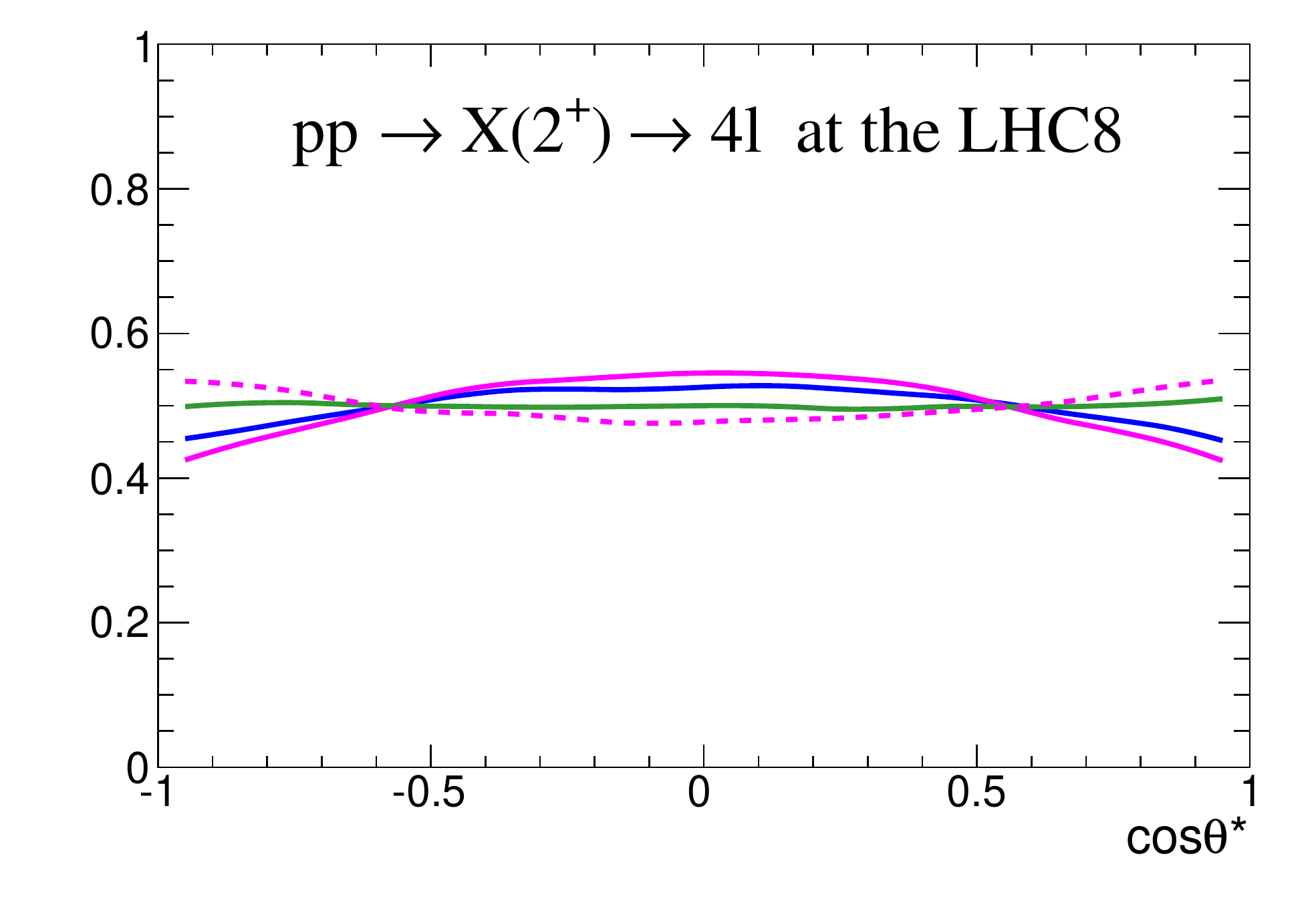}
 \includegraphics[width=0.40\textwidth,clip=true]{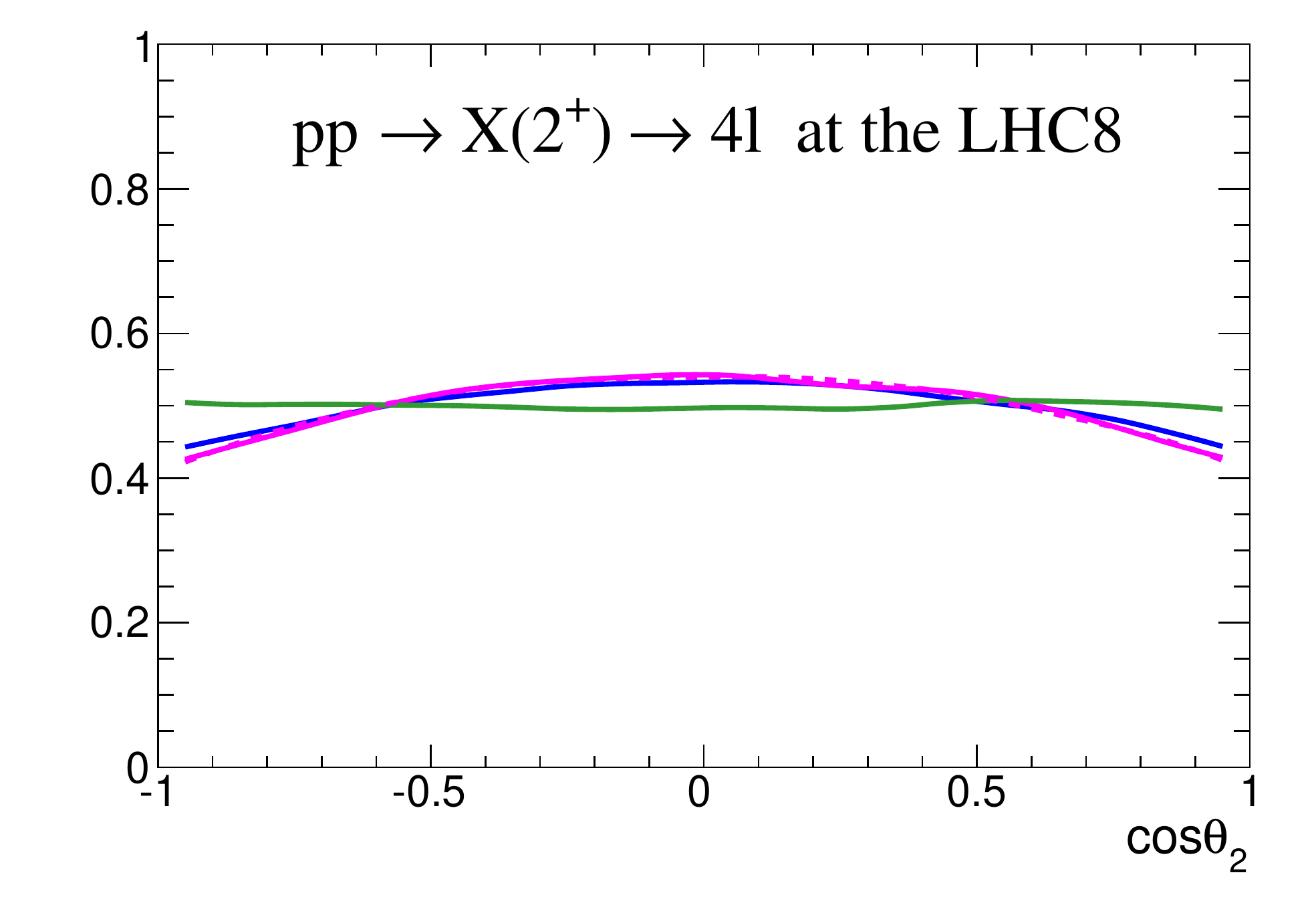}\quad
 \includegraphics[width=0.40\textwidth,clip=true]{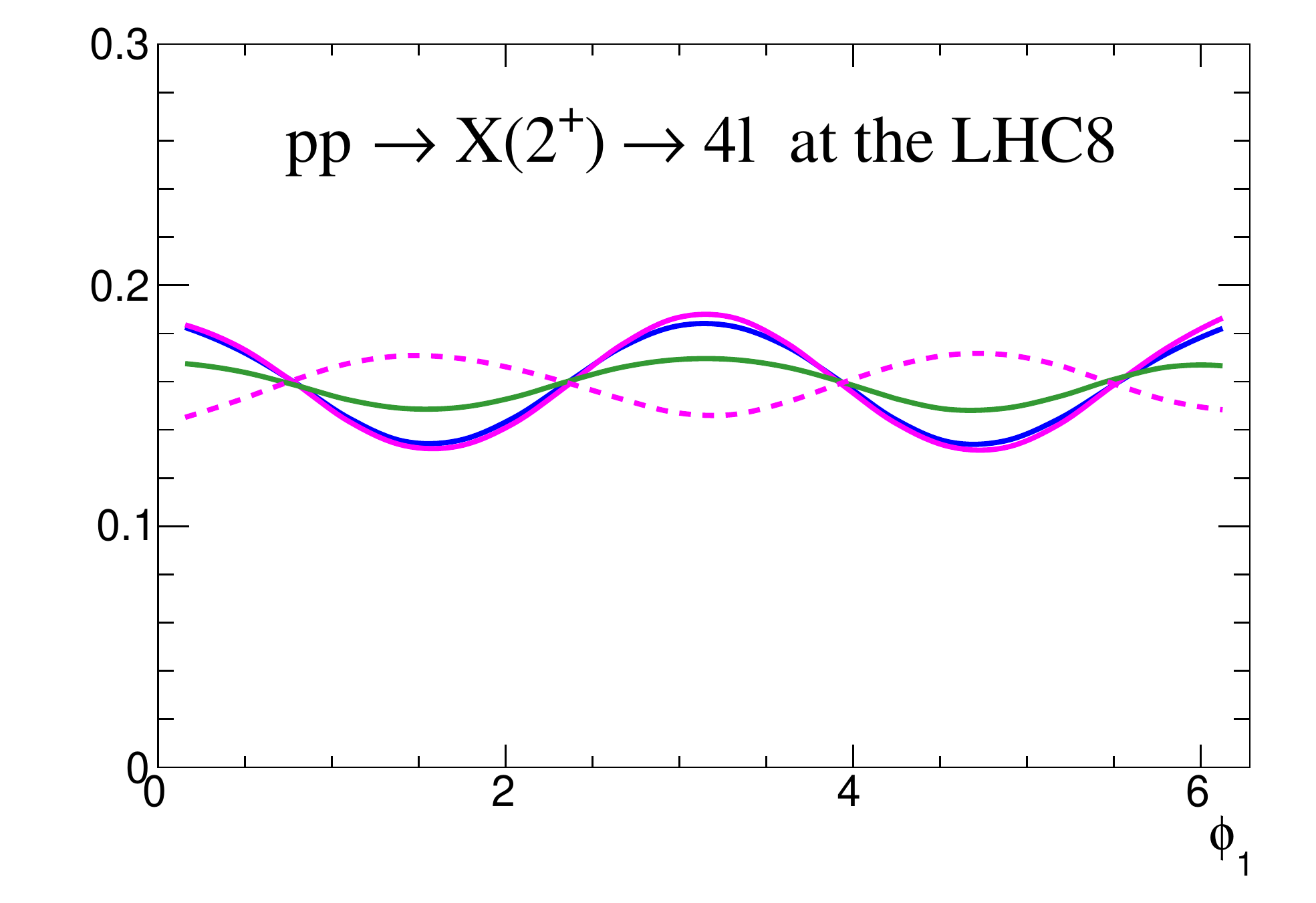}
 \caption{Normalised distributions in $pp\to X_2\to\mu^+\mu^-e^+e^-$ for
   different $\kappa_{\ell}$ values: the invariant masses of the two lepton
   pairs $m_1$, $m_2$ (with $m_1>m_2$), $\cos \theta^*$, $\cos\theta_2$, and
 $\phi_1$, as defined in
   ref.~\cite{Bolognesi:2012mm}.  Event simulation performed at the leading
   order, parton level only (no shower/hadronisation).}
 \label{fig:x2decay_dis}
\end{figure}

For spin 2, our minimal approach consists of sticking to the minimal
five-dimensional interaction and of imposing the invariance of ${\cal L}_2$
under the gauge symmetries of the SM. As a result, in the case of universal
couplings to SM particles $X_2$ is equivalent to a minimal RS-graviton.  As
it will be discussed in the following, a spin-2 state with non-universal
couplings to SM particles might have a very different behaviour with
respect to that of an RS-graviton, especially at high energies.  In order to
further this point, in fig.~\ref{fig:x2decay_dia} we show some of the diagrams
involved in the decay $X_2 \to 4 \ell$. Were the resonance above twice the $Z$
mass, one could certainly only consider the first diagram, which would be by
far the dominant one. For a mass around 125 GeV, however, one of the
$Z$-bosons is not on-shell and diagrams such as the second and third one
become relevant and need to be included. In fig.~\ref{fig:x2decay_dis} the
dependence on the coupling $\kappa_{\ell}$ that appears in eq.~\eqref{eq:grav}
of key distributions, i.e. the invariant mass of the lepton pairs 
$m_1$ and $m_2$ (with $m_1>m_2$),
$\cos \theta^*$, $\cos\theta_2$, and $\phi_1$ distributions (see
ref.~\cite{Bolognesi:2012mm} for their definition), is shown.  The magenta lines
are the case for $\kappa_Z=\kappa_{\ell}\ne 0$ with the $gg$ initial state
($\kappa_g\ne 0$, solid) and the $q\bar q$ ($\kappa_q\ne0$, dotted).
The most striking differences are seen for the case where the 
spin-2 coupling to the fermions is enhanced by a factor 
of 10 (green line). 

\subsection{Higher orders in QCD\label{sec:HO}}
The LO predictions previously discussed can be systematically
improved by including the effects due to the emission of QCD partons;
this can be done by considering both tree-level and full-NLO matrix elements,
and their matching with parton showers.

The ME+PS simulations are based on tree-level matrix elements for production
and decays, and allow one to retain all spin correlations.  Extra jet
radiation can be realistically taken into account by merging matrix elements
with different parton multiplicities with parton shower programs, such as 
{\sc HERWIG}~\cite{Corcella:2000bw,Bahr:2008pv}
or {\sc Pythia}~\cite{Sjostrand:2006za,Sjostrand:2007gs}. 
The {\sc MadGraph\,5} platform features an interface with
{\sc Pythia6.4}~\cite{Sjostrand:2006za} that makes use of the 
MLM-$\kt$~\cite{Mangano:2001xp,Alwall:2007fs}, and of the 
shower-$k_T$~\cite{Alwall:2008qv} merging prescriptions.  The two
matching schemes have been tested in several cases and shown to give
equivalent results (see e.g.~refs.~\cite{Alwall:2008qv,Alwall:2011cy}).
Studies presented in this work are performed using the MLM-$\kt$ matching
scheme.

{\sc aMC@NLO} is an event generator that implements the matching of 
any NLO QCD computation with parton showers according to the
MC@NLO formalism~\cite{Frixione:2002ik}, and which is embedded 
in the {\sc MadGraph\,5} framework. 
It is based on two main building blocks, each devoted to the
generation and evaluation of a specific contribution to an NLO-matched
computation. {\sc MadFKS}~\cite{Frederix:2009yq} deals with the Born and
real-emission amplitudes, and in particular it performs, according to 
the FKS prescription~\cite{Frixione:1995ms,Frixione:1997np}, the
subtraction of the infrared singularities that appear in the latter
matrix elements; moreover, it is also responsible for the generation of 
the so-called Monte Carlo subtraction terms, namely the contributions 
that prevent any double-counting in the {\sc MC@NLO} cross sections.  
{\sc MadLoop}~\cite{Hirschi:2011pa} computes the one-loop amplitudes, 
using the OPP~\cite{Ossola:2006us} integrand-reduction method and its
implementation in {\sc CutTools}~\cite{Ossola:2007ax}. 
These procedures are fully automated (hence, they do not require
any coding by the user, with the relevant computer codes being 
generated on-the-fly), provided that a basic knowledge is available
about the underlying theory and the interactions of its particles with
QCD partons. For {\sc MadFKS} this amounts to the ordinary Feynman rules;
for {\sc MadLoop}, to Feynman rules, UV counterterms, and special tree-level
rules necessary to, and defined by, the OPP method, which are called $R_2$.
While Feynman rules are automatically computed given the lagrangian
(via {\sc FeynRules}), this is not yet possible for the UV counterterms
and $R_2$ rules\footnote{However, a preliminary version of {\sc FeynRules} 
exists which does exactly this.}. The solution adopted thus far is that of 
coding by hand these pieces of information, for all cases where the relevant
analytical computations had already been carried out, namely
QCD and EW corrections in the SM~\cite{Draggiotis:2009yb,
Garzelli:2009is,Garzelli:2010qm,Shao:2011tg} and for QCD corrections in SUSY 
models~\cite{Shao:2012ja}.

The upshot of this is the following: for the Higgs characterisation model,
all the ingredients entering the MC@NLO cross sections can be computed
automatically, except for (some of) the one-loop matrix elements. 
In order to amend
the latter issue, one can choose either of the following two strategies:
that of computing analytically the relevant UV counterterms and $R_2$
rules, and of implementing them in the appropriate {\sc UFO} module;
or that of computing directly the relevant one-loop matrix elements.
While the former strategy has a broader scope, the (considerable) effort
it entails is not justified in view of the progress with {\sc FeynRules} 
mentioned above. Hence, the latter strategy is quicker to pursue,
and less error-prone in the short term. This is because it can rely
on results readily available in the literature. For $pp \to X_0\,+$ anything,
the one-loop matrix elements for both the $0^+$ and $0^-$ states have been
known since a long time~\cite{Dawson:1990zj,Kauffman:1993nv}. 
Results for the production of a $CP$-mixed state can also be easily
obtained, even though this scenario is not yet implemented. The case
of $pp \to X_1\,+$ anything is exactly the same as Drell-Yan. For the
inclusive production of a spin-2 boson, the analytic results for the virtual 
amplitudes of refs.~\cite{Mathews:2005bw,Kumar:2008pk,Kumar:2009nn,
Agarwal:2009xr,Agarwal:2009zg,Agarwal:2010sn,Agarwal:2010sp,Frederix:2012dp} 
have been extended to allow for non-universal couplings to quarks and gluons.
Their implementations in {\sc aMC@NLO} includes spin correlated decays to 
$\gamma\gamma$, $W^+ W^-$, and $ZZ$ to four leptons. All the three classes
of matrix elements mentioned here have been implemented by hand in 
{\sc aMC@NLO}, and used for the simulations presented in this paper.
We point out that, in the case of vector boson fusion (VBF) and of
vector-boson associated 
production, all NLO computations can be done automatically and in full
generality\footnote{Because the corresponding one-loop amplitudes are trivial,
and do not necessitate any UV or $R_2$ information from the 
Higgs-characterisation lagrangian -- in the case of VBF, this assumes
that the pentagon contributions are discarded, as is customary in the SM.}, 
with the exception of the spin-2 case, which is feasible provided that
one assumes vanishing couplings with QCD particles.  Studies, such as those presented in
refs.~\cite{Englert:2012xt,Englert:2012ct,Englert:2013opa} could therefore be performed at NLO accuracy. 
Finally, we mention  that Level-2 studies in $t\bar t$ associated production can be performed for
spin-0~\cite{Frederix:2011zi} and spin-1~\cite{Frederix:2009yq} in a fully
automatic way.

\subsubsection{Inclusive production $pp\to X(J^P)$: ME+PS vs. aMC@NLO}

\begin{figure}[h!]
\center
 \includegraphics[width=0.525\textwidth,clip=true, trim = 80 180 50 210]{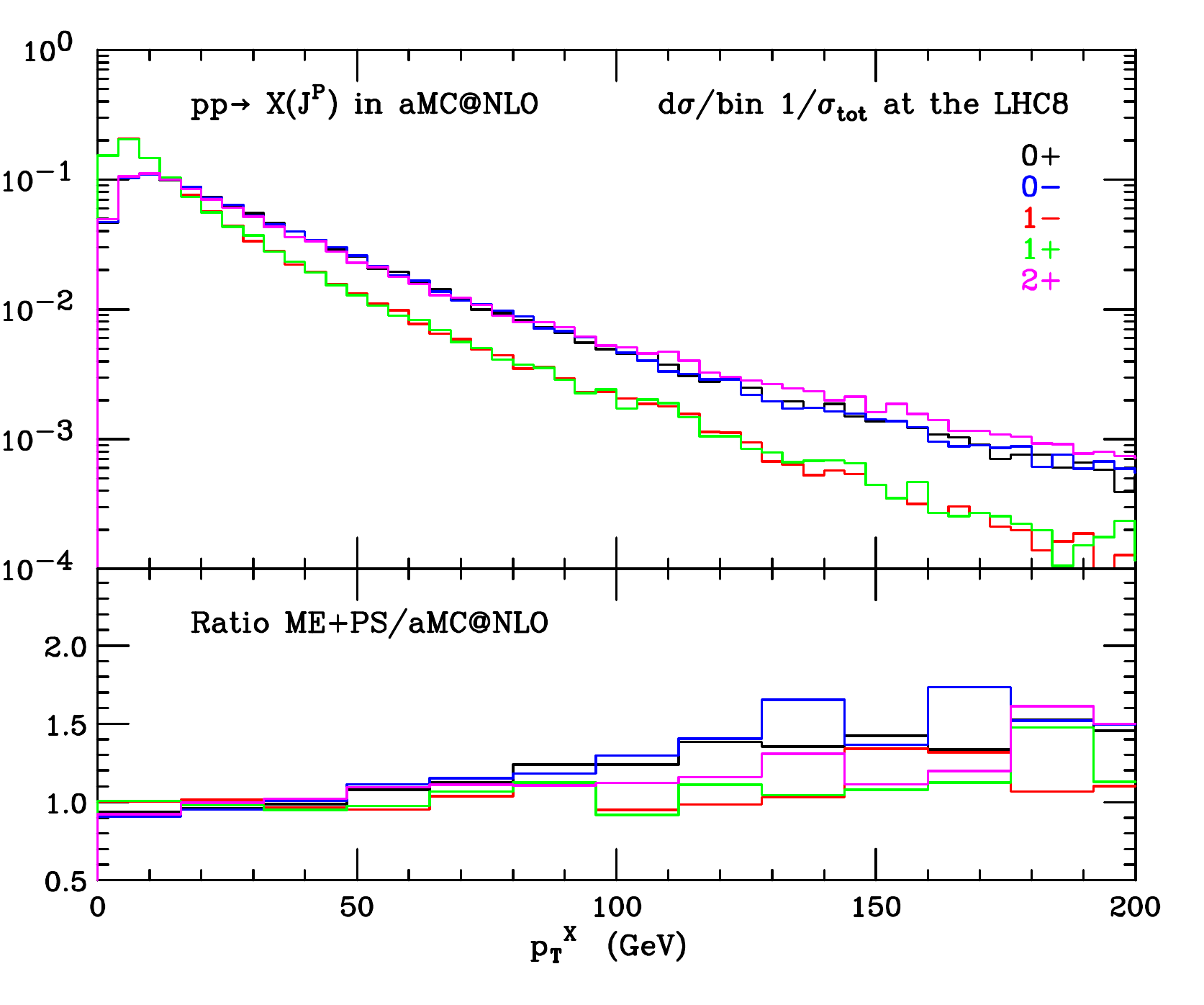} 
 \includegraphics[width=0.525\textwidth,clip=true, trim = 80 180 50 210]{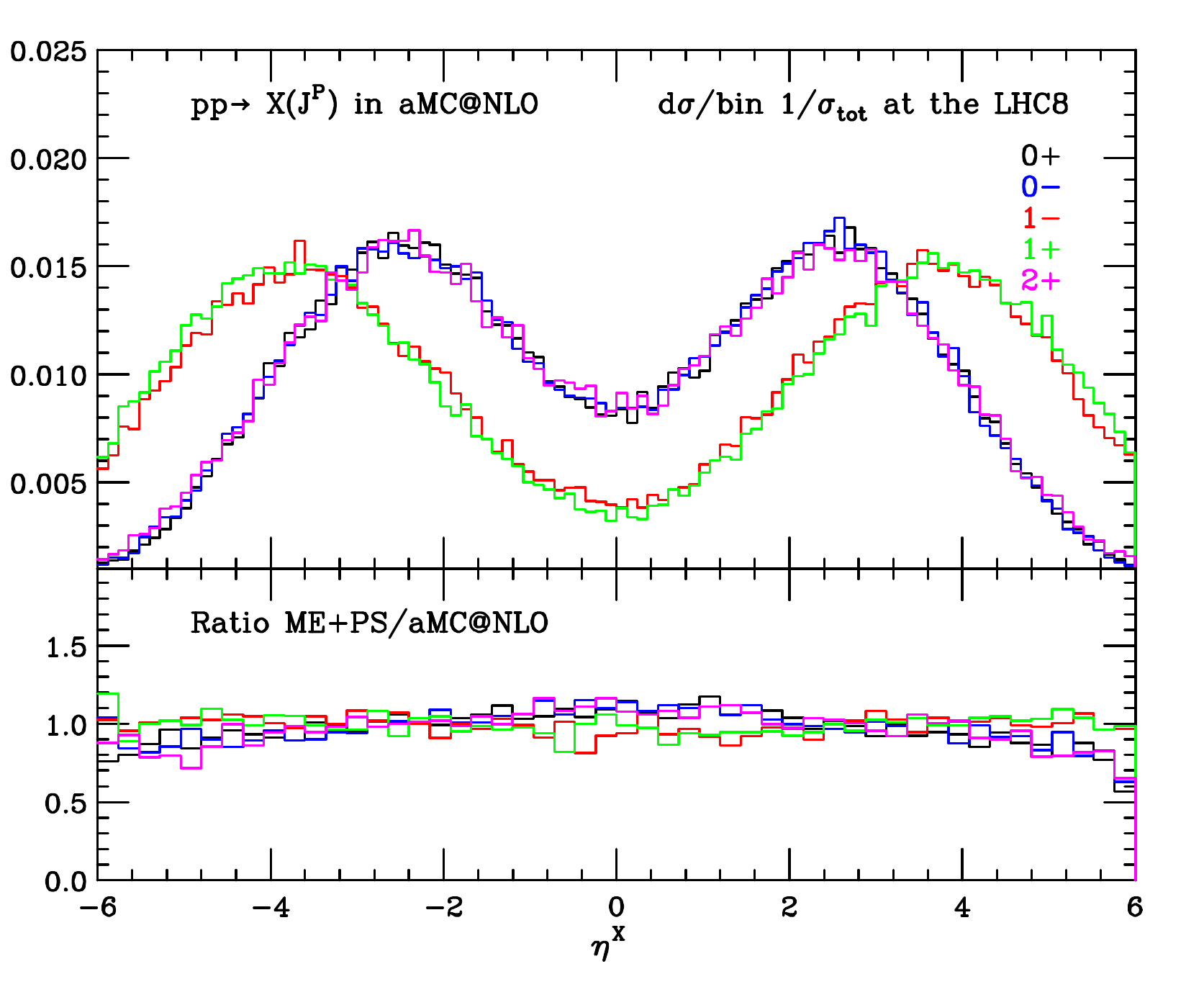} 
 \includegraphics[width=0.525\textwidth,clip=true, trim = 80 180 50 210]{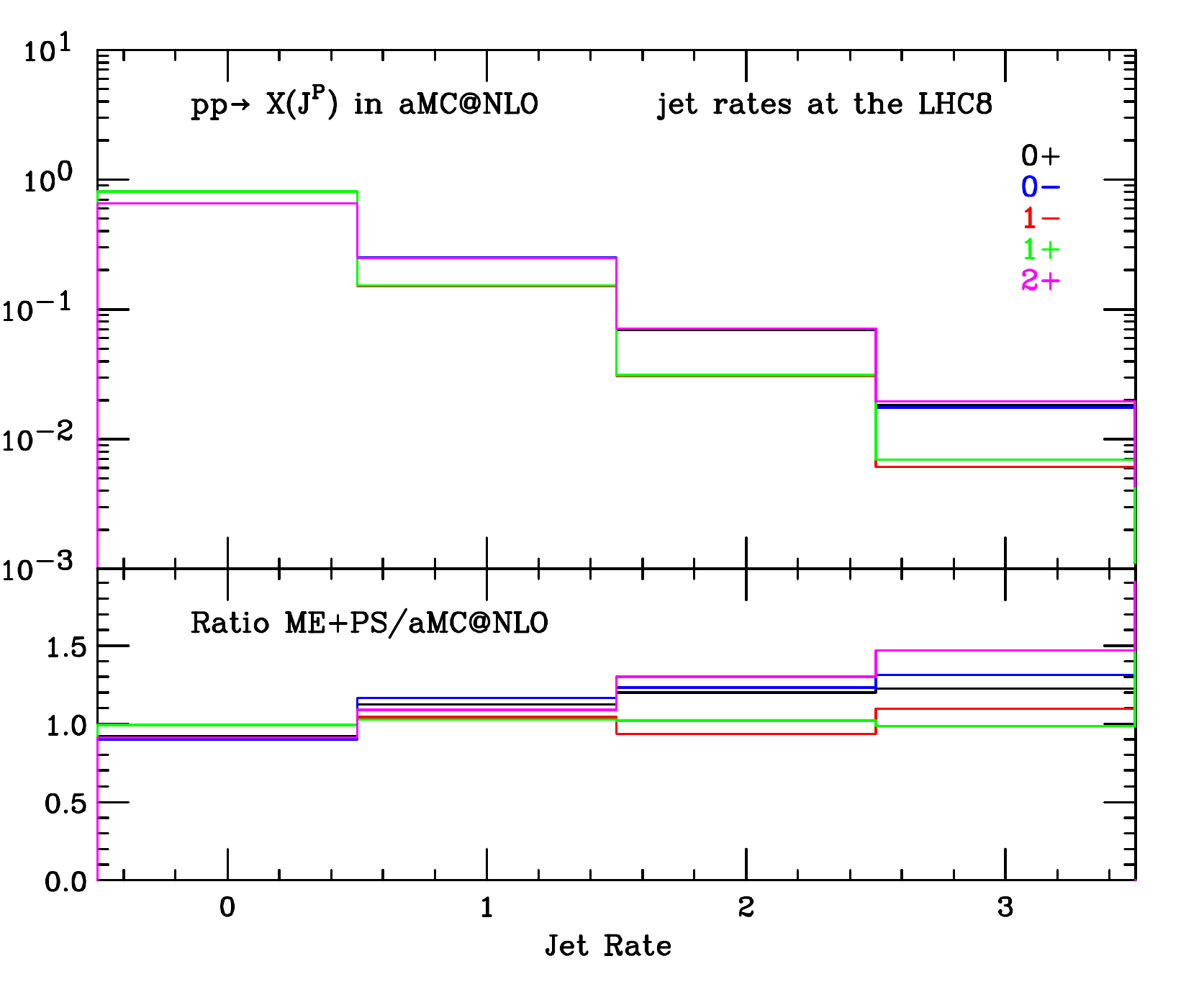} 
 \caption{The transverse momentum $p_{T}^X$, pseudorapidity $\eta^X$, and jet
   rates of the new boson $X(J^P)=0^+,0^-,1^+,1^-,2^+$ as obtained from {\sc
     aMC@NLO}. The lower inset shows the bin-by-bin ratio of the same
   distribution obtained via ME+PS merging and that of {\sc aMC@NLO}. }
 \label{fig:xpt}
\end{figure}

\begin{figure}
\center
 \includegraphics[width=0.48\textwidth,clip=true, trim= 50 180 50 200]{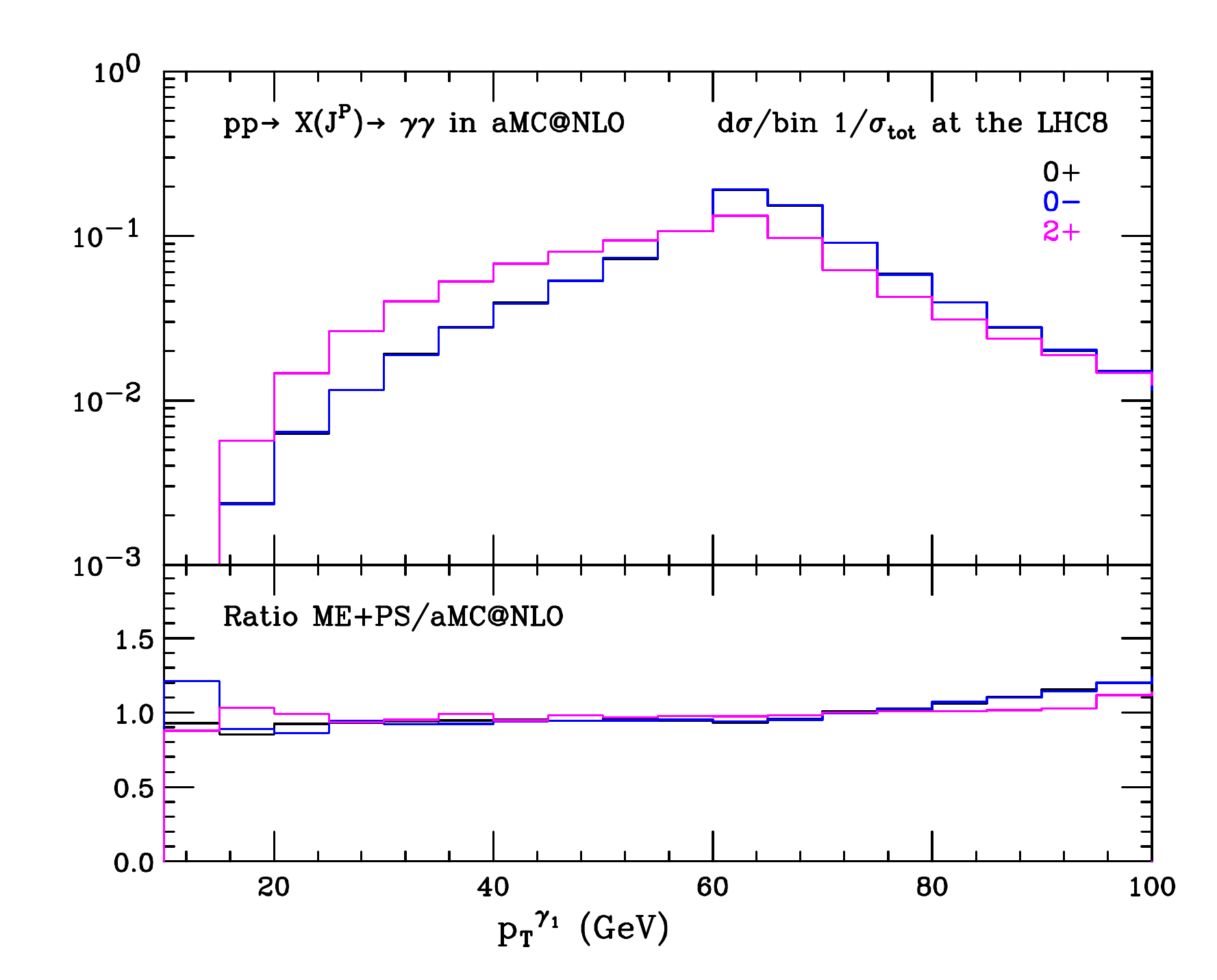} 
 \includegraphics[width=0.48\textwidth,clip=true, trim = 50 180 50 200]{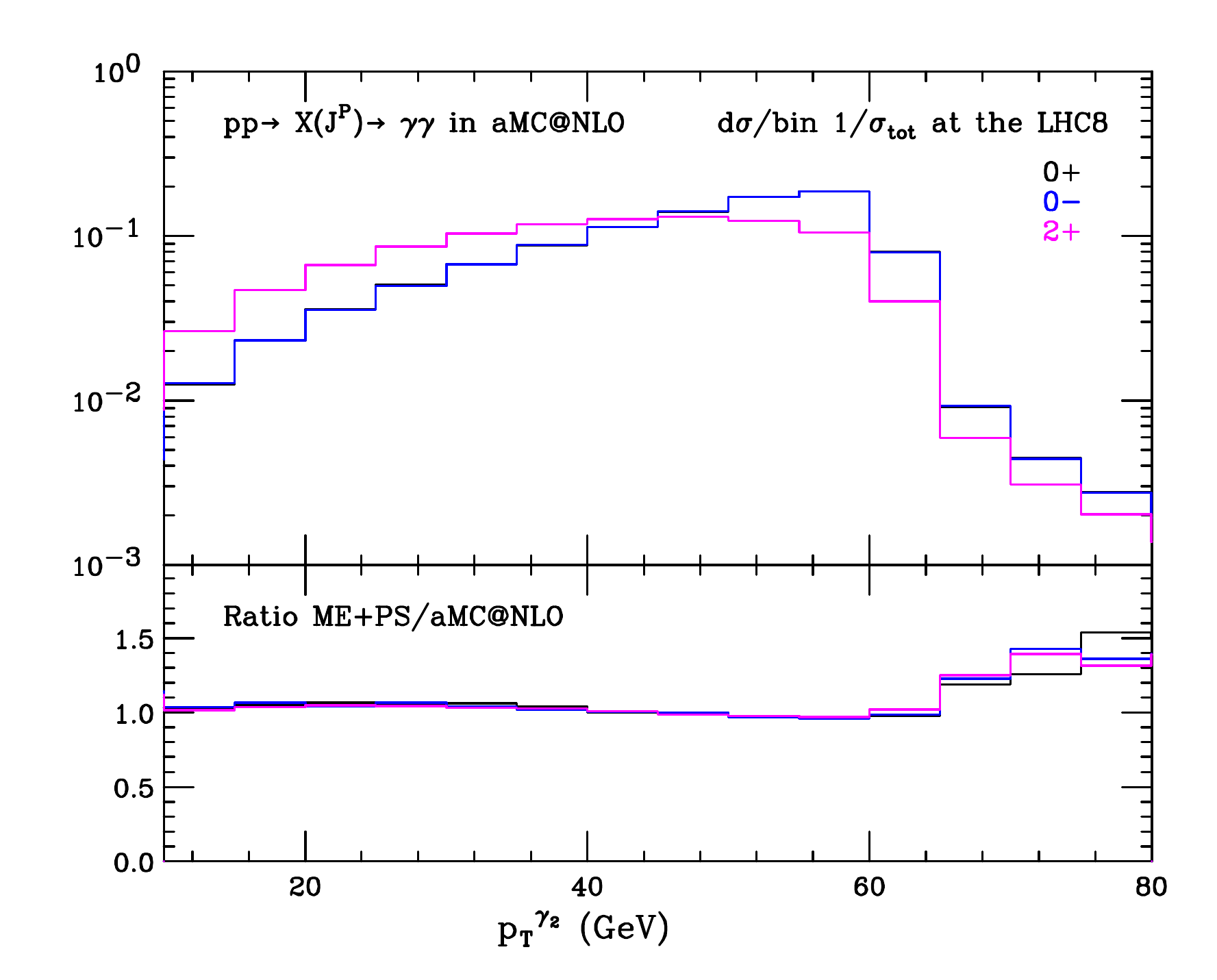}  
 \includegraphics[width=0.48\textwidth,clip=true, trim = 50 180 50 200]{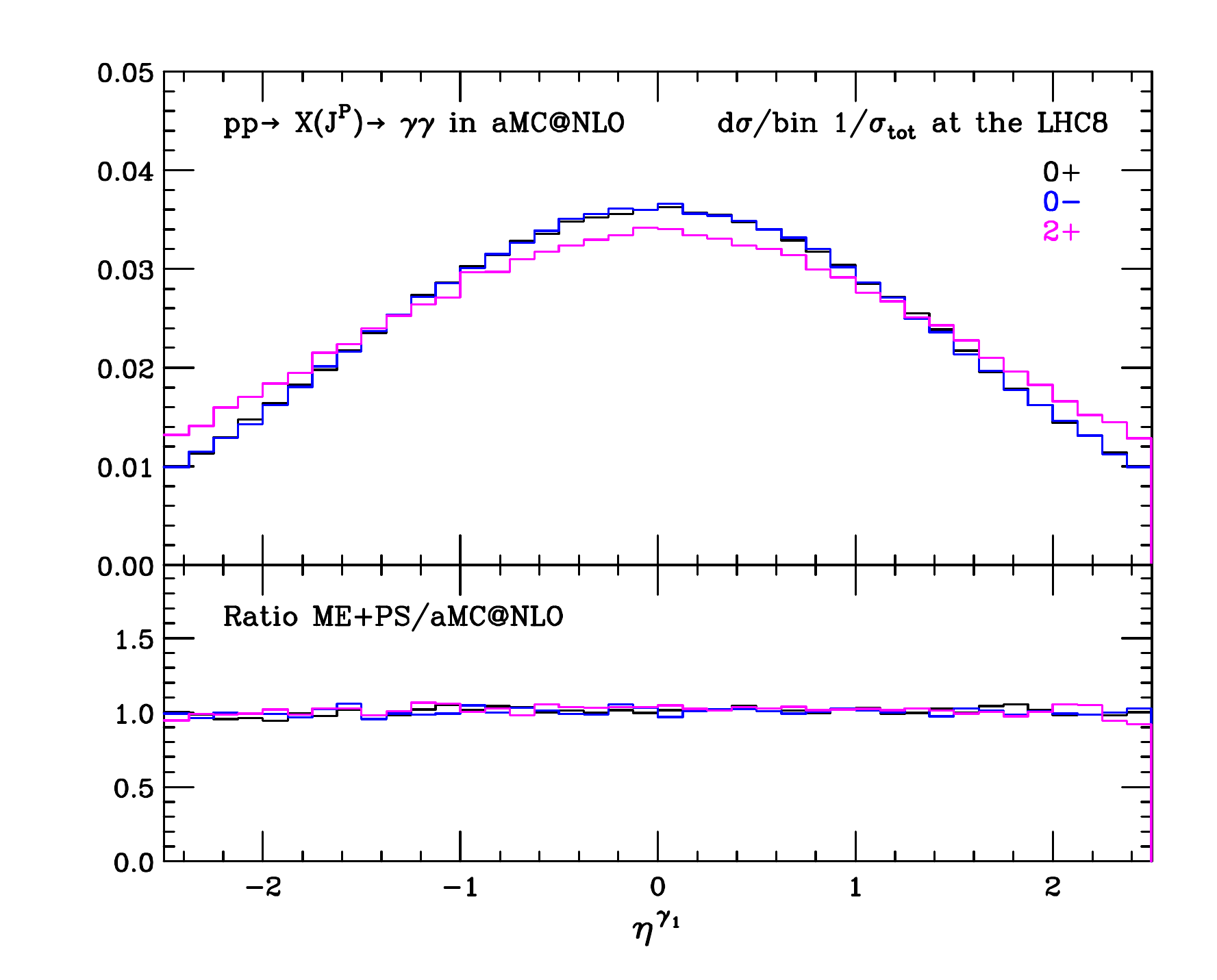}  
 \includegraphics[width=0.48\textwidth,clip=true, trim = 50 180 50 200]{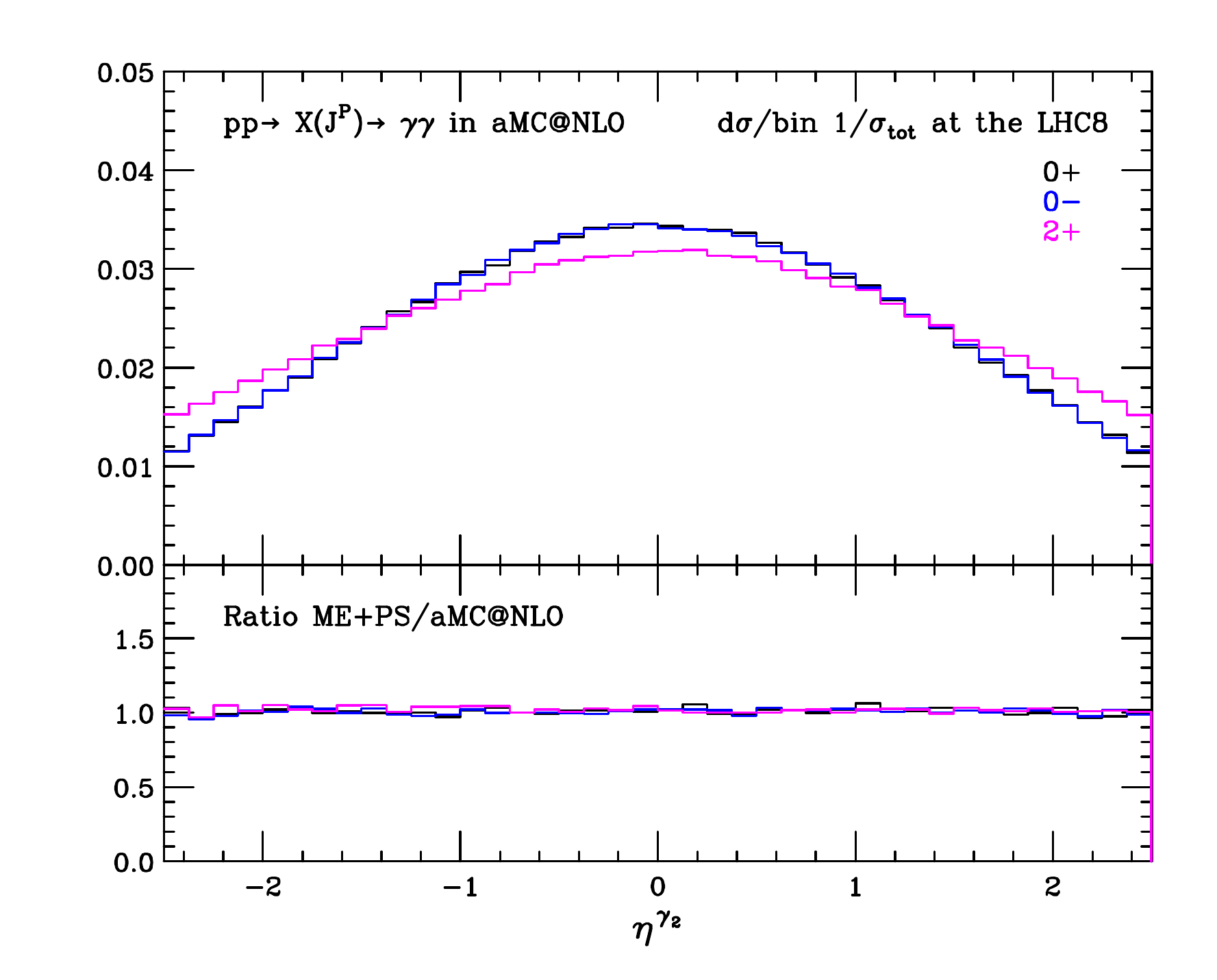}  
 \caption{Distributions in $X\to\gamma\gamma$: (a) and (b) the transverse
   momentum of the leading and subleading photon, $p_{T}^{\gamma_1}$ and
   $p_{T}^{\gamma_2}$, (c) and (d) the rapidity of the leading and subleading
   photon, $\eta^{\gamma_1}$ and $\eta^{\gamma_2}$.}
 \label{fig:ppaa}
\end{figure}

As is well known, the ME+PS and MC@NLO approaches often give complementary
benefits. In those phase-space regions where both of them are sensible, it is
interesting to compare their predictions, as a way towards their validation
through a mutual consistency check.  To this end, in this section we present
the results of a comparison between ME+PS and {\sc aMC@NLO} for the case of
inclusive $X(J^P)$ production, with $J^P=0^+$, $0^-$, $1^+$, $1^-$, and $2^+$.
In the following analyses, we generate events at the LHC with a
center-of-mass energy $\sqrt{s}=8$~TeV and assume the mass of the new boson to
be $m_X=125$ GeV. ME+PS merged samples consist of events at the matrix element
level for $pp\to X+0,1,2$ partons obtained with {\sc MadGraph\,5}, with
parameters $Q^{\rm ME}_{\rm min}=40$~GeV and $Q^{\rm jet}_{\rm min}=25$~GeV,
and with CTEQ6L1~\cite{Pumplin:2002vw} PDFs. These samples are then showered
with {\sc Pythia} 6.4 ($p_T$-ordered) by using the MLM-$\kt$ scheme for
merging.  The {\sc aMC@NLO} samples are obtained by setting the
renormalisation and factorisation scales equal to $m_X$ and by employing
MSTW2008 NLO PDFs~\cite{Martin:2009iq} for the short-distance calculation.
Both LO and NLO samples are showered with the default parameter settings in
{\sc Pythia} (including the PDFs, CTEQ5L~\cite{Lai:1999wy}), in order to be mostly
sensitive to differences arising at the level of matrix elements. After shower
and hadronisation, final state particles are clustered into jets using the
anti-$k_T$ algorithm~\cite{Cacciari:2008gp} (as implemented in {\sc
  FastJet}~\cite{Cacciari:2011ma}) with radius parameter $\Delta R =
0.4$. Jets are required to have a transverse momentum $p_T^j>25$~GeV.

We start by presenting distributions for the transverse momentum and
pseudorapidity of the new boson, as well as for the exclusive jet
multiplicity (see fig.~\ref{fig:xpt}). Both the $p_T^X$ and $\eta^X$ 
distributions roughly fall into two classes, determined by the dominant
production mechanism at the LO ($gg$ or $q\bar q$) -- gluon fusion ($q\bar{q}$
annihilation) accounts for 100\% (0\%), 0\% (100\%), and 96\% (4\%) in
the case of the production of a spin-0, spin-1, and spin-2 with universal 
couplings  (i.e., an RS graviton) state, respectively. Processes 
dominated by $gg$ fusion display a harder $p_T$ spectrum and are more central
than the $q \bar q$-dominated ones. The rapidity difference is easily
understood by the fact that at a $pp$ collider the $q$ are valence quarks
while the $\bar q$ are from the sea and therefore configurations with
asymmetric Bjorken $x$'s for the two partons are more frequent.  Another
important observation is that the inclusive distributions for a spin-0 and
spin-2 are indeed very similar, i.e. the spin has no real relevance for these
observables.  In the lower insets the bin-by-bin ME+PS over {\sc aMC@NLO}
ratios are shown. These ratios, computed by first normalising the 
corresponding distributions to unity, only convey shape information. It is
manifest that the two methods give very similar predictions, 
both in $p_T^X$ and $\eta^X$. Differences in the $p_T^X$
spectra start to be significant above $m_X$, and in particular the merged
samples produce a bit harder spectra for very large $p_T^X$'s. This is obviously
the effect of the larger amount of hard radiation in the ME+PS samples, which
is in turn due to the presence there of the $pp\to X+2$~partons matrix
elements, which are not included in the {\sc aMC@NLO} predictions. This is
also the reason why the exclusive jet multiplicities, shown in the plot at the
bottom of fig.~\ref{fig:xpt}, are larger when $n>1$ jets in the case of the
merged samples than when computed with {\sc aMC@NLO}.

\begin{figure}
\center
 \includegraphics[width=0.48\textwidth,clip=true, trim = 60 190 50 190]{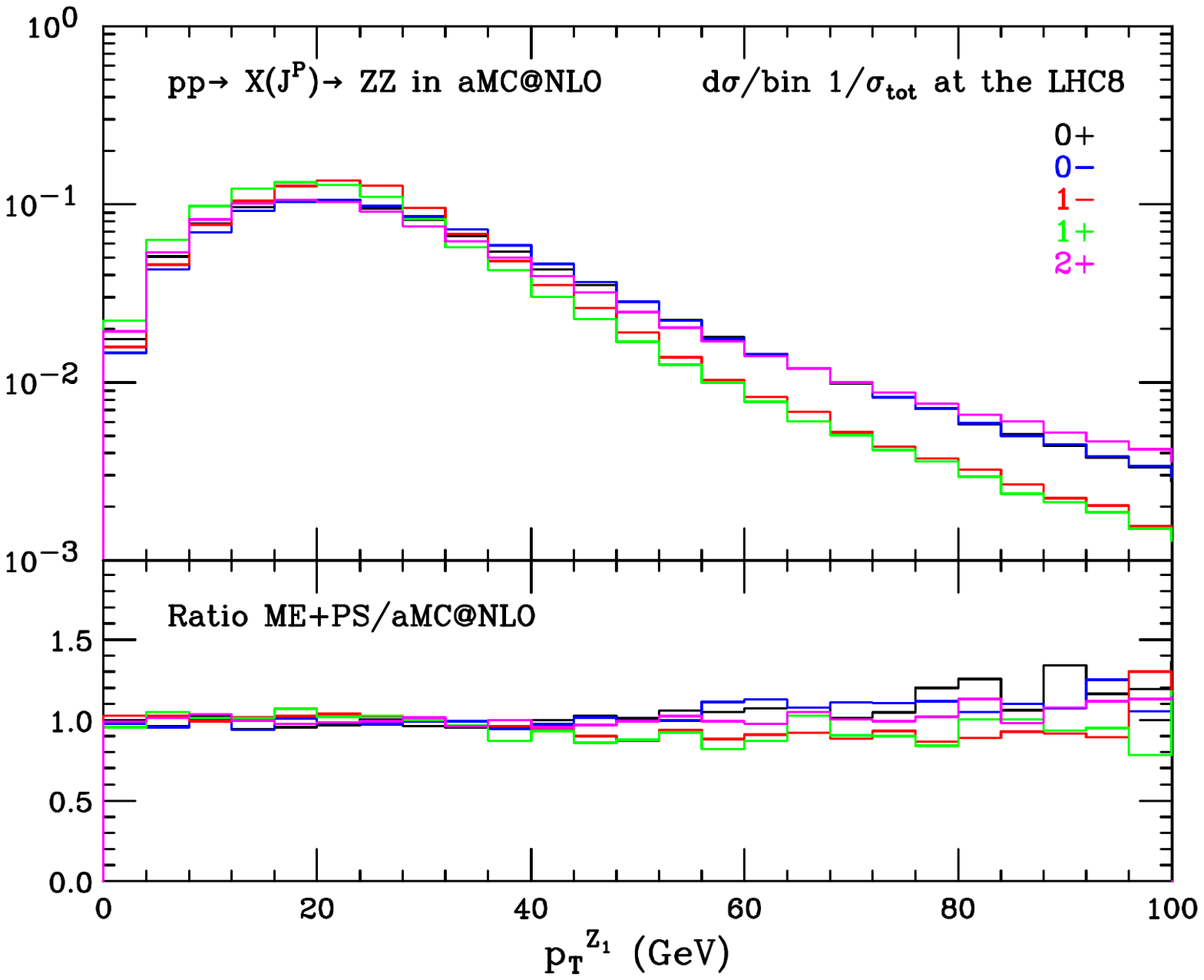} 
  \includegraphics[width=0.48\textwidth,clip=true, trim = 60 190 50 190]{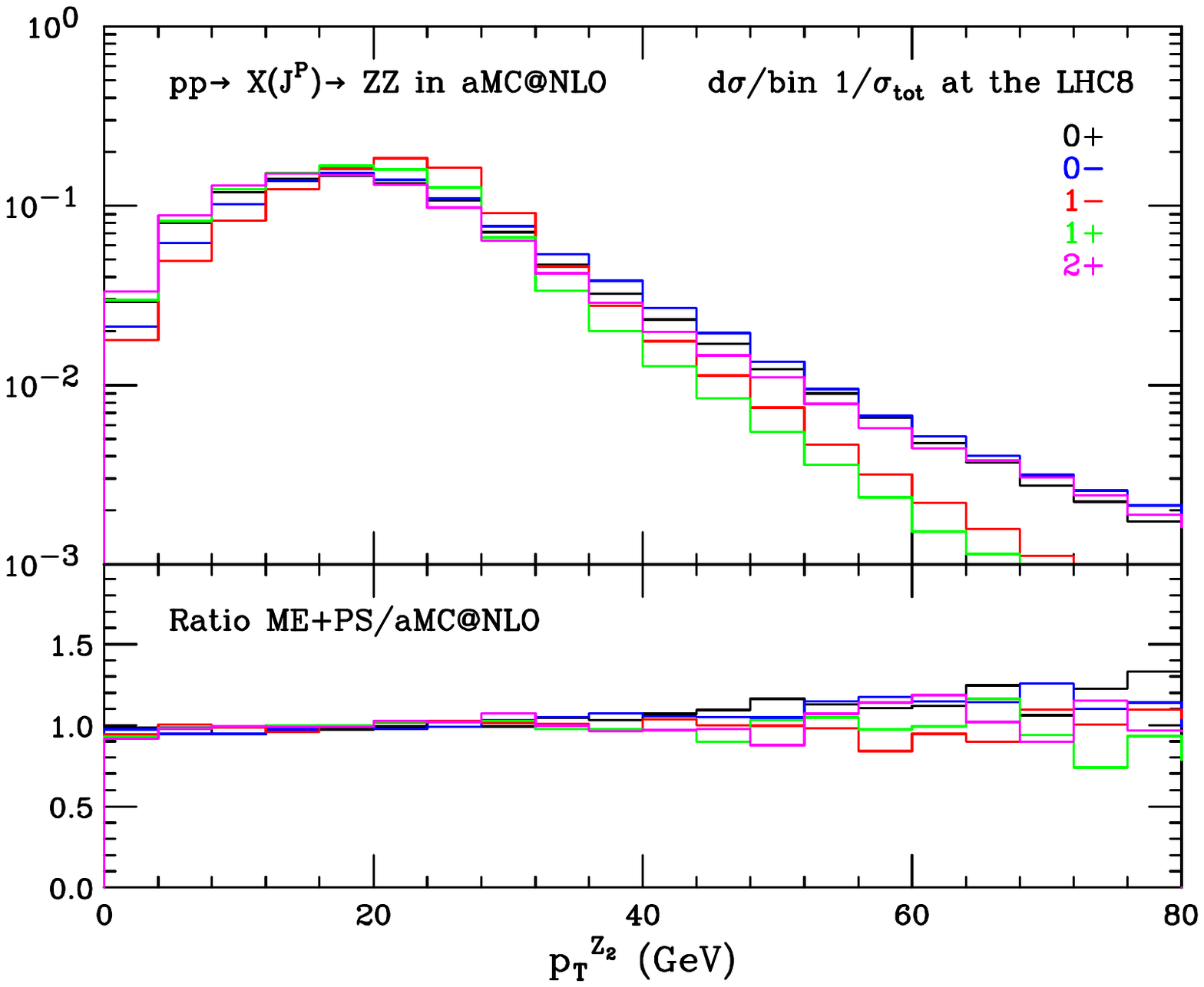} 
 \includegraphics[width=0.48\textwidth,clip=true, trim = 60 190 50 190]{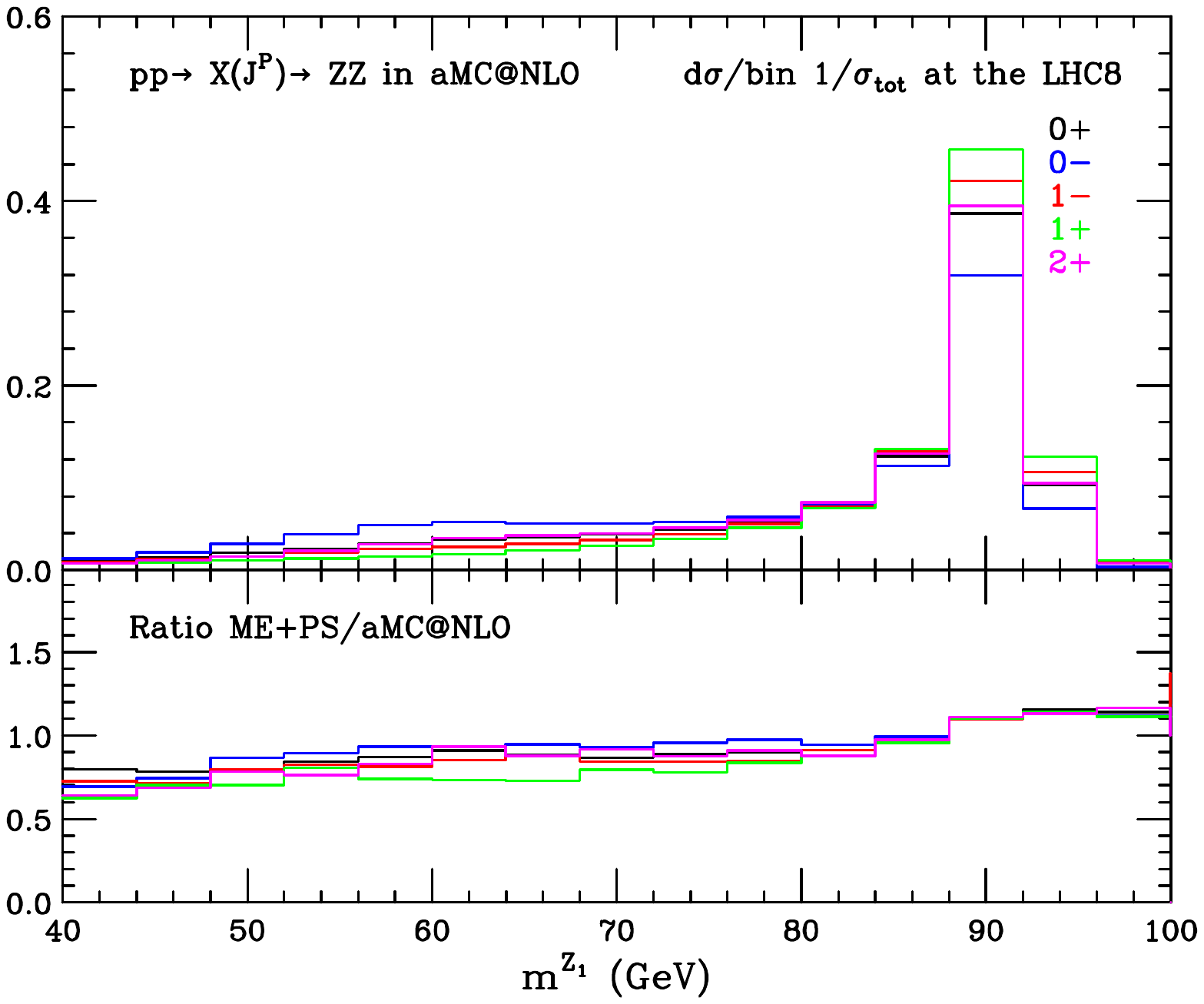} 
  \includegraphics[width=0.48\textwidth,clip=true, trim = 60 190 50 190]{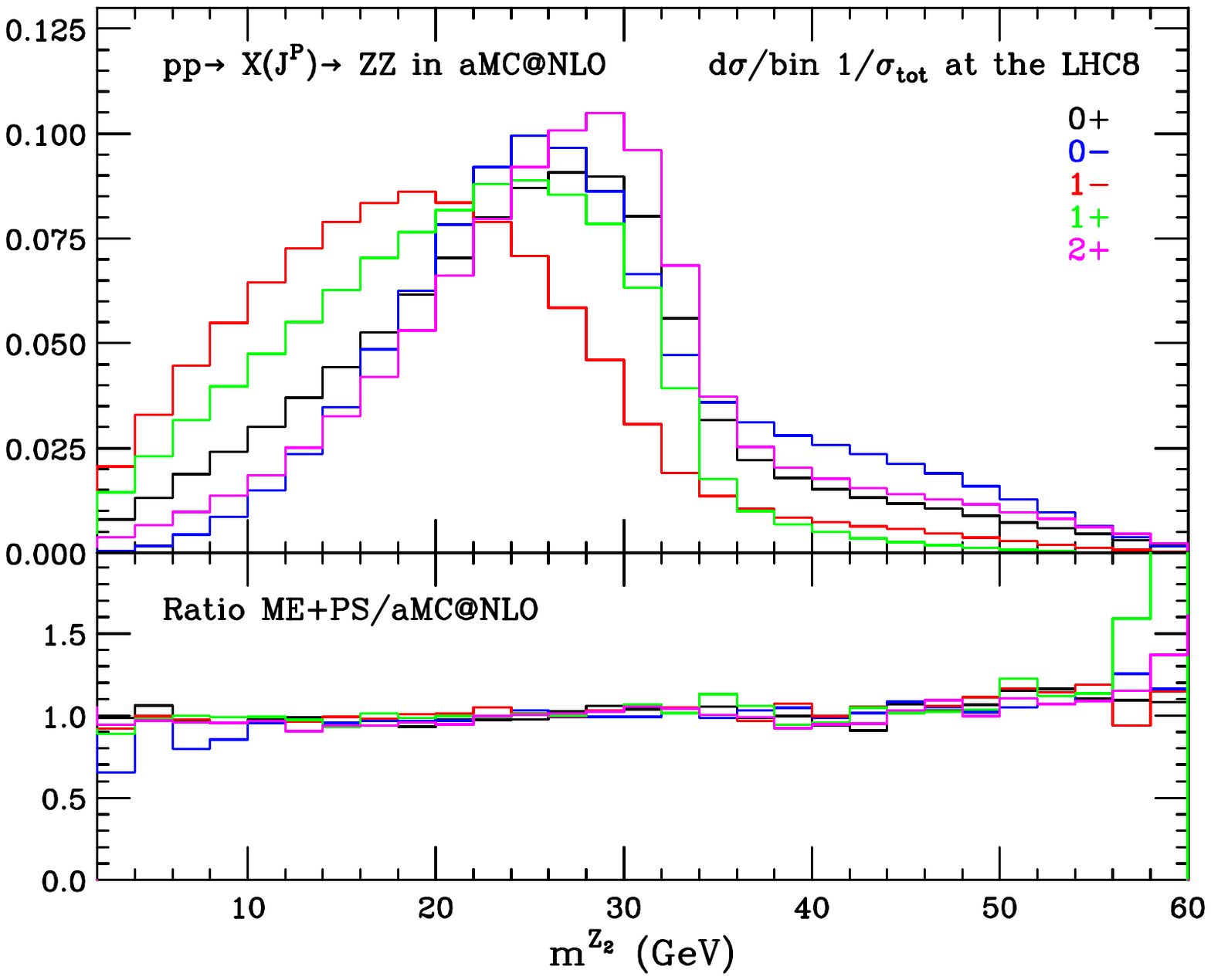}
 \caption{Distributions of the $Z$ bosons in 
 $X(\to ZZ^*)\to\mu^+\mu^-e^+e^-$:  
 (a) and (b) the transverse momentum of the $Z$ boson with the highest
 and lowest reconstructed mass, $p_{T}^{Z_1}$ and $p_{T}^{Z_2}$,  
 (c) and (d) the invariant mass of the two leptons $m_{\ell\ell}$
 corresponding to $Z_1$ and $Z_2$.}
 \label{fig:ppzz}
\end{figure}

\begin{figure}
\center
 \includegraphics[width=0.48\textwidth,clip=true, trim = 60 190 70 190]{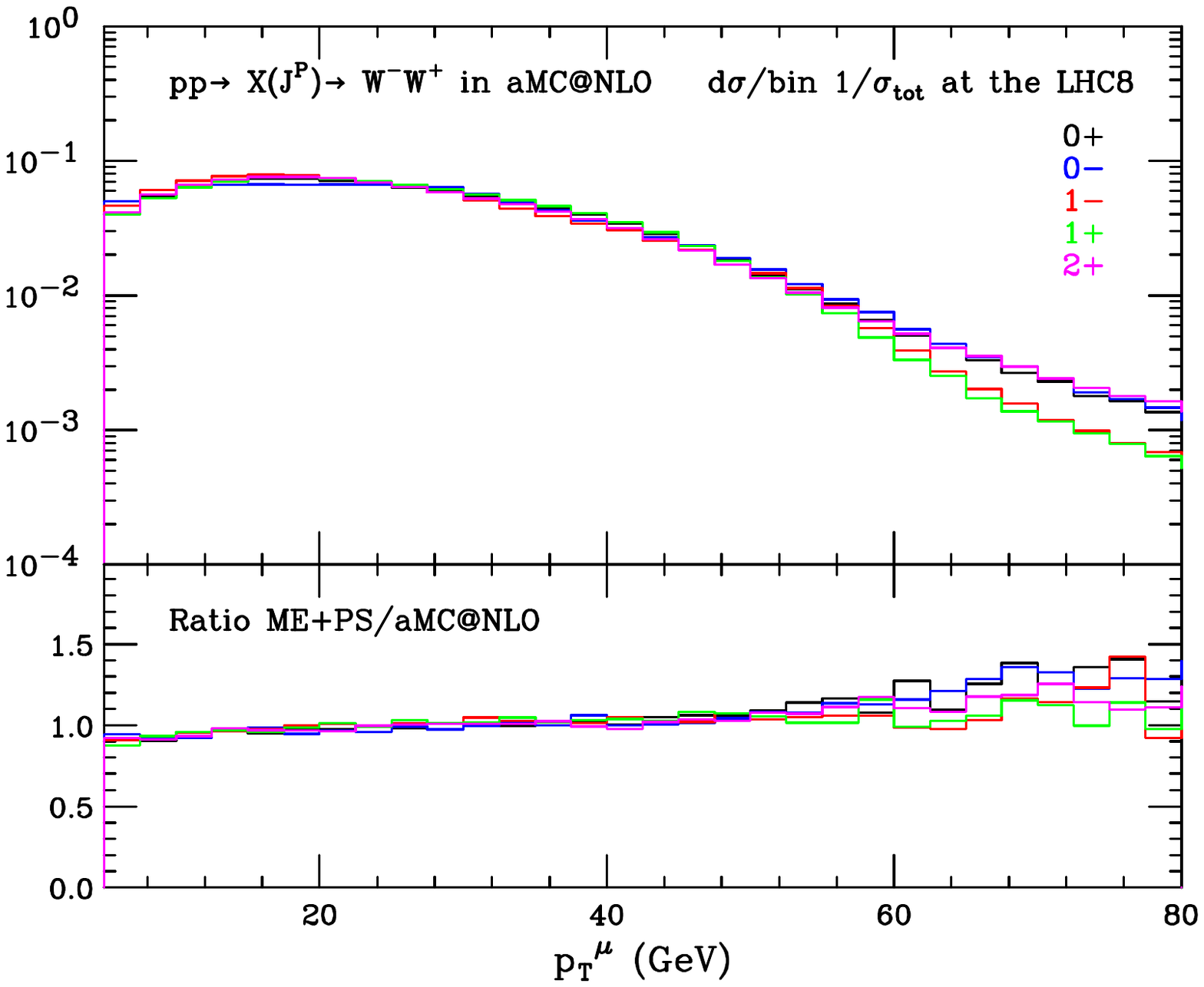} 
 \includegraphics[width=0.48\textwidth,clip=true, trim = 60 190 70 190]{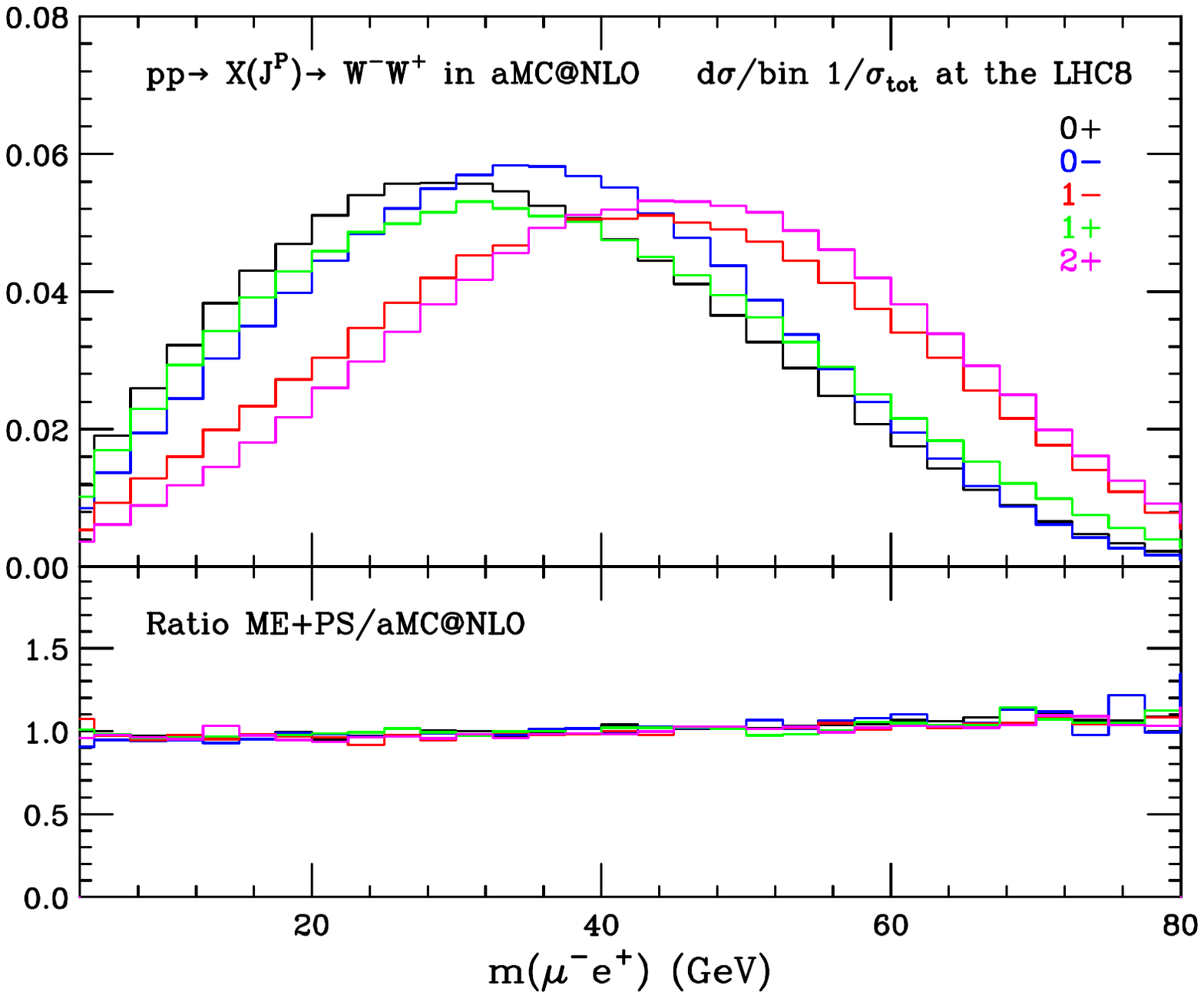} 
 \caption{Distributions of the leptons in 
 $X(\to WW^*)\to\mu^-\bar\nu_{\mu}e^+\nu_e$:  
 (a) the transverse momentum of the muon,  $p_{T}^{\mu}$, 
 (b) the invariant mass of the two leptons $m(\mu^- e^+)$.}
 \label{fig:ppww}
\end{figure}

In the context of an automated approach, taking into account of the
spin correlations relevant to the $X(J^P)$ decay products simply amounts
to generating the process with those decay products as final states
(the presence of $X(J^P)$ as an intermediate particle can also be imposed).
From the general discussion given above, it should be clear that this
is always feasible in the case of the ME+PS approach, while {\sc aMC@NLO}
may be limited by the availability of the one-loop matrix elements.
However, spin 0 is obviously a trivial case (a spinless particle does not
induce spin correlations). On the other hand, in the spin-1 and spin-2 cases 
the spin-correlated virtuals have been calculated; this is rather easy
to do, since their expressions factorise the underlying Born matrix elements.
We have then compared many key distributions as predicted by ME+PS
and {\sc aMC@NLO}, and have always found a satisfactory agreement.  
For the sake of illustration we show in figs.~\ref{fig:ppaa}, \ref{fig:ppzz}
and~\ref{fig:ppww} the results for a few selected final states of special
interest, i.e. $X\to\gamma\gamma$, $X(\to ZZ^*)\to 4\ell$, and 
$X(\to WW^*)\to 2\ell 2\nu$. We have imposed minimal acceptance cuts 
on the photons and charged leptons, namely:
\begin{align}
 p_T^{\gamma,\ell}>5~{\rm GeV}\,,\quad
 |\eta^{\gamma,\ell}|<2.5\,.
\end{align}
The $\gamma\gamma$-case plots (fig.~\ref{fig:ppaa}) suggest that 
a good discriminating power between the spin-0 and spin-2 cases can be 
obtained from the $p_T$ distributions.
Figure~\ref{fig:ppzz} illustrates the different shapes in $p_T$ and invariant
mass of the two reconstructed $Z'$s ($Z_1$ being the one with the largest
invariant mass) for the different spin and parity hypotheses. As already noted
in the literature~\cite{Choi:2002jk,Bolognesi:2012mm} the lowest pair
invariant mass is particularly sensitive to both spin and parity 
assignments.  Finally, the transverse momentum of one of the charged leptons
and the invariant mass distributions of the two charged
leptons in the $WW^*$ channel are shown in fig.~\ref{fig:ppww}.
The lepton $p_T$ distribution is
sensitive to initial state radiation and it is harder at large $p_T$'s in the
case of the spin-0 and spin-2 hypotheses, reflecting the different $p_T^X$ 
shapes of such cases w.r.t.\ that resulting from $X(1^\pm)$ production.

The overall agreement between the predictions of ME+PS and {\sc aMC@NLO}
is rather good for all those observables that are not sensitive to hard 
radiation of at least two extra partons with respect to the Born kinematics. 
Other visible differences are mostly related to the harder $p_T^X$ spectra of
the ME+PS samples (as documented in the upper plot of fig.~\ref{fig:xpt}), 
which result in the enhancement in $p_T^{\gamma_2}$ (and to less extent also
in $p_T^{\gamma_1}$) above the kinematic threshold $m_X/2$ (see the first two
plots of fig.~\ref{fig:ppaa}), and in $p_T^{Z_{1,2}}$ as well (see the first
two plots of fig.~\ref{fig:ppzz}).

\section{Applications}\label{sec:apps}

\subsection{Unitarity-violating behaviour of models with a spin-2 state}
\label{sec:unitarity}

In this section we discuss the behaviour of a spin-2 state with non-universal
couplings to SM particles, i.e. with different $\kappa_i$ in the ${\cal L}_2$
lagrangian (in other words, eq.~(\ref{eq:RSspin2}) does {\em not} hold
here). The interest for this case comes from the fact that a model that
features an RS-graviton with a mass of 125 GeV and universal couplings
has been already excluded at the 
Tevatron~\cite{Abazov:2010xh,Aaltonen:2010cf,Aaltonen:2011xp}. In
addition, the current measured 
branching ratios and cross sections impose a very clear pattern in the values
of couplings~\cite{Ellis:2012jv,Ellis:2012mj,Ellis:2013ywa}. It is therefore
important to investigate the effects of setting the couplings to non-equal
values (non-universal scenario), in particular for what concerns the stability
of the effective field theory with respect to higher order corrections.  The
first important point to realize is that couplings can be changed without
breaking any of the gauge symmetries of the SM, as one can explicitly check by
inspecting the E-M tensor for QED (eqs.~(\ref{EMf}) and~(\ref{EMa})), which is
manifestly invariant under a gauge transformation of the fermion and $A^\mu$
fields. In so doing, however, the spin-2 current is not conserved anymore,
which can also be easily checked. In the case at hand, i.e.~of a theory with a
massive spin-2 state, this poses no problem of principle. It has, on the other
hand, important effects in the behaviour of the scattering amplitudes at high
energy, as we shall now explicitly show.

As was already mentioned, in the case of an RS graviton (universal
couplings) the LO cross section is dominated by the $gg$ production
channel (96\% vs.  4\% due to the $q\bar{q}$ contribution). It 
is tempting to explore the case where this hierarchy is 
inverted, by tuning the parameters $\kappa_q$ and $\kappa_g$ that enter 
in the couplings of the graviton with the E-M tensor of quarks and gluons:
\begin{align}
 {\cal L} = -\frac{1}{\Lambda}\kappa_q T^q_{\mu \nu} X^{\mu \nu}_2
            -\frac{1}{\Lambda}\kappa_g T^g_{\mu \nu} X^{\mu \nu}_2\,.
\end{align}
Note that while $T^g_{\mu\nu}$ contains only gauge fields, the
first term $T^q_{\mu\nu}$ involves a coupling of fermionic fields with the
gauge field through the covariant derivative.  $T^q_{\mu\nu}$ and
$T^g_{\mu\nu}$ are separately $SU(3)_C$ gauge invariant.

At the NLO, Born and virtual $2 \to 1$ and real $2 \to 2$ contributions need
to be taken into account.  As it has been already noted in several papers
(see e.g.~\cite{Mathews:2004xp,Kumar:2008pk,Karg:2009xk,Frederix:2012dp}), when
$\kappa_q=\kappa_g$ all the UV divergences present in the intermediate stages
of an NLO calculation cancel with the standard UV counterterms, and no 
additional overall renormalisation is required. This property is a 
consequence of the conservation of the E-M tensor. For non-universal 
couplings this is not the case anymore: 
UV divergences appear and therefore loop amplitudes need to be renormalised.
The details of this procedure are 
given in appendix~\ref{app:ren}.
As far as the real emission contributions are concerned, they are associated
with the processes $g g \to X_2 g$ and $q \bar q \to X_2 g$ (plus their
crossings).  The $gg \to X_2 g$ amplitude depends only on $\kappa_g$, and
therefore there is no impact on this amplitude from the non-conservation of
the spin-2 current.  On the other hand, a unitarity-violating behaviour stems
from the $q \bar q \to X_2 g$ amplitude (and its crossings). In fact, this
amplitude contains three diagrams proportional to $\kappa_q$ and one
proportional to $\kappa_g$.  A calculation of such an amplitude gives:
\begin{align}
|{\cal M}|^2 = \frac{N}{\Lambda^2 s\, t\, u \,m^4} 
 &\Big\{  3 \kappa_g^2 m^4  \left[2 m^4-2 m^2 (t+u)+t^2+u^2\right] 
           \left[m^4-m^2   (t+u)+4 t u\right]  \nonumber \\
 &+ (\kappa_q-\kappa_g) \, 6 \, \kappa_g  m^4  s 
           \left[m^6+m^2 s (s+2 u)-2 s u (s+u)\right]\nonumber \\
 &+ (\kappa_q-\kappa_g)^2  s 
   \big[ 6 m^{10}-6 m^8 (t+u)+3 m^6 (t^2+u^2)-12 m^4 t u (t+u) \nonumber \\ 
 &\hspace*{1.5cm}+ 2 m^2 t u (t^2+12 t u+u^2)-2 t u (t^3+t^2 u+tu^2+u^3) 
 \big] \Big\} \,,
\label{eq:qqXg}
\end{align}
where $N$ is a dimensionless function of the couplings, and $m$ is the
mass of the $X_2$ state.  Firstly, we note that for $\kappa_q=\kappa_g$ the
expression above reduces to the well-known result for graviton production in
extra-dimension scenarios~\cite{Han:1998sg,Giudice:1998ck,Mirabelli:1998rt}.
Secondly, we stress that the soft and collinear limits of the amplitude
in eq.~(\ref{eq:qqXg}) are the same as those of the universal-coupling case,
$\kappa_q=\kappa_g$; in other words, the terms which arise when 
$\kappa_q \ne \kappa_g$ do not modify the IR behaviour w.r.t.~the RS case,
and thus the resulting divergences factorise over the corresponding 
$gg\to X_2$ or $q\bar q \to X_2$ amplitudes.

\begin{figure}
\center
 \includegraphics[width=0.75\textwidth,clip=true, trim = 100 230 60 200 ]{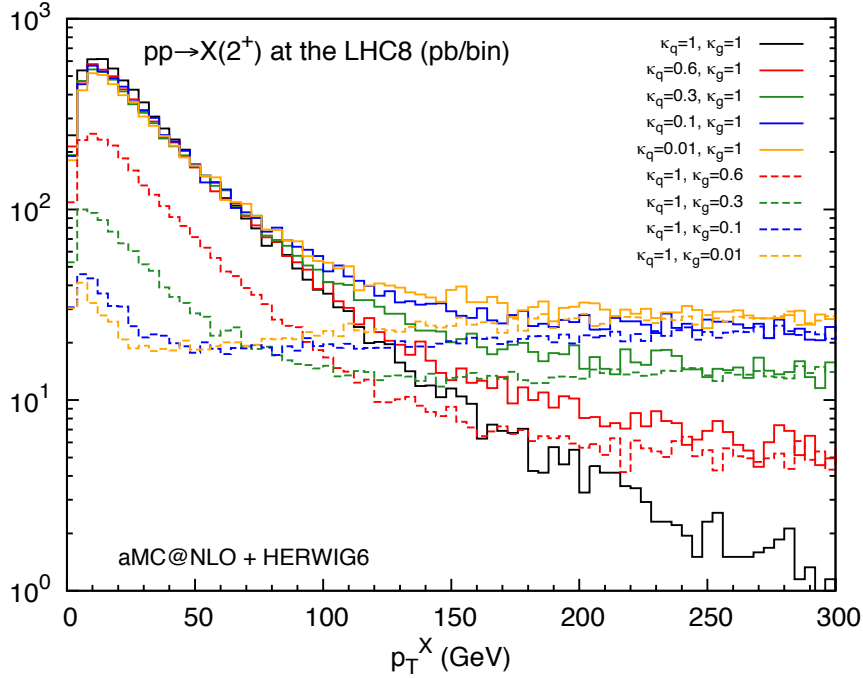} 
 \caption{The transverse momentum $p_{T}^X$ of a spin-2 state with non universal couplings to quarks and gluons $\kappa_q\neq \kappa_g$ as obtained from {\sc aMC@NLO}. }
 \label{fig:xpt2}
\end{figure}

For $\kappa_q=\kappa_g$ the amplitude in eq.~(\ref{eq:qqXg}) grows with energy
as $s/\Lambda^2$, i.e. with the scaling expected from a dimension-five
interaction.  On the other hand, the term proportional to
$(\kappa_q-\kappa_g)^2$ grows as fast as $s^3/m^4\Lambda^2$. This term can be
traced back to the longitudinal parts of the graviton polarisation tensor that
decouple only when the graviton current is conserved, i.e. when
$\kappa_q=\kappa_g$. It is easy to verify that with the couplings needed to
reproduce the Higgs signal at the LHC energy the amplitude does not yet
violate the unitarity bound, even though the $p_T^X$ spectrum is significantly
affected.  Such a growth is not present in the $2\to 1$ amplitudes simply
because at the leading order the two contributions $q \bar q \to X_2$ and $gg
\to X_2$ are completely independent. In order to study these effects in a
consistent way, we have extended the NLO calculation of
ref.~\cite{Frederix:2012dp} to the non-universal case and have implemented it
in {\sc aMC@NLO}. As a striking example of the non-universality effects on the
spin-2 production, we display in fig.~\ref{fig:xpt2} the  $p_T^X$
distributions of the spin-2 state for various choices of the quark/gluon
couplings. The rather flat tails in several of the distributions are an
evident sign of the increased unitarity-violating behaviour of the scenarios
with $\kappa_q\neq \kappa_g$. We note that the cases where one assumes
that the spin-2 state is being produced either in the $gg$- or in the
$q\bar q$-initiated process give very different results w.r.t.~those of 
the RS graviton scenario.  
As a further confirmation that the unitarity-violating behaviour is induced by
short-distance cross sections with at least one final-state QCD parton, we
have verified that the spectra obtained with ME+PS display the same behaviour
as those of {\sc aMC@NLO} shown in fig.~\ref{fig:xpt2}, and in particular that
yet higher parton multiplicities do not alter significantly the
unitarity-violating behaviour of the $2\to 2$ amplitudes.

\subsection{Higher order QCD effects on spin observables for a spin-2 state}

A generally interesting question is that of whether higher-order (QCD)
corrections have a sizable impact on observables constructed to be particularly
sensitive to spin-correlation effects. The expectation that they do not,
owing to the fact that kinematics effects such as the recoil of the
primary system against QCD radiation largely factor out in spin-correlation
observables, may simply be too naive. One must in fact account for
the possibility that matrix elements with larger (than Born) final-state
multiplicities give rise to new helicity configurations that may significantly
affect spin correlations.

In this section, we address this question specifically for 
Higgs production. The case of spin-0 state is trivial; we have just 
used it in order to check that our observables are correctly defined. 
On the other hand, the spin-1 case is a possibly interesting one. However, 
the effects we aim at studying can hardly be seen in inclusive production,
since $X_1$ is dominantly produced through the $q\bar q$ channel, and only 
the tiny mass of the initial state quarks and the virtuality of the
gluon initiated quarks can generate 
the helicity-zero state. Some effects could be visible in subdominant 
production mechanisms, such as VBF, $V X_1$ or $t \bar t X_1$ associated 
production, but we shall not investigate them here. We are thus left with
the case of a spin-2 particle, which we shall deal with in the following,
by considering its $\gamma\gamma$ and $ZZ\to 4 \ell$ decay channels.
We shall present results obtained with the ME+PS approach. As was the
case of sect.~\ref{sec:hosim}, we have verified that {\sc aMC@NLO}
predictions are fairly close to those of ME+PS.

\begin{figure}[t]
\center
 \includegraphics[width=0.75\textwidth,clip=true]{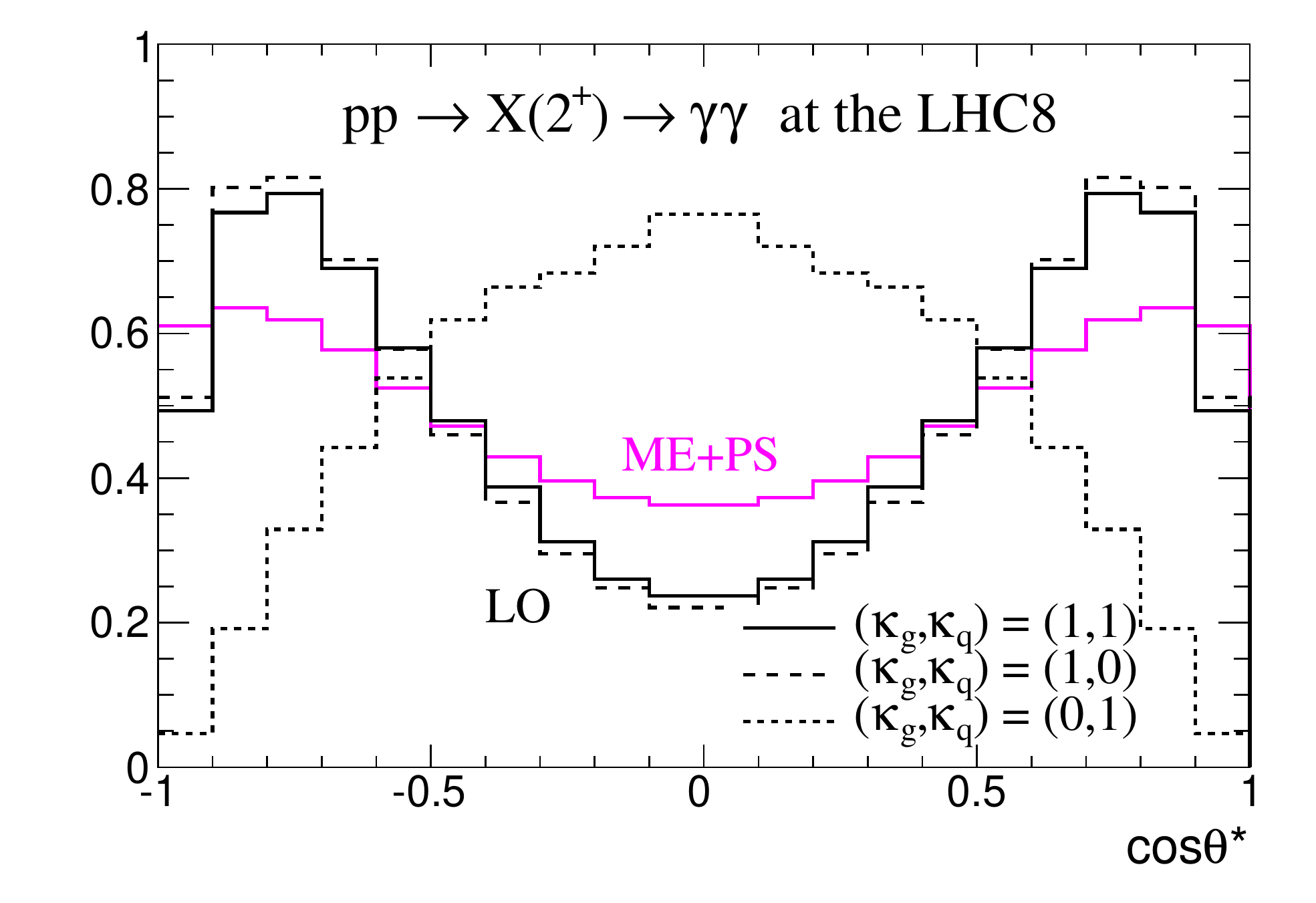}\quad
 \caption{Normalised distribution of $pp \rightarrow X_2 \rightarrow \gamma \gamma$ events with respect to $\cos \theta^\star$
resulting from different approaches: LO with $(\kappa_g,\kappa_q)=(1,1)$ (solid black line),
LO with $(\kappa_g,\kappa_q)=(1,0)$ (dashed black line)
LO with $(\kappa_g,\kappa_q)=(0,1)$ (dotted black line)
and ME+PS merging approach with $(\kappa_g,\kappa_q)=(1,1)$ (solid magenta line).}
 \label{fig:spin2mlm}
\end{figure}

\begin{figure}[t]
\center
 \includegraphics[width=0.45\textwidth,clip=true]{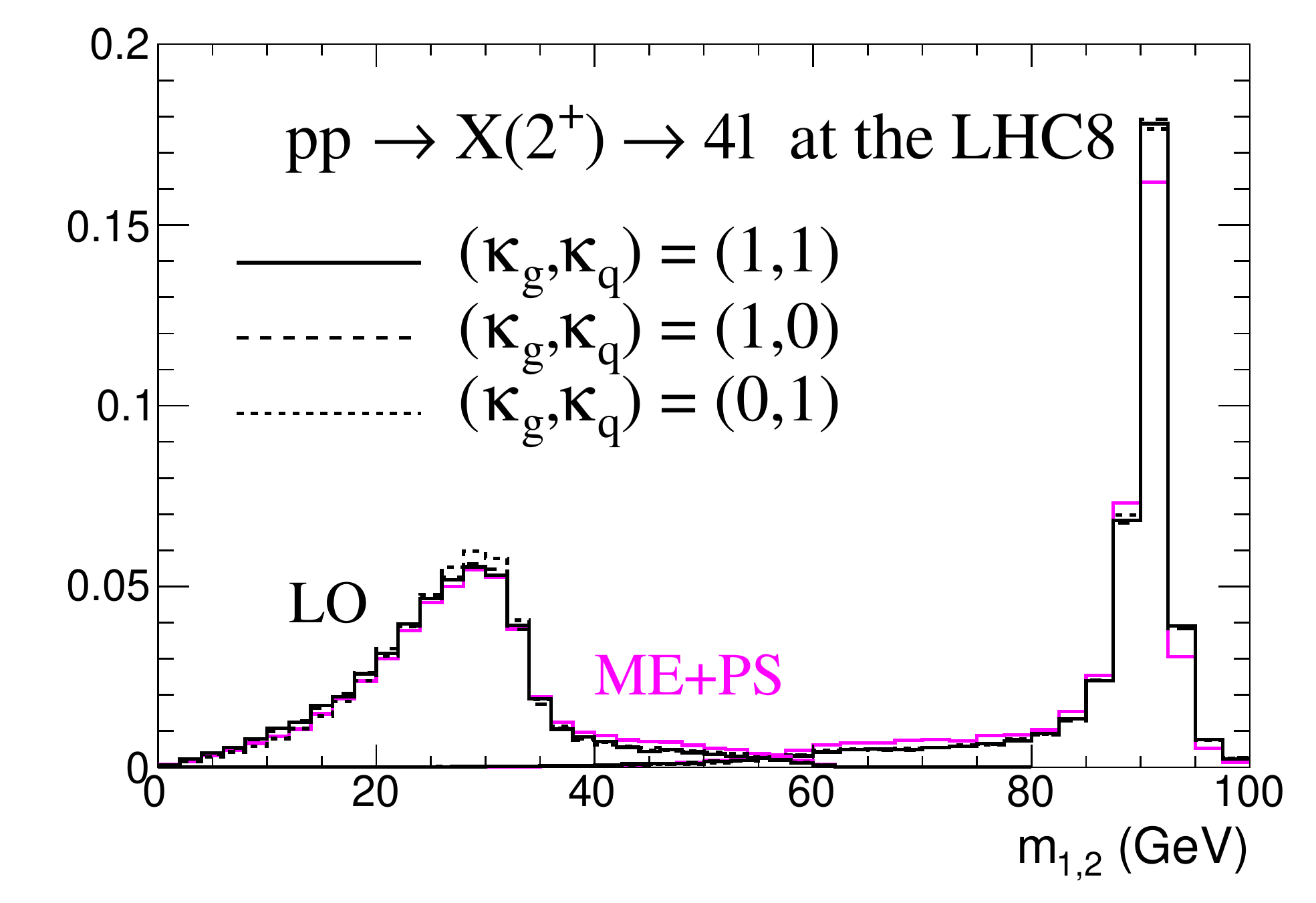}\quad
 \includegraphics[width=0.45\textwidth,clip=true]{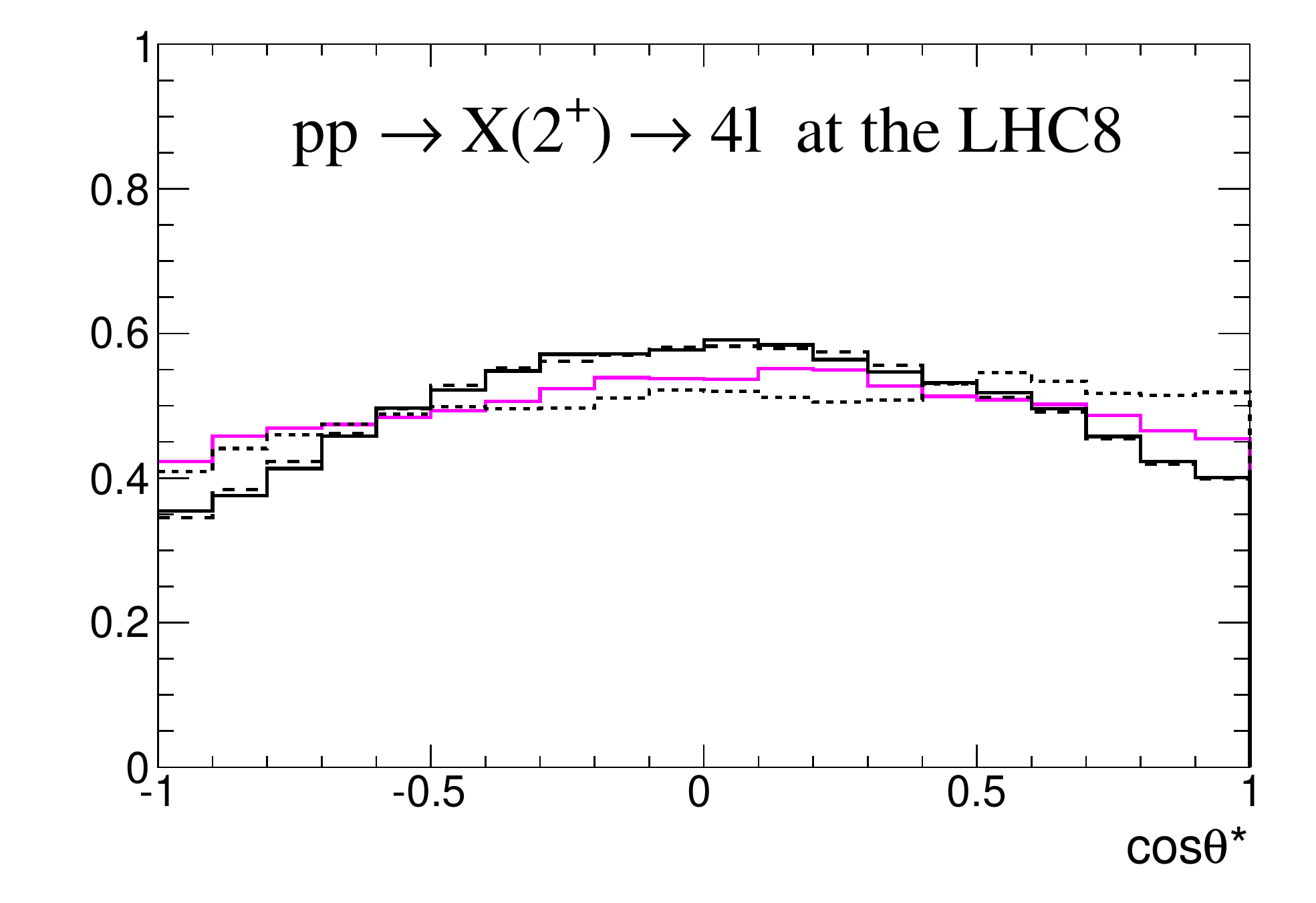} 
 \includegraphics[width=0.45\textwidth,clip=true]{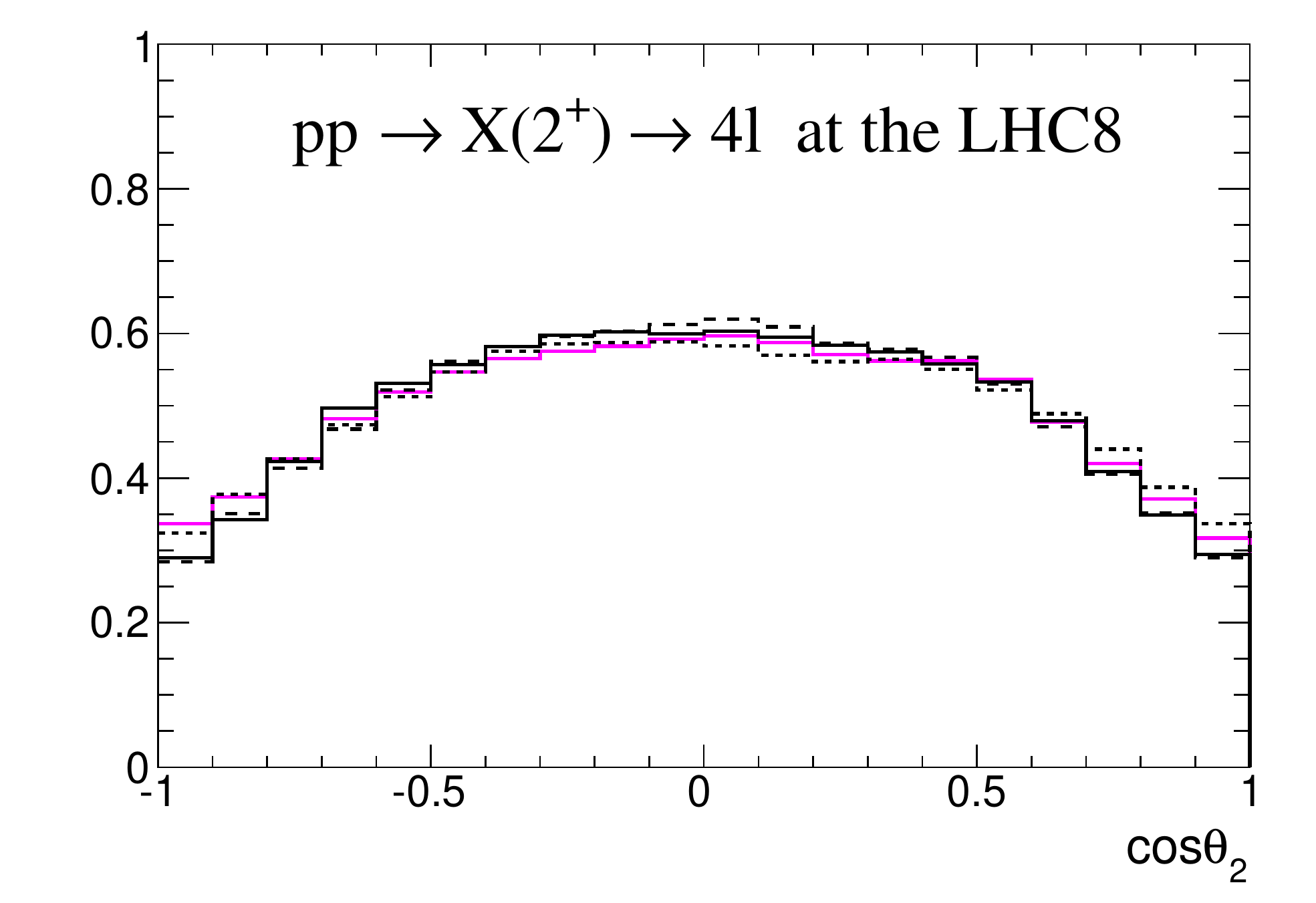}\quad
 \includegraphics[width=0.45\textwidth,clip=true]{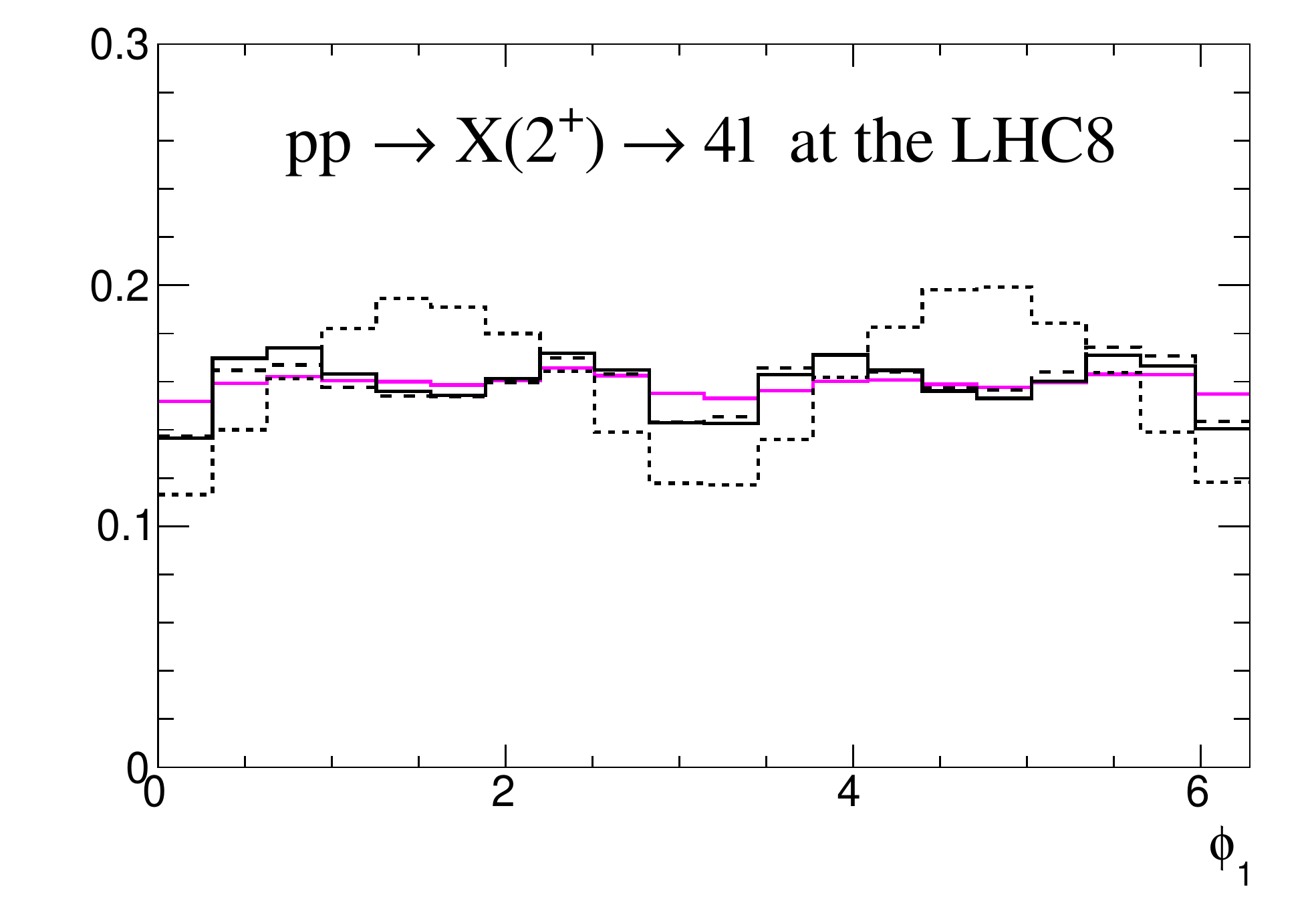}
 \caption{Normalised distributions in $pp \to X_2\to\mu^+\mu^-e^+e^-$ at
   LO and in presence of extra QCD radiation as described by a ME+PS merged
   sample. The solid curve corresponds to the $(\kappa_g,\kappa_q)=(1,1)$
   case. }
 \label{fig:spin2zzmlm}
\end{figure}

In order to introduce the argument in a simplified way, let us 
consider the production of a spin-2 boson in the universal coupling
scenario, at the Born level (i.e. without any final-state partons),
and in the partonic rest frame. In this way, the polarisation of $X_2$ 
lies along the beam axis and takes the values of $\pm 2$ ($\pm 1$) for the 
$gg$-channel ($q\bar{q}$-channel) contribution.  The distributions in the 
decay angle $\theta^*$ can be expressed in terms of some $d$ functions,
whose forms depend on the initial- and final-state particle helicities. 
The two different production modes lead to totally different $\theta^*$ 
distributions; specifically, one has:
\begin{align}
 \frac{d\sigma(gg)}{d\cos\theta^*} &\propto 
   |d^2_{22}(\theta^*)|^2+|d^2_{2-2}(\theta^*)|^2
  =\frac{1}{8}(1+6\cos^2\theta^*+\cos^4\theta^*)\,, \\
 \frac{d\sigma(q\bar q)}{d\cos\theta^*} &\propto
   |d^2_{12}(\theta^*)|^2+|d^2_{1-2}(\theta^*)|^2
  =\frac{1}{2}(1-\cos^4\theta^*)\,.  
\end{align}
The dominance of either the $gg$ or $q\bar{q}$ channels can be clearly
seen in fig.~\ref{fig:spin2mlm} -- the former leading to enhanced cross
sections at the end points ($\cos\theta^*=\pm 1$ -- however, right on the 
end points there is a kinematical-driven depletion), which are on the
other hand associated with a suppressed production in the latter case.
Unfortunately, the clarity of this picture is blurred by the inclusion
of higher-order effects, which we present here only for the 
universal-coupling scenario. This is clearly the effect of the much
richer helicity configurations of matrix elements with larger 
multiplicities, and of the more involved parton-luminosity structure
at higher orders, whose role is therefore essential for proper 
phenomenological studies\footnote{Note that, 
in the presence of extra radiation, the angle $\theta^*$ is defined as the
angle between the momentum of $X_2$ in the laboratory frame and that of 
the photon in the $X_2$ rest frame.}.

In the case of the decay of $X_2$ to four leptons more observables can be
studied. In fig.~\ref{fig:spin2zzmlm} we show the distributions of the
invariant masses of the two lepton pairs $m_1$, $m_2$ (with $m_1>m_2$), 
$\cos\theta^*$, $\cos\theta_2$, and $\phi_1$.
While differences in the invariant mass distribution
of the lepton pairs are minor, the angle distributions, and especially 
the $\cos\theta^*$ one, are affected by higher order corrections.

\subsection{Determination of the $CP$-mixing of a spin-0 state with
the matrix element method }

In this section we illustrate how the availability of the Higgs 
Characterisation Model in {\sc FeynRules} and {\sc MadGraph\,5} opens 
the way to using advanced analysis tools such as 
{\sc MadWeight}~\cite{Artoisenet:2010cn}.

The matrix element method (MEM)~\cite{Kondo:1988yd} has been successfully
employed in the context of the Higgs boson discovery and spin
determination~\cite{Gao:2010qx,Gainer:2011xz,Stolarski:2012ps,Bolognesi:2012mm,
Alves:2012fb,Avery:2012um}.
Recently, the MEM has been used by both the ATLAS~\cite{ATLAS-CONF-2013-013}
and CMS~\cite{CMS-PAS-HIG-13-002} experiments to test the hypothesis of a
SM-like scalar boson against other possible $J^{P}$ assignments. The CMS
experiment has also considered the possibility that the coupling of the
newly-discovered resonance to the $Z$ boson is a mixture of the $CP$-even
operator $Z_\mu Z^\mu$ and the $CP$-odd operator 
$Z_{\mu \nu} \tilde{Z}^{\mu\nu}$.

We now show how, by using the MEM and its automatic implementation in 
{\sc MadWeight}, the analysis of the
properties of the new resonance can be further extended by considering a
specific example, namely the discrimination of a SM-like coupling to the $Z$
boson against the hypothesis of a coupling involving a superposition of the
higher-dimension operators $Z_{\mu \nu} {Z}^{\mu \nu}$ ($CP$-even) and $Z_{\mu
  \nu} \tilde{Z}^{\mu \nu}$ ($CP$-odd) (see the fourth line of 
eq.~(\ref{eq:0vv})). For the sake of illustration, we
consider here only a simplified analysis by: i) neglecting the presence of
background events; ii) neglecting any resolution effects associated with the
reconstruction of the leptons; and iii) considering only the channel $X_0
\rightarrow \mu^+ \mu^- e^+ e^- $.  We stress, however, that our approach and
techniques are general enough to allow one to perform more complete studies, 
including background and resolution effects.
Samples of events at $\sqrt{s}=8$~TeV are generated with the ME+PS approach that was presented in
sect.~\ref{sec:hosim}. We select events where each of the four leptons
satisfies $p_T > 7$ GeV and $|\eta|<2.4$.  We generate twelve samples of
$3\times 10^4$
events with different coupling parameters: the first sample is generated with
$\kappa_{SM}=c_\alpha=1$, $\kappa_{\sss HZZ}=\kappa_{\sss AZZ}=0$, and
corresponds to the SM (referred to as the SM hypothesis hereafter),
whereas the eleven other samples are generated with $\kappa_{SM}=0$,
$\kappa_{\sss HZZ}=\kappa_{\sss AZZ}=1$ and with $c_\alpha$ ranging from 0 to
1 in steps of 0.1. They correspond to the assumption that the yield originates
from the contribution of higher-dimension operators with a parity-mixing
parameter $c_\alpha$ (referred to as the HD($c_\alpha$) hypothesis
hereafter). All events in the twelve samples are passed to {\sc
  MadWeight}~\cite{Artoisenet:2010cn} for the automatic evaluation of the
weights.

\begin{figure}
\center
\includegraphics[width=0.45\textwidth,clip=true]{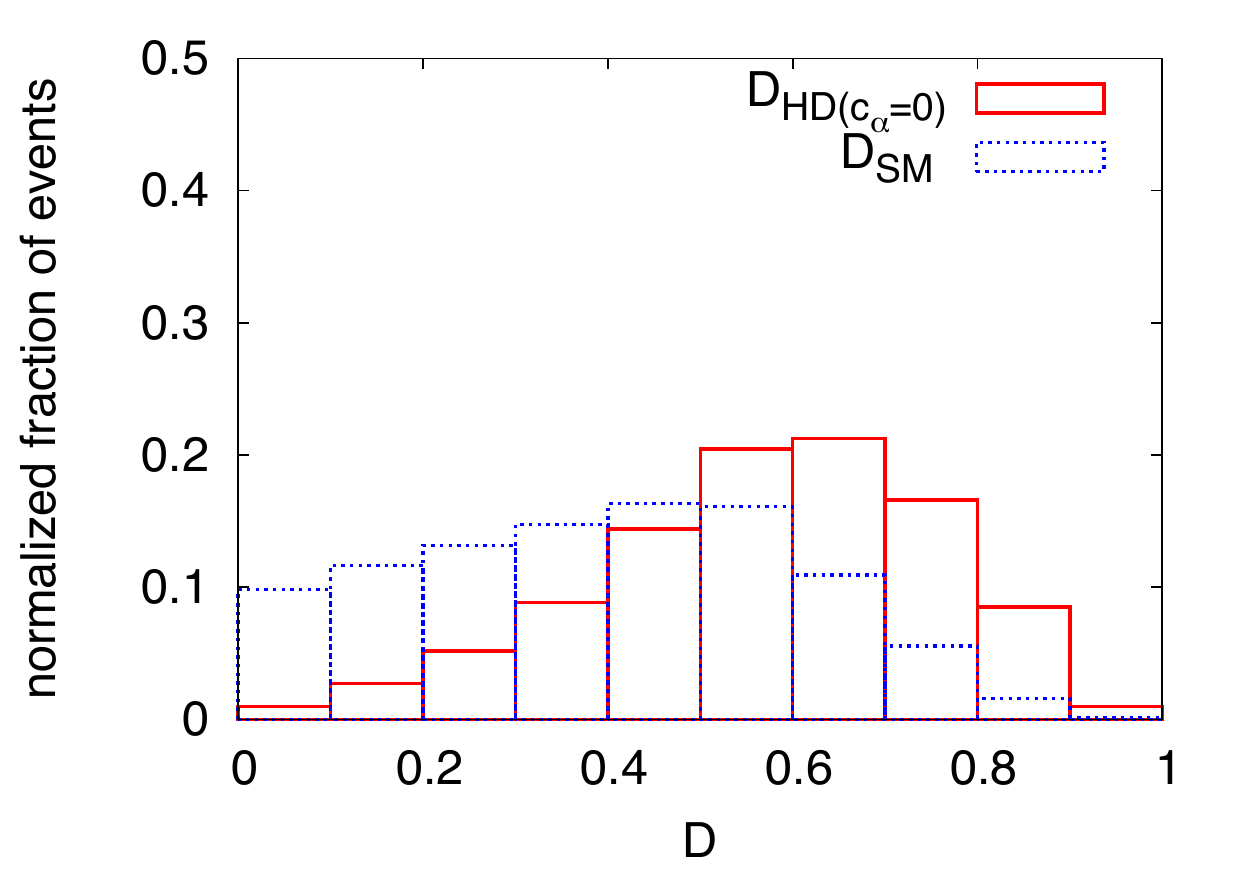}\quad
\includegraphics[width=0.45\textwidth,clip=true]{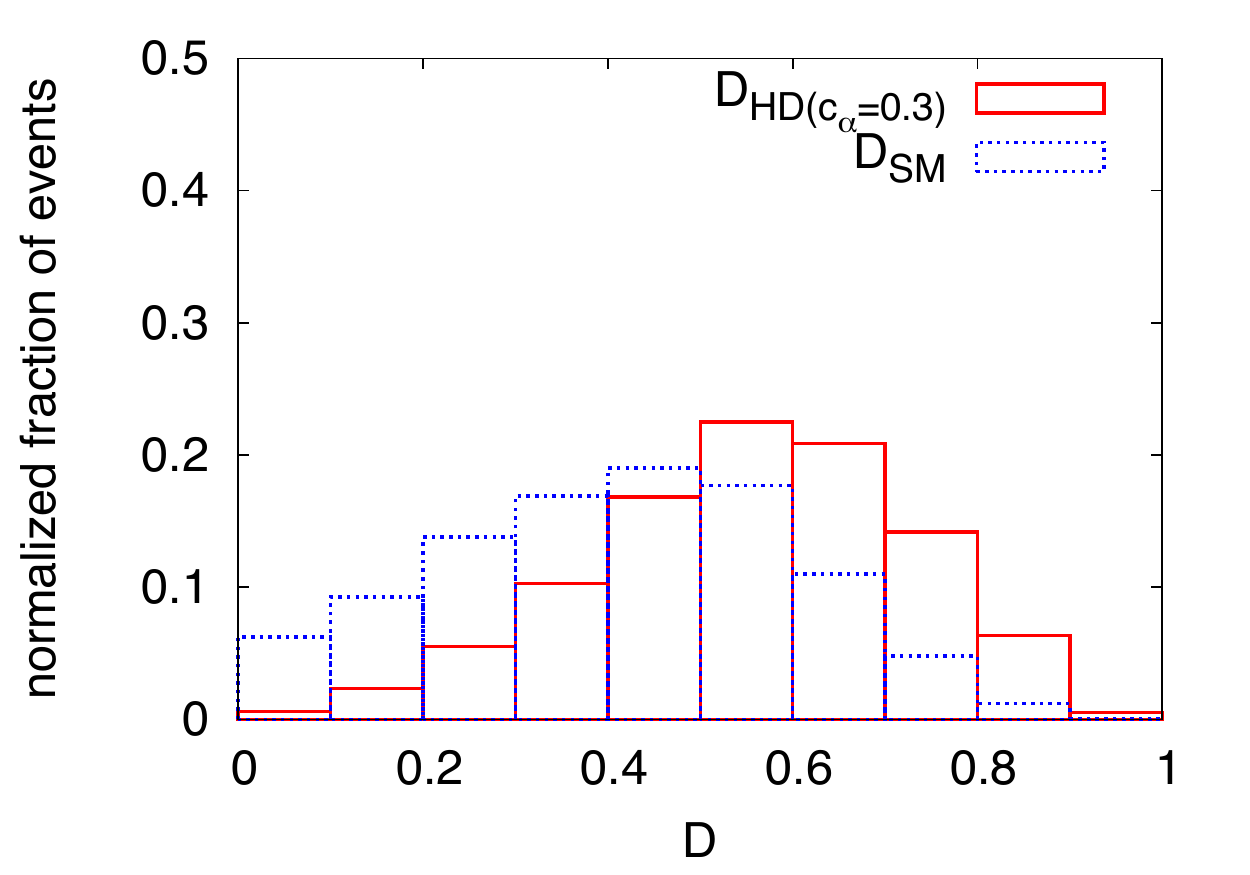}\\
\includegraphics[width=0.45\textwidth,clip=true]{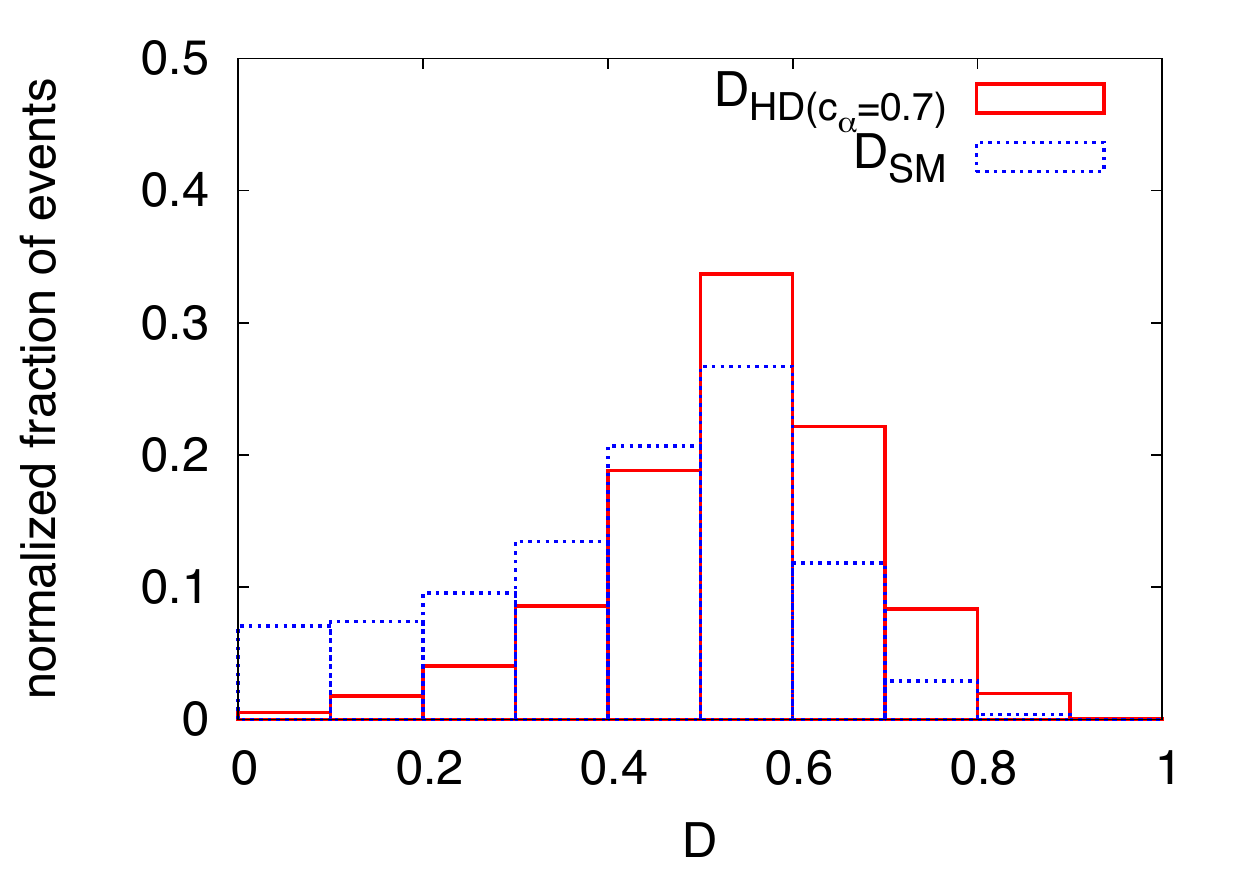}\quad
\includegraphics[width=0.45\textwidth,clip=true]{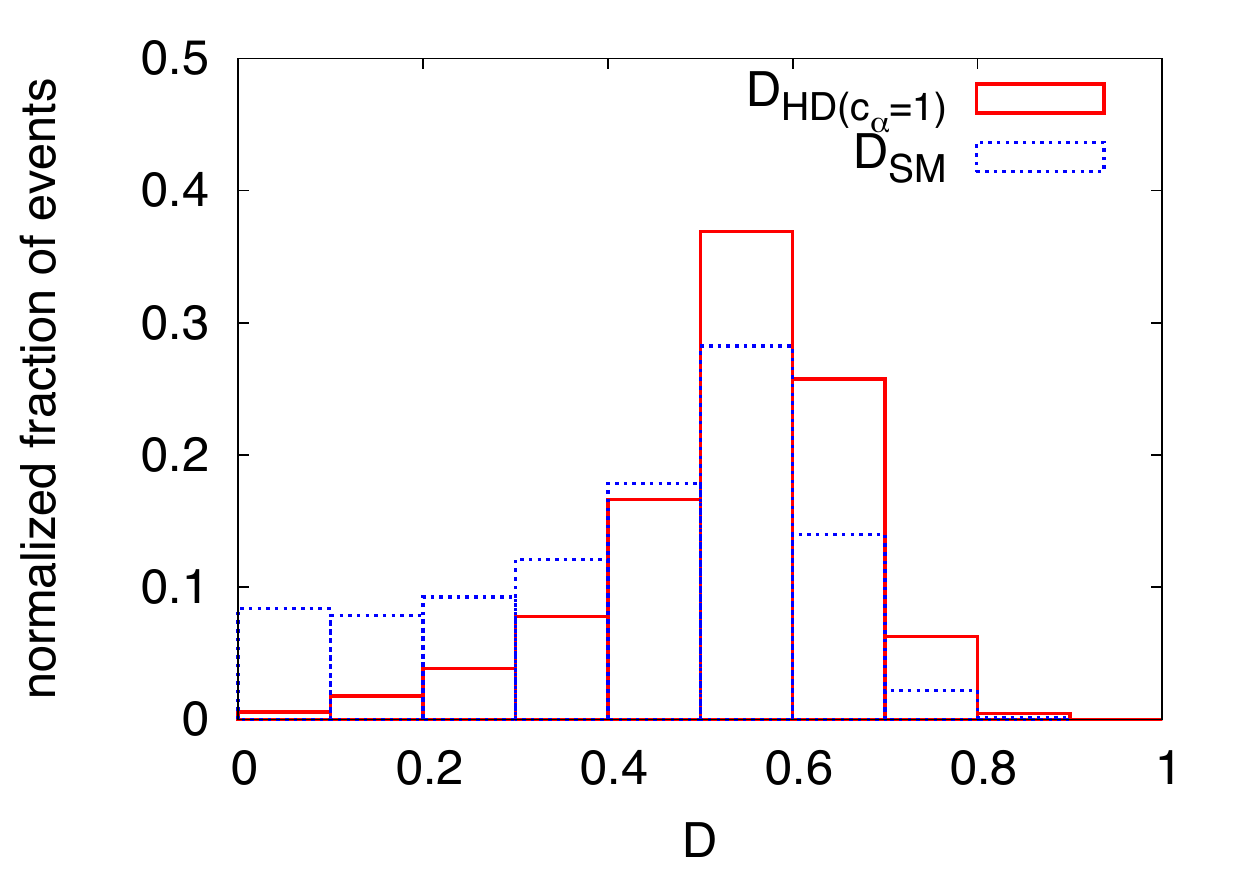}
\caption{Normalised distributions per event with respect to the MEM-based discriminant $D$, 
for specific values of the mixing parameter of $c_\alpha$: 0, 0.3, 0.7 and 1.0. }
\label{discriminant}
\end{figure}

Following the approach of ref.~\cite{Artoisenet:2013vfa}, for a generic event
$i$ with kinematics $\bs x_i$ the MEM-based observable $D_i$ for testing the
SM against the HD($c_\alpha$) hypotheses is evaluated as follows:
\begin{align}
D_i=\frac{P[\bs x_i|\textrm{HD}(c_\alpha)]}{P[\bs x_i|\textrm{HD}(c_\alpha)]+
P[\bs x_i |\textrm{SM}]}\,.
\end{align}
Expected (normalised) distributions of SM and HD$(c_\alpha)$ events in this 
observable are denoted by  $D_{\textrm{SM}}$ and $D_{\textrm{HD}(c_\alpha)}$, 
and are shown in fig.~\ref{discriminant} for some specific values 
of $c_\alpha$.

In order to assess the significance that can be achieved at the LHC
to reject the hypothesis HD($c_\alpha$) if the SM
hypothesis is realised, we consider a large number of pseudo-experiments, each
with a given number $N$ of $X_0 \rightarrow \mu^+ \mu^- e^+ e^- $ events.  We
set $N=10$, which is close to the number of events (in the SM hypothesis)
expected to be reconstructed in the ATLAS~\cite{ATLAS-CONF-2013-013} and
CMS~\cite{CMS-PAS-HIG-13-002} detectors at $\sqrt{s}=8$~TeV, in this specific decay
channel and with an integrated luminosity of 20 fb$^{-1}$.  For each event,
the corresponding $D_i$ value is generated according to the probability law
$D_{\textrm{SM}}$ (in the case of a SM pseudo-experiment) or
$D_{\textrm{HD}(c_\alpha)}$ (in the case of a $\textrm{HD}(c_\alpha)$
pseudo-experiment) which are shown in fig.~\ref{discriminant}.  This procedure
is used to generate $10^6$ pseudo-experiments under each hypothesis, SM or
$\textrm{HD}(c_\alpha)$.

\begin{figure}
\center
\includegraphics[width=0.45\textwidth,clip=true]{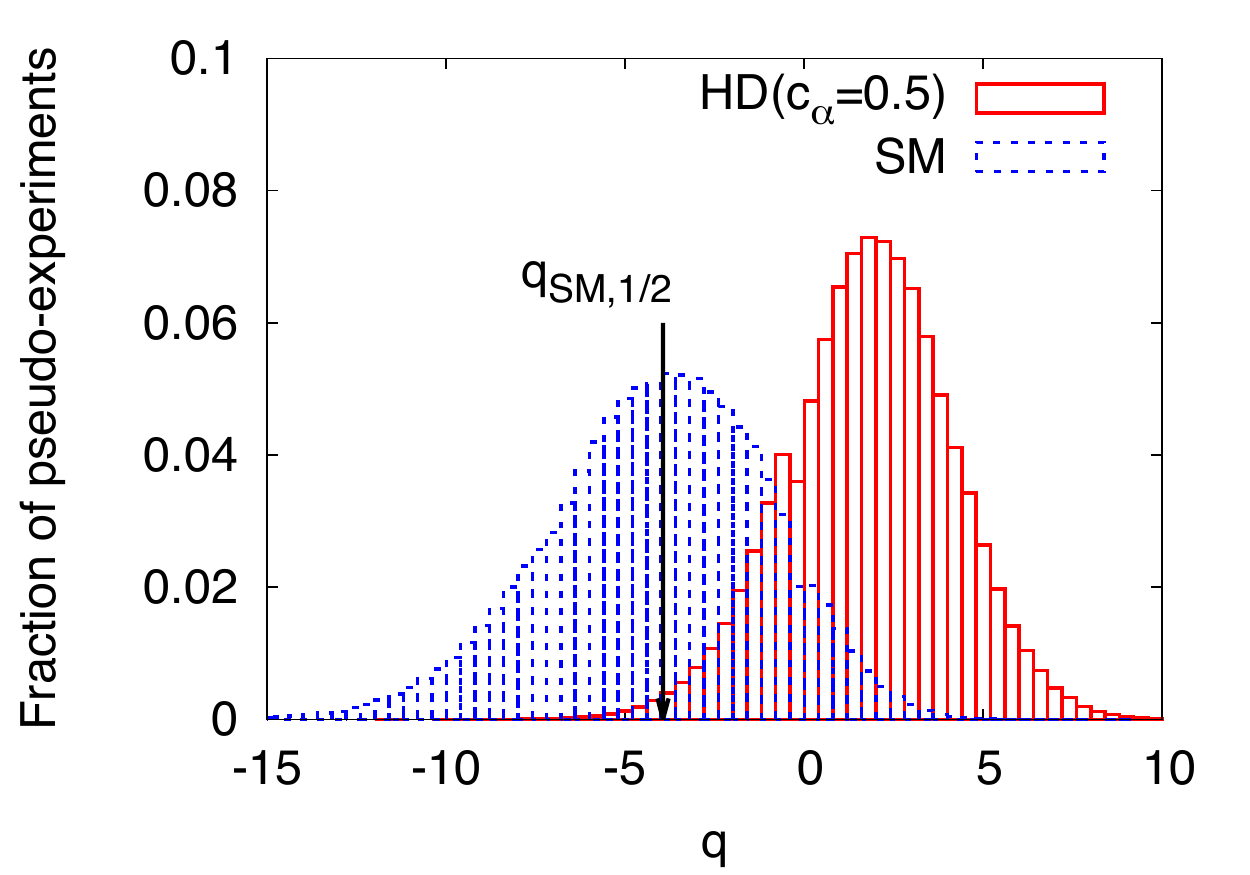}\quad
\includegraphics[width=0.48\textwidth,clip=true, trim = 0 16 0 5]{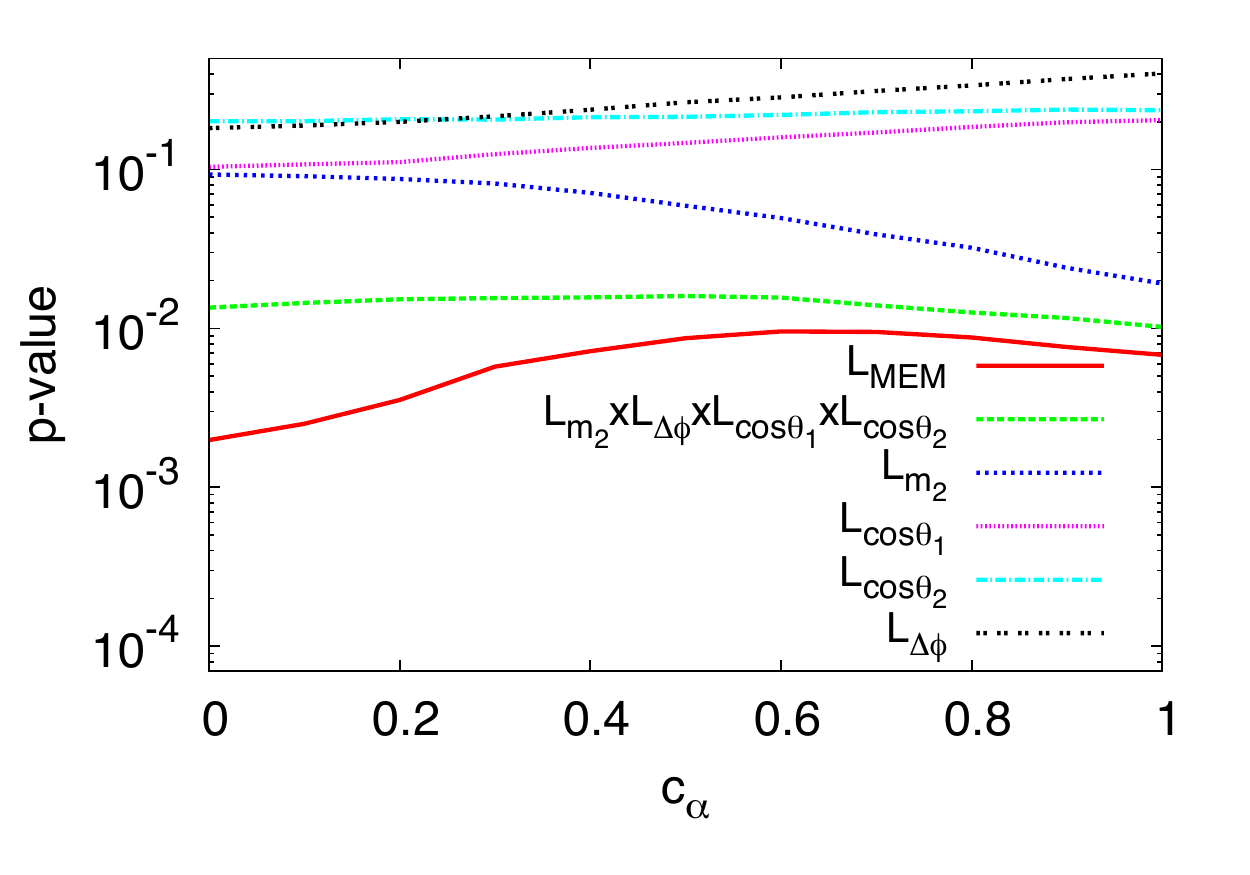}
\caption{Left: distributions of pseudo-experiments with respect to $q=\ln (L_{\textrm{MEM}})$
for the case of $c_\alpha = 0.5$.
Right: expected $p$-value at which hypothesis HD$({c_\alpha})$ is rejected
if hypothesis SM is realised, as a function of $c_\alpha$ and for different choices
of the likelihood function.}
\label{significance}
\end{figure}

For each pseudo-experiment the likelihood ratio
$L$ is calculated as follows:
\begin{align}
 L_{\textrm{MEM}} =  \prod_{i}^{N} \frac{P[\bs x_i| \textrm{HD}(c_\alpha)]}  
{P[\bs x_i |\textrm{SM} ]}
\label{Lshape} 
  = \prod_i^{N}   \frac{  D_i }{1-D_i}  \,.
\end{align}
The resulting SM and $\textrm{HD}(c_\alpha)$ distributions of
pseudo-experiments in $q= \ln \left( L_{\textrm{MEM}} \right)$
are shown in fig.~\ref{significance} (left) for the specific case of
$c_\alpha=0.5$. The significance is estimated by calculating the median
$q_{\textrm{SM},1/2}$ of the SM distribution and by counting the fraction of
pseudo-experiments in the $\textrm{HD}(c_\alpha)$ distribution with
$q<q_{\textrm{SM},1/2}$. Such a fraction of events provides us with an
estimate of the $p$-value associated with the statistical test for rejecting
hypothesis $\textrm{HD}(c_\alpha)$ if the SM hypothesis is realised.  The
$p$-value as a function of $c_\alpha$ is shown in fig.~\ref{significance}
(right).

The power of the MEM can be illustrated by comparing the significance that is
achieved when using the MEM-based likelihood function $L_{\textrm{MEM}}$ with
the significance resulting from a likelihood function built upon the cross
section differential in the observable $\mathcal{O}$:
\begin{align}
 L_{\mathcal{O}} =  \prod_{i}^{N} \frac{\sigma^{-1}_{ \textrm{HD}(c_\alpha)} \frac{d \sigma_{ \textrm{HD}(c_\alpha)}}{d\mathcal{O}}( \mathcal{O}_{i})  }  
{
\sigma^{-1}_{\textrm{SM}} \frac{d \sigma_{ \textrm{SM} }}{d\mathcal{O}}( \mathcal{O}_{i})
}\,.
\label{Lsingle} 
\end{align}
In this specific example, $\mathcal{O}$ is chosen in the set of spin/parity
observables $\{ m_2, \Delta \phi,$ $\cos \theta_1,$ $\cos \theta_2 \}$ defined
in ref.~\cite{Bolognesi:2012mm}.  The discriminant power of each of these four
variables taken separately can be assessed by using the same Monte Carlo
procedure as before, with $L_{\textrm{MEM}}$ replaced by
$L_{\mathcal{O}}$. The resulting $p$-values as a function of $c_\alpha$ are
also shown in fig.~\ref{significance} (right). Even when the likelihood
function is set equal to the product $ L_{m_2}\times L_{\Delta\phi} \times
L_{\cos \theta_1}\times L_{\cos \theta_2}$, one observes that the significance
is smaller than the one obtained by the MEM-based likelihood analysis,
presumably because all correlations among reconstructed variables can be kept
only in the latter case.

\section{Summary and outlook}\label{sec:summary}

The determination of the properties and interactions of the newly-discovered
boson will be one of the top priorities of the experimental and theory
communities in the forthcoming years, through which a definite answer
will be given to the question of whether this is, or is rather not, 
the Higgs boson predicted by the Standard Model.

In this paper, we have advocated the use of an effective-theory approach
as a powerful way to tackle this and related issues. We have also shown
how such an approach becomes an extremely flexible and multifaceted
tool when its lagrangian is embedded into the {\sc FeynRules} and 
{\sc MadGraph\,5} frameworks 
(through what we have called the Higgs Characterisation model),
owing to the capability of the latter to include higher-order QCD
corrections, both at the tree-, multi-parton level (ME+PS) and with
next-to-leading order accurate calculations ({\sc aMC@NLO}) matched
to parton showers. Indeed, we have found evidence of the fact that
such corrections are a very important ingredient for performing sensible
phenomenology studies.

In the spirit of an automated approach, we could only give here
a glimpse of the possibilities of the Higgs Characterisation model,
which can be fully exhausted only in the context of 
complete physics analyses such as those performed by the LHC 
experiments. In particular, we have restricted ourselves to the
case of inclusive production, and have considered two directions.
Firstly, we have validated our approach in different ways, prominent 
among which is the observation that the ME+PS and {\sc aMC@NLO} results 
are fairly consistent with each other. Secondly, we have presented three
sample applications, selected because they summarise well the flexibility
and the potential for accuracy of our approach. In particular:
{\em a)} We have shown that, in the case of the production of a
spin-2 state with non-universal couplings (i.e., the only spin-2 case
still phenomenologically viable), leading-order simulations give vastly
inadequate predictions for both rate and shapes, being in particular
unable to account for a unitarity-violating behaviour at large transverse
momenta. {\em b)} We have given examples of how higher-order QCD corrections 
can significantly affect spin-correlation variables that may help in
the discrimination of a spin-2 state from other spin hypotheses.
{\em c)} We have proven that the Higgs Characterisation model 
allows one to use effectively advanced analysis tools such as 
{\sc MadWeight}, by presenting a study on the determination of
the amount of $CP$ mixing of a spin-0 resonance based on matrix-element
methods.

Improvements or further developments of our framework could be achieved 
on two main directions. From the model point of view, we have built an effective lagrangian 
that is general enough to include all the effects coming from the (gauge-invariant) set
of dimension-six operators that affect the three-point Higgs interactions, with exactly one Higgs particle. 
One could therefore complete the effective lagrangian to include
the full set of operators which involve modifications or new four-point interactions,
including those featuring two Higgs particles. This is straightforward and work in progress. 

Regarding the possibility of accurate simulation we remark that
the automation of the ME+PS techniques employed in this paper is
complete, and thus such techniques can be used regardless of the
process and/or applications one considers. On the other hand, the
{\sc aMC@NLO} predictions have been obtained by partly using 
analytically-computed virtual contributions, due to the present
limitations in the calculations of one-loop matrix elements stemming
from a user-defined lagrangian. While we remark that, for certain
types of production mechanisms such as VBF or associated production, the
current framework is already sufficient for automatic one-loop 
computations (owing to the structures of the virtuals in such production
processes), we also point out that the outlook is quite positive, 
given the recent progress in {\sc FeynRules} which will lift the
limitations mentioned above. Among other things, this will also
provide one with the possibility of using the FxFx 
NLO-merging~\cite{Frederix:2012ps} framework, which has the advantages
of both the ME+PS and {\sc aMC@NLO} approaches.
Regardless of this near-future developments, it is important to keep
in mind that the ME+PS and {\sc aMC@NLO} results have complementary benefits,
the former being better in those corners of the phase space which receive
significant contributions from multi-leg matrix elements, while the latter
being able to give realistic estimates of perturbative uncertainties.

\section*{Acknowledgments}

In primis, we would like to thank all the members of Higgs Cross Section
Working Group for the encouragement in pursuing the work presented here. We
also thank Jean-Marc G\'erard, Christophe Grojean, Kaoru Hagiwara, Gino
Isidori, Margarete
M\"uhlleitner, Riccardo Rattazzi, Veronica Sanz, and Reisaburo Tanaka for many
stimulating discussions.

FM thanks KITP for hospitality during the last phases of this work. SF is on
leave of absence from INFN, sezione di Genova.  This work has been supported
in part by the ERC grant 291377 ``LHCtheory: Theoretical predictions and
analyses of LHC physics: advancing the precision frontier'', by the
Forschungskredit der Universit\"at Z\"urich, by the Swiss National Science
Foundation (SNF) under contract 200020-138206 and by the Research Executive
Agency (REA) of the European Union under the Grant Agreement number
PITN-GA-2010-264564 (LHCPhenoNet). This work has been realised in the
framework of the BELSPO Belgium/India collaboration project
HT\&LHC-BL/10/IN05.  The work of FM, FD, MZ is supported by the IISN
``MadGraph'' convention 4.4511.10, the IISN ``Fundamental interactions''
convention 4.4517.08.  PA is supported by a Marie Curie Intra-European
Fellowship (PIEF-GA-2011-299999 PROBE4TeVSCALE). PdA and KM are supported in
part by the Belgian Federal Science Policy Office through the Interuniversity
Attraction Pole P7/37, in part by the ``FWO-Vlaanderen" through the
project G.0114.10N, and in part by the Strategic Research Program ``High Energy
Physics" and the Research Council of the Vrije Universiteit Brussel.  SS is
supported by the Senior Reach Fellowship of UGC, New Delhi.  MKM and VR
acknowledge the support from RECAPP@HRI.

\appendix

\section{Spin-1 hypothesis and two-photon final states}
\label{app:spin1}

In this appendix we comment on the possibility that a spin-1 resonance
$X_1$ might lead to a peak structure in the
$\gamma \gamma$ invariant mass spectrum, as suggested,  for example, in ref.~\cite{Ralston:2012ye}.

The Landau-Yang theorem~\cite{Landau:1948kw,Yang:1950rg} states that a spin-1 state cannot couple to two identical massless vectors. The theorem is based on Bose and Lorentz symmetry and assumes the massive spin-1 state to be on-shell, i.e. to have only three degrees of freedom. The question
that arises, though, is whether  $gg \to X_1 \to \gamma \gamma$
scattering might occur if the particle is off-shell and, were that the case, whether the interference
with the background $gg\to \gamma \gamma$ via a box loop might still give rise to a structure in the invariant mass spectrum of the
two photons around the $X_1$ mass.

To analyse this possibility one can proceed in different ways. 
To show that the question itself is relevant and to make our argument as 
simple and concrete as possible, we consider  $gg \to Z^{(*)} \to \gamma \gamma$ scattering in the SM, where the 
$gg \to Z^{(*)}$ and $Z^{(*)}\to \gamma \gamma$ transitions happen via fermion triangle loops. 
Extending the results to the most general interactions with a generic vector state $X_1$ is straightforward.

The computation of the triangle loop is straightforward. 
First, only axial-vector coupling $\gamma^{\mu} \gamma_5 $  
between $Z$ and fermion line contributes,   due to Furry's theorem ($C$-invariance). 
Second, the diagrams are anomalous, yet the anomalies cancel  in the SM once all the contributing fermions in a generation 
are considered (including therefore the charged lepton in the $\gamma\gamma$ case). 
In fact, the cancellation is exact and the result vanishes if the masses
of the internal fermions are neglected, as normally done for the first and second generations. 
It is enough then to consider only the contributions from top, bottom and tau.  
The amplitude $Z\gamma\gamma$  via a $W$ loop is zero. 

The $g(p_1) g(p_2)\to Z(P)$ vertex, taking place via a quark loop, can be computed solving the following integral
\begin{align}
	i \mathcal{V}^{ \,\mu\alpha\beta , ab }_{gg \to Z} &( p_1, p_2 )  =  
	\frac{ 8 \pi \alpha_s m_W I_q }{v \cos \theta_W } \nn\\
 &\times\delta^{ ab } \int \frac{d^4 k}{(2\pi)^4} 
	\frac{\mathrm{Tr} \big[ \gamma^{\mu} \gamma_5 \big( \slashed{k} +m_q \big) 
	\gamma^{\alpha} \big( \slashed{k}+\slashed{p}_1 +m_q \big) 
	\gamma^{\beta} \big( \slashed{k}+\slashed{p}_1+\slashed{p}_2 +m_q \big)  \big] }{\big[k^2-m_q^2 \big]
	\big[ (k+p_1 )^2-m_q^2 \big]  \big[ ( k+p_1+p_2 )^2-m_q^2 \big] }\,.
\end{align}
We keep full dependence on the mass of the quark and we put the incoming gluons on their mass shell.
Our result, in agreement with the result in appendix~A of ref.~\cite{Kniehl:1989qu}, is
\begin{align}
\label{eq:ggZvertex}
	i \mathcal{V}^{ \,\mu\alpha\beta, ab }_{gg \to  Z} ( p_1, p_2 ) & =  
	\delta^{ ab } \frac{\alpha_s \, m_W \, I_q}{2 \pi v \cos \theta_W } 
	\bigg[ 1 - \frac{4m_q^2}{s}\, f \Big( \frac{4m_q^2}{s} \Big) \bigg]  \epsilon^{\alpha \beta \rho \sigma}\,
	\frac{2 \, p_{1 \rho} \,p_{2 \sigma} }{s} \, P^\mu\,,
\end{align}
where
\begin{align}
 f(x) = 
 \begin{cases}
 \Big[ \arcsin \big( \frac{1}{\sqrt{x}} \big) \Big]^2		\qquad & \textrm{if} \quad x \ge 1\,, \\[2mm]
 -\frac{1}{4} \Big[ \ln \Big( \frac{ 1+\sqrt{1-x} }{ 1-\sqrt{1-x} } \Big) - i \pi \Big]^2 		
				\qquad & \textrm{if} \quad x < 1\,.  \\
 \end{cases} \label{eq:effepiccoladitau}
\end{align}
Since $I_t = - I_b$, it is evident from eq.~\eqref{eq:ggZvertex} that if the fermions in an isospin doublet have the same mass,
their contributions sum to zero. 
Thus, as we previously stated, only top and bottom quarks give a non-negligible contribution to the vertex, and 
it effectively depends on  $m_t^2-m_b^2$. 
The expression for the  $Z \to \gamma\gamma$ vertex is analogous, including also the tau loop. 

The most important feature of eq.~(\ref{eq:ggZvertex}) is that the effective vertex is proportional to the $Z$ momentum 
$P^\mu$.  
The calculation of the $gg\to Z \to \gamma\gamma$ amplitude therefore entails a contribution of the type
\begin{align}
 P^{\mu}\Pi_{\mu \nu}P^\nu\,,
\end{align} 
where $\Pi_{\mu \nu} = - g_{\mu \nu} + \frac{P_{\mu}P_{\nu} }{m_{Z}^2}$
is the numerator of the $Z$ propagator in the unitary gauge.
If we contract one of the two vertices, say the one with $P^\mu$, 
with the projector $\Pi_{\mu \nu}$
we find an expression proportional to 
\begin{align}
 (s-m_Z^2) P_\nu \,, 
\end{align} 
where $s=P^2$ is the usual Mandelstam variable.
This entails that the amplitude squared for the on-shell decay (or production) $ Z \to gg$ or $ Z \to \gamma\gamma$ is zero,  
in agreement with the Landau-Yang theorem\footnote{
The original theorem holds only for the decay of a massive spin-1 state to two photons; since, however, the $Z$ colour structure
is trivial, the same theorems is valid in this particular case also for the decay to two gluons.}.

When, instead, one wants to compute the transition amplitude $gg \to Z \to \gamma \gamma$, one has to pay attention, 
because a blind application of Feynman rules leads to a wrong result. 
To illustrate this, we now proceed in the same way as in ref.~\cite{Ralston:2012ye}. 
Once contracted with $P^{\mu}P^{\nu}$, the $Z$ propagator in the unitary
gauge gives 
\begin{align}
 \frac{i s}{m_Z^2} \,  \frac{s - m_{Z}^2}{s - m_{Z}^2  + i \Gamma_{Z} m_{Z} } \,.
\end{align} 
This expression complies with the Landau-Yang theorem, as it is zero when the $Z$ is on-shell; 
however, it also displays a non-trivial structure at $s=m_Z^2$, i.e. a dip. 
It is then natural to wonder whether such contribution 
might lead to a peak or a dip-peak structure at $s=m_Z^2$ when interfered with the $gg \to \gamma \gamma$ continuous background, 
as suggested in ref.~\cite{Ralston:2012ye}. An explicit calculation, which we do not report here, shows that this is indeed the case. 

However, the derivation above is not correct, as it relies on having put a non-zero width for the $Z$ in the propagator not in a consistent way\footnote{We thank Kaoru Hagiwara for enlightening discussions on this point.}.

In fact, there is no pole at  ${s=m_Z^2}$ in the $gg \to Z \to \gamma \gamma$ amplitude, as numerator and denominator exactly cancel
when we use the propagator
\begin{align}
 P^{\mu}  \frac{i \Pi_{\mu \nu}}{s-m_Z^2}  P^\nu = \frac{i s}{m_Z^2}   
\end{align} 
and there is no need to introduce a width in the denominator in the first place.  The same result can be obtained by introducing the width in a consistent way, i.e., using the complex mass scheme and replacing $m_Z^2 \to m_Z^2 - i \Gamma_{Z} m_{Z}$  everywhere in the expression of the amplitude.

The results for the helicity amplitudes of the process therefore read
\begin{align}\label{eq:ggZaaPolarAmpl}
  \mathcal{M}_{--++}^{g g \rightarrow Z \rightarrow \gamma \gamma} & = 
  S_{--++}\, \delta^{ab}\,   
  \mathcal{M}_{gg \rightarrow Z}\, \frac{ s  }{ m_{Z}^2 }\, 
  \mathcal{M}_{Z \rightarrow \gamma \gamma} \,,
\intertext{and}
  \mathcal{M}_{++++}^{g g \rightarrow Z \rightarrow \gamma \gamma} & = 
  S_{++++} \, \delta^{ab}  \,
  \mathcal{M}_{gg \rightarrow Z} \,  \frac{ s +2t }{ m_{Z}^2 }\, 
  \mathcal{M}_{Z \rightarrow \gamma \gamma} \,,
\end{align}
where $S_{\pm\pm++}$ are spinor phases and
\begin{align}
  \mathcal{M}_{gg \rightarrow Z} &  =  
  \frac{\alpha_s \, m_W}{2 \pi v \cos \theta_W }  \sum_{q=t,b} I_q \bigg[ 1 - \frac{4 m_q^2}{s}
  f \Big( \frac{4m_q^2}{s} \Big) \bigg] \,, \label{eq:PartialAmplggZ}\\[2mm]
  \mathcal{M}_{Z \rightarrow \gamma\gamma} &  =   
    \frac{\alpha \, m_W}{\pi v \cos \theta_W } \sum_{f=t,b,\tau}
 N_c^{(f)} \, Q_f^2 \, I_f 
 \bigg[ 1 - \frac{4 m_f^2}{s} 
    f \Big( \frac{4m_f^2}{s} \Big) \bigg] \,. \label{eq:PartialAmplZaa}
\end{align}
Such amplitudes do not display any enhancement or zero at the $Z$ pole and therefore cannot lead to any peak or dip-peak structure in the $\gamma \gamma$ invariant mass spectrum around the $Z$ mass. 

It is interesting, however, to note that the amplitudes are not zero and can be interpreted as coming from
a contact  $gg\gamma\gamma$ interaction. 
An analogous result can be obtained, for example, calculating the amplitude $gg \to Z \to t \bar t$ which is
also non-vanishing, proportional to $m_t (m_t^2 -m_b^2)$ and without any structure at $s=m_Z^2$.

It is easy to see that any possible effective vertex that can be written for a generic vector $X_1$ and 
two massless identical gauge vectors, $gg$ or $\gamma \gamma$, due to the Landau-Yang theorem either gives a vanishing contribution
to $gg \to X_1 \to  \gamma \gamma$ or leads to the cancellation of the propagator, effectively leaving a $gg\to \gamma \gamma$ contact interaction. 

For example, an expression analogous to the SM one  for the $ gg \,\to\, Z $ vertex can be deduced from the dimension-six operator
\begin{align}
\label{eq:ggX1operator}
	{\cal L}_{ggZ} =  \frac{1}{\Lambda^2}  \left( \partial_\mu Z^\mu \right) 
	G^a_{\alpha\beta} \tilde G^{a, \alpha\beta} \,,
\end{align}
which makes manifest that the non-vanishing result for the amplitude  is due to the non-conservation of the neutral axial-current in the SM due to the fermion masses.

\section{Divergences in the $pp \to X_2$ computation at NLO}
\label{app:ren}

As mentioned in sect.~\ref{sec:unitarity}, 
when $\kappa_q=\kappa_g$ all the UV divergences
present at the intermediate stages of an NLO calculation cancel with the
standard counterterms, thanks to the fact that the E-M tensor is conserved.  
When $\kappa_q \ne \kappa_g$ such cancellations are not there any longer, and the two couplings need to be renormalised.
In this appendix we illustrate how the renormalisation is performed in this case. 

One starts from the renormalisation mixing matrix
$Z_{ij}$ with $i,j=q,\overline q,g$, which can be easily computed  from  
the quark and gluon contributions to $T_{\mu\nu}^{q,g}$ at one loop.
These contributions are both UV and IR divergent; studying the UV behaviour of such contributions, one can obtain the 
renormalisation matrix $Z_{ij}$. Defining
\begin{align}
 Z_{ij} = 1 + {\alpha_s (\mu_R^2)\over 4 \pi} {Z^{(1)}_{ij} \over \epsilon}\,,
\end{align}
one obtains
\begin{align}
 Z^{(1)}_{qq} = -{16\over 3}C_F\,,\quad 
 Z^{(1)}_{qg} =  {16\over 3}C_F\,,\quad
 Z^{(1)}_{gg} = -{8\over 3}n_f T_F\,,\quad 
 Z^{(1)}_{gq} =  {8\over 3}n_f T_F\,.
\end{align}
From $Z_{ij}$ it is possible to obtain the overall renormalisation constant for
the operator $T^q_{\mu\nu}+T^g_{\mu\nu}$, which, as expected, is the identity.  
Due to the operator mixing at ${\cal O}(\alpha_s)$ correction, 
the couplings $\kappa_{q,g}$ develop a
scale dependence. The coupled renormalisation group equations are controlled
by the anomalous dimension matrix defined by 
\begin{align}
\gamma = Z^{-1} \mu_R^2 {d Z\over d \mu_R^2}\,, 
\end{align}
and hence these couplings run with the scale $\mu_R$.

The information on the $Z_{ij}$ matrix can be exploited to compute the NLO corrections to the
$p p \to X_2$ process. The loop contributions will have
diagrams in which the spin-2 state couples both to quarks and gluons. In the case in which $\kappa_q \ne \kappa_g$,
UV divergences can be renormalised by using $Z_{ij}$. After renormalisation, as expected,
the resulting expressions contain only IR divergences and finite
terms.  The IR divergences are proportional to $\kappa_q^2$ and $\kappa_g^2$
separately (no $\kappa_q \kappa_g$ terms), having double and single poles
in $\epsilon$ ($D=4+\epsilon$ is the space time dimension in dimensional
regularisation).  This confirms the universality of soft and collinear
singularities of the virtual amplitudes. We find that the double and 
single pole terms contain the appropriate universal coefficients to cancel against
those coming from real emission processes and mass factorisation counterterms
hence providing a check of our  computation with $\kappa_q\neq\kappa_g$.

\bibliography{library}
\bibliographystyle{JHEP}

\end{document}